\documentclass[%
reprint,
nofootinbib,
 amsmath,amssymb,
 aps,
prc,
floatfix,
]{revtex4-2}

\usepackage{graphicx}% Include figure files
\usepackage{dcolumn}% Align table columns on decimal point
\usepackage{bm}% bold math
\usepackage{float}
\usepackage{xcolor}
\usepackage{pifont}% http://ctan.org/pkg/pifont
\usepackage{textcomp, gensymb}
\usepackage{support-caption}
\usepackage{caption}
\usepackage{subcaption}
\usepackage{hyperref}
\usepackage{physics}
\usepackage{ragged2e}

\usepackage{tikz}
\usetikzlibrary{external}
\usetikzlibrary{decorations.pathmorphing}
\usetikzlibrary{decorations.markings}
\usetikzlibrary{shapes.arrows}
\usetikzlibrary{arrows}

\usetikzlibrary{positioning,fit,calc}
\usetikzlibrary{arrows.meta}
\usetikzlibrary{calc,fadings,decorations.pathreplacing}
\usetikzlibrary{decorations.pathmorphing}

\usepackage[compat=1.1.0]{tikz-feynman}
\tikzset{external/system call={lualatex
      \tikzexternalcheckshellescape -halt-on-error -interaction=batchmode
-jobname "\image" "\texsource"}}
\newcommand{\xmark}{\ding{55}}%

\begin{document}

\preprint{APS/123-QED}

\title{Sensitivity of Double Deeply Virtual Compton Scattering observables to Generalized Parton Distributions}% Force line breaks with \\

\author{J. S. Alvarado}
\email{alvarado-galeano@ijclab.in2p3.fr}
\author{M. Hoballah}
%\email{mostafa.hoballah@ijclab.in2p3.fr}
\author{E. Voutier}%
%\email{voutier@ijclab.in2p3.fr}
\affiliation{Université Paris-Saclay, CNRS/IN2P3, IJCLab, 91405 Orsay, France       }

\date{\today}% It is always \today, today,

\begin{abstract}

Double Deeply Virtual Compton Scattering (DDVCS) is a promising channel for Generalized Parton Distribution (GPD) studies as it is a generalization of the Deeply Virtual Compton Scattering (DVCS) and Timelike Compton Scattering (TCS) processes. Contrary to DVCS and TCS, the GPD phase space accessed through DDVCS is not constrained by on-shell conditions on the incoming and outgoing photons thus allowing unrestricted GPD extraction from experimental observables. Considering polarized electron and positron beams directed to a polarized proton target, we study the sensitivity of the DDVCS cross-section asymmetries to the chiral-even proton GPDs from different model predictions. The feasibility of such measurements is further investigated in the context of the CLAS and SoLID spectrometers at the Thomas Jefferson National Accelerator Facility and the future Electron-Ion Collider at the Brookhaven National Laboratory.

% \begin{description}
% Secondary publications and information retrieval purposes.
% You may use the \texttt{description} environment to structure your abstract;
% use the optional argument of the \verb+\item+ command to give the category of each item. 
% \end{description}
\end{abstract}

\maketitle

%\tableofcontents

\section{Introduction}
The internal structure of a hadron is encoded, in the most general case, in Wigner distributions $W(x,k_{\perp},r_{\perp})$ \cite{wigner1, wigner2} describing the transverse position ($r_{\perp}$) and momentum (longitudinal $x$ and transverse $k_{\perp}$) distributions of partons. As not all degrees of freedom are relevant for studying deep exclusive processes, simpler structure functions are defined by integrating away some dependencies of the Wigner distribution. In particular, integration over the transverse momentum of partons defines Generalized Parton Distributions $GPD(x,r_{\perp})$ \cite{GPD0,GPD1,GPD2} which correlate the transverse position and longitudinal momentum of partons inside the nucleon. Consequently, GPDs provide a femto-tomography of the nucleon \cite{GPD3} and access its mechanical properties \cite{GPD4}. For instance, the orbital angular momentum contribution of partons to the hadron total spin is accessed through Ji’s sum rule \cite{GPD5}, and shear forces and pressure distribution of quarks are indirectly accessed through gravitational form factors \cite{GFF1,GFF2,GFF3}.

\begin{figure}
\includegraphics[width=0.7\linewidth]{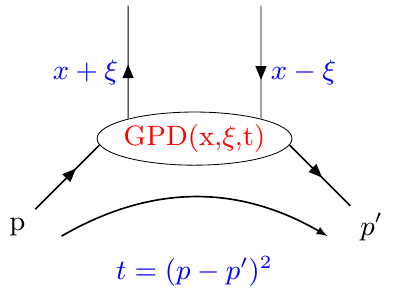}
\caption{\justifying{Parameterization of GPDs in terms of momentum fractions.}}
\label{GPDs}
\end{figure}

The kinematics of GPDs is depicted in Fig. \ref{GPDs}, where the three-dimensional dependence of GPDs is given by the fraction $x$ of the average momentum of the nucleon carried by the parton, the nucleon momentum transfer $t$, and the fraction $\xi$ of the nucleon longitudinal momentum transfer carried by the parton. Currently, direct GPD extraction is obtained from deep exclusive processes such as the electroproduction of a real photon $eN\rightarrow eN\gamma$ through Deeply Virtual Compton Scattering (DVCS) or the photoproduction of a lepton pair $\gamma N\rightarrow N\ell^{+}\ell^{-}$ through Time-like Compton Scattering (TCS). Another privileged channel for GPD measurements is the electroproduction of a lepton pair $eN\rightarrow eN\ell^{+}\ell^{-}$ through Double Deeply Virtual Compton Scattering (DDVCS) \cite{belitsky, guidal, victor}. The latter is a generalized Virtual Compton Scattering (VCS) process with DVCS and TCS as limiting cases. Such VCS processes, when probing a spin $1/2$ hadron, access the four quark helicity-conserving (chiral-even) GPDs denoted by $H$, $E$, $\widetilde{H}$ and $\widetilde{E}$ through cross-section and asymmetries depending on the spin and charge of the interacting particles.

% \begin{figure}[ht]
%     \centering
%     \includegraphics[width=0.3\textwidth]{GPD_nice.png}
%     \caption{Kinematics of a GPD dependent process.}
%     \label{GPDs}
% \end{figure}

While previous experimental studies of GPDs at HERMES \cite{HERMES}, COMPASS \cite{COMPASS} and CEBAF \cite{CEBAF}\cite{CEBAFold} focused on investigating the DVCS and DVMP processes, Jefferson Lab (JLab)\cite{CEBAF} and the future Electron-Ion Collider (EIC) \cite{EICYellow} present unique opportunities for DDVCS measurements \cite{nearfut}. On the one hand, the $\mu$CLAS detector \cite{CLAS12} with a foreseen luminosity upgrade to $10^{37} \text{cm}^{-2}\cdot\text{s}^{-1}$ \cite{LOIDDVCS} and the SoLID spectrometer \cite{solid,solid2}, complemented with a muon detector (SoLID$\mu$), will access GPDs in the valence region. On the other hand, DDVCS measurements at EIC will access the sea-quark contribution to GPDs at values as low as $x_{B}\sim 10^{-4}$ thanks to the large center-of-mass energy \cite{victor}. As the DDVCS cross-section increases at small $x_{B}$, the process might be measurable at EIC \cite{EIC, EIC1} with the expected luminosity of $10$-$100\;\mathrm{fb}^{-1}\cdot \mathrm{year}^{-1}$.

DDVCS measurements require a large integrated luminosity as the cross-section is at least 100 times smaller than DVCS/DVMP. Moreover, the DDVCS event selection has additional complications if the time-like photon decays into an $e^{+}e^{-}$ pair due to the indistinguishability of the scattered electron from the produced one. Then, data interpretation would require a complex antisymmetrization procedure considering all processes with a $e^{+}e^{-}$ final state which ultimately dilutes the DDVCS signal. As an alternative, considering muon pairs in the final state would overcome the ambiguity of the final state identification and relieve the need for antisymmetrization. 

Moreover, the physics accessible with positron beams is a subject of increasing interest in the community. In particular, positron beams let us access the real part of Compton Form Factors (CFFs) through Beam Charge Asymmetries \cite{zhaoWP}. To date, there are several feasibility studies for a positron beam at JLab \cite{positron} while positron beams and muon detection capabilities at EIC are being discussed. 

In this paper, we study the merits of a DDVCS program in terms of novel GPD information, model sensitivity and the feasibility of measuring spin-dependent observables at JLab and the EIC. In section \ref{DDVCSsec} we provide a more detailed introduction to the DDVCS process and its experimentally accessible observables. In section \ref{models}, we present the different GPD models used to predict the experimental observables. Section \ref{exps} is devoted to the main results and discussion about experimental observables at JLab and EIC configurations.

\section{The DDVCS process} \label{DDVCSsec}
\begin{figure}
\includegraphics[width=0.7\linewidth]{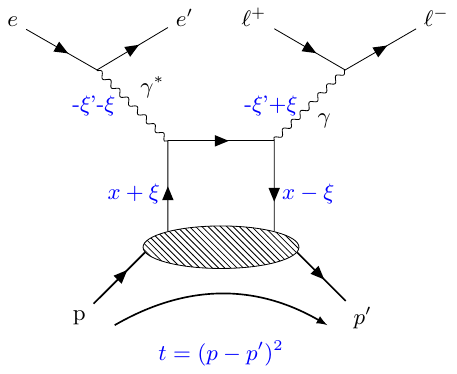}
\caption{\justifying DDVCS handbag diagram (direct term). $\xi'$ and $\xi$ are the skewness parameters defined in Eqs. \ref{xipeq} and \ref{xieq}.}
\label{DDVCS}
\end{figure}

\begin{figure}
\begin{subfigure}[b]{0.2\textwidth}
\includegraphics[width=0.7\linewidth]{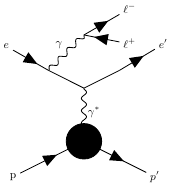}
\end{subfigure}
\begin{subfigure}[b]{0.2\textwidth}
\includegraphics[width=0.7\linewidth]{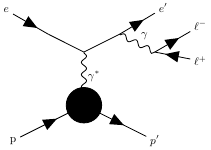}
\end{subfigure}
\begin{subfigure}[b]{0.2\textwidth}
\includegraphics[width=0.7\linewidth]{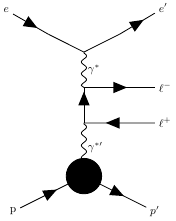}
\end{subfigure}     
\begin{subfigure}[b]{0.2\textwidth}
\includegraphics[width=0.7\linewidth]{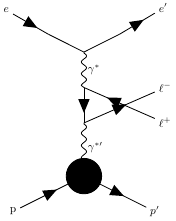}
\end{subfigure}     
\caption{\justifying Bethe-Heitler contributions to $ep\rightarrow ep\ell^{+}\ell^{-}$. Top diagrams represent initial and final state radiation processes. Bottom diagrams show a di-lepton pair production in the nuclear field.}
     \label{DDVCS-BH}
\end{figure}

The processes entering into the electroproduction of a lepton pair $ep\rightarrow ep\ell^{+}\ell^{-}$ amplitude include, among other contributions, the DDVCS process illustrated in Fig. \ref{DDVCS}. Unlike DVCS and TCS, such a process removes the on-shell condition on the final (DVCS) or initial (TCS) photon. This introduces the two virtualities $Q^{2}$ and $Q^{\prime 2}$ as kinematic parameters that can be varied independently. The DDVCS process interferes with the Bethe-Heitler mechanisms shown in Fig. \ref{DDVCS-BH}, which comprise the emission of a space-like virtual photon from the initial/final state electron followed by its decay into a lepton pair (top diagrams) and the lepton pair production from vacuum excitation (bottom diagrams).

Similar to the DVCS and TCS cases, the cross-section dependence on GPDs is expressed in terms of CFFs, given at Leading Order (LO) and twist-2 accuracy by:

\begin{eqnarray}
\mathcal{F}(\xi',\xi,t)=&&\mathcal{P}\int_{0}^1dx~F^{+}(x,\xi,t)\bigg[\frac{1}{x-\xi'}\pm \frac{1}{x+\xi'}\bigg]
\nonumber\\
&&-i\pi F^{+}(\xi',\xi,t).
\label{eq1}
\end{eqnarray}

\noindent
where $\mathcal{P}$ represents the Cauchy's principal value and $F^{+}$=$F(\xi',\xi,t)\mp F(-\xi',\xi,t)$ represents the singlet GPD combination for $F$=$H,E,\widetilde{H},\widetilde{E}$. The top and bottom signs apply, respectively, to the unpolarized GPDs ($H$, $E$) and to the polarized GPDs ($\widetilde{H}$, $\widetilde{E}$). The skewness parameters $\xi$ and $\xi'$ are given at leading twist by:

\begin{align}
    \xi'&=\frac{Q^{2} - Q^{\prime 2}}{2Q^{2}/x_{B} - Q^{2} - Q^{\prime 2} + t}, \label{xipeq}\\
    \xi&=\frac{Q^{2} + Q^{\prime 2}}{2Q^{2}/x_{B} - Q^{2} - Q^{\prime 2} + t}. \label{xieq}
\end{align}

\begin{figure}[ht]
    \centering
    \includegraphics[width=\linewidth]{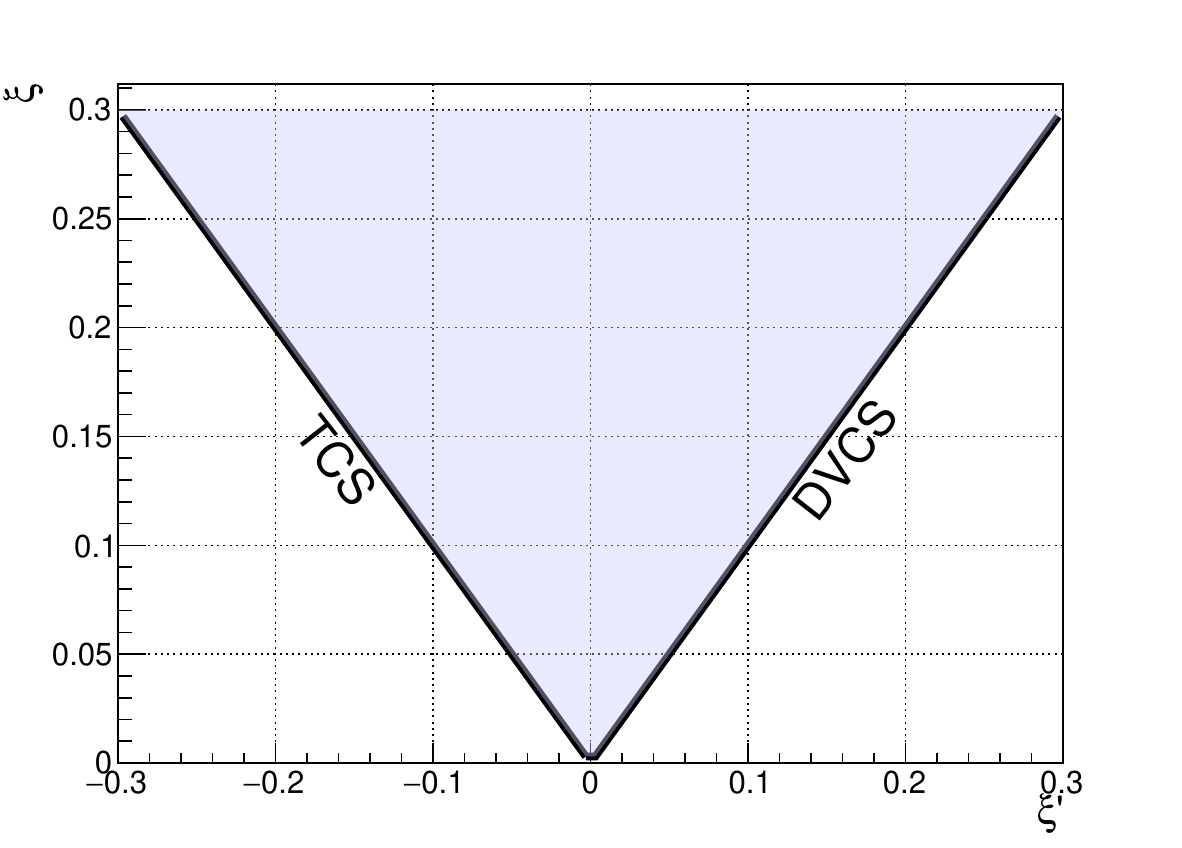}
    \caption{\justifying $\xi$ vs $\xi'$ phase space of the DDVCS process. The $\xi=\pm\xi'$ trajectories correspond to the DVCS/TCS limits.}
    \label{DDVCS-PS}
\end{figure}

The real and imaginary parts of CFFs are experimentally accessible and provide information about GPDs. Although the GPD information accessible from the real part is restricted by the convolution integral, the imaginary part offers a direct measurement of the singlet GPD combination $F^{+}$ at independent $x$=$\xi'$ and $\xi$ values, overcoming the DVCS and TCS limitation at $x$=$\pm\xi$ and allowing a general GPD exploration in the phase space shown in Fig. \ref{DDVCS-PS}.

\begin{figure}
    \centering
    \includegraphics[width=\linewidth]{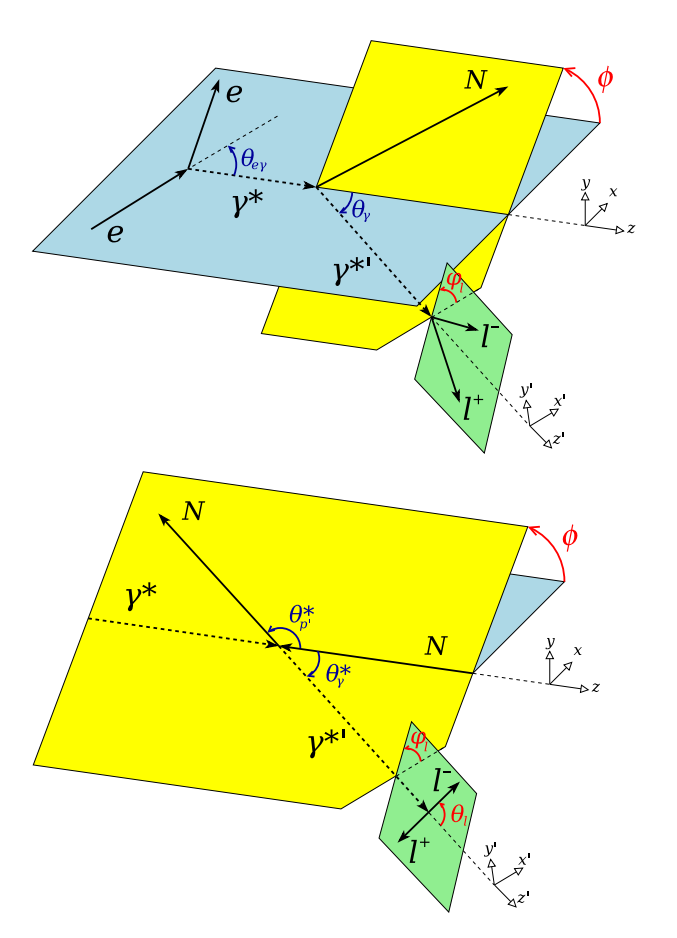}
    \caption{\justifying Top: DDVCS in the target rest frame (laboratory frame). Bottom: DDVCS in the muon center-of-mass frame. $\varphi_{\ell}$ is defined in the muon pair center-of-mass frame and $\phi$ in the target rest frame \cite{zhaoWP}.}
    \label{planes}
\end{figure}

The DDVCS 7-fold cross-section can be decomposed into contributions independent and dependent on the target polarization and the beam helicity/charge,
\begin{align}
    \frac{d^{7}\sigma}{dx_{B}dQ^{2}dQ^{\prime 2}dtd\phi d\Omega_{\ell}} &\equiv d^{7}\sigma_{P_{b},P_{t}}^{Q_{b}} \label{7fold}
\end{align}
\noindent
being $d^{7}\sigma_{P_{b},P_{t}}^{Q_{b}}$ a short-cut for the 7-fold cross-section, $P_{b}$ denotes the beam polarization, $P_{t}$ the target polarization, $Q_{b}$=$\pm 1 $ the beam charge, $\phi$ is the angle between the leptonic and hadronic planes as defined by the Trento convention \cite{trento} and $d\Omega_{\ell}$=$\sin\theta_{\ell}d\theta_{\ell}d\varphi_{\ell}$ where $\theta_{\ell}$ and $\varphi_{\ell}$ are the angles of the muon pair in their center-of-mass frame in the Berger-Diehl-Pire convention\cite{BDP}. See Fig. \ref{planes} for a sketch of the kinematic planes of the DDVCS reaction. We construct DVCS-like experimental observables in terms of the 5-fold differential cross-section $d^{5}\sigma$\cite{SYthese,positron} obtained by a specific integration over $\theta_{\ell}$ and $\varphi_{\ell}$ :

\begin{align}
    d^{5}\sigma &\equiv \frac{d^{5}\sigma}{dx_{B}dQ^{2}dQ^{\prime 2}dtd\phi} \label{d5sigma}\\
    &= \int_{0}^{2\pi}d\varphi_{\ell}\int_{\pi/4}^{3\pi/4}d\theta_{\ell} \frac{d^{7}\sigma}{dx_{B}dQ^{2}dQ^{\prime 2}dtd\phi d\Omega_{\ell}}, \nonumber
\end{align}
\noindent
where the integration over the muon-pair polar angle suppresses the BH squared amplitude as noted in \cite{belitsky}. Moreover, we construct TCS-like observables considering the 5-fold differential cross-section $d^{5}\Sigma$ obtained by integration over $\theta_{\ell}$ and $\phi$.

\begin{align}
    d^{5}\Sigma &\equiv \frac{d^{5}\Sigma}{dx_{B}dQ^{2}dQ^{\prime 2}dtd\varphi_{\ell}} \label{d5Sigma}\\
    &= \int_{0}^{2\pi}d\phi\int_{\pi/4}^{3\pi/4}d\theta_{\ell} \frac{d^{7}\sigma}{dx_{B}dQ^{2}dQ^{\prime 2}dtd\phi d\Omega_{\ell}}. \nonumber
\end{align}

Using the different contributions to the cross-section, we construct the following twist-2 observables
\begin{itemize}
    \item[\textbullet] Beam Spin Asymmetry (BSA) \\ \begin{equation}A_{LU}^{\pm}=\frac{(d^{5}\varsigma^{\pm}_{++} + d^{5}\varsigma^{\pm}_{+-}) - (d^{5}\varsigma^{\pm}_{-+} + d^{5}\varsigma^{\pm}_{--}) }{d^{5}\varsigma^{\pm}_{++} + d^{5}\varsigma^{\pm}_{+-} + d^{5}\varsigma^{\pm}_{-+} + d^{5}\varsigma^{\pm}_{--} }.\end{equation}
    \item[\textbullet] Target Spin Asymmetry (TSA) \\ \begin{equation}A_{UL}^{\pm}=\frac{(d^{5}\varsigma^{\pm}_{++} + d^{5}\varsigma^{\pm}_{-+}) - (d^{5}\varsigma^{\pm}_{+-} + d^{5}\varsigma^{\pm}_{--}) }{d^{5}\varsigma^{\pm}_{++} + d^{5}\varsigma^{\pm}_{+-} + d^{5}\varsigma^{\pm}_{-+} + d^{5}\varsigma^{\pm}_{--} }.\end{equation}
    \item[\textbullet] Double Spin Asymmetry (DSA) \\ \begin{equation}A_{LL}^{\pm}=\frac{(d^{5}\varsigma^{\pm}_{++} - d^{5}\varsigma^{\pm}_{+-}) - (d^{5}\varsigma^{\pm}_{-+} - d^{5}\varsigma^{\pm}_{--}) }{d^{5}\varsigma^{\pm}_{++} + d^{5}\varsigma^{\pm}_{+-} + d^{5}\varsigma^{\pm}_{-+} + d^{5}\varsigma^{\pm}_{--}}.\end{equation}
    \item[\textbullet] Unpolarized Beam Charge Spin Asymmetry (BCA) \\ \begin{equation}A_{UU}^{C}=\frac{d^{5}\varsigma_{UU}^{+} - d^{5}\varsigma_{UU}^{-}}{d^{5}\varsigma_{UU}^{+} + d^{5}\varsigma_{UU}^{-}}.\end{equation}
\end{itemize}

\noindent
where, $\varsigma=\sigma, \Sigma$ and the subscripts $U$ and $L$ stand for unpolarized and longitudinally polarized respectively. Besides, $d^{5}\varsigma_{UU}^{\pm}$=$(d^{5}\varsigma_{++}^{\pm} + d^{5}\varsigma_{+-}^{\pm} + d^{5}\varsigma_{-+}^{\pm} + d^{5}\varsigma_{--}^{\pm})/4$ is the unpolarized cross-section with $e^{\pm}$ beams. The dominant contribution to the asymmetry amplitude, hence driving the CFF dependence, is given by the respective $\sin\Phi$ or $\cos\Phi$ moments ($\Phi$=$\phi, \varphi_{\ell}$) \cite{belitsky}

\begin{align}
    A_{LU}^{\sin(\phi)}
     \propto & \mathfrak{Im}\bigg[\left(F_{1}\mathcal{H} - \frac{t}{4M^{2}}F_{2}\mathcal{E}\right) \label{BSA}\\
    &\quad\quad\quad\quad\quad\quad\quad\quad\quad\quad + \xi'(F_{1} + F_{2})\widetilde{\mathcal{H}}\bigg], \nonumber\\
    A_{UU}^{C\; \cos(\phi)}
 \propto& \mathfrak{Re}\bigg[\frac{\xi'}{\xi}\left(F_{1}\mathcal{H} - \frac{t}{4M^{2}}F_{2}\mathcal{E}\right)\\
    &\quad\quad\quad\quad\quad\quad\quad\quad\quad\quad  + \xi(F_{1} + F_{2})\widetilde{\mathcal{H}} \bigg], \nonumber \\
    A_{UL}^{\sin(\phi)}
 \propto& \mathfrak{Im}\left[\xi'(F_{1} + F_{2})\left(\mathcal{H} + \frac{\xi}{1+\xi}\mathcal{E} \right) \right.  \label{AULeq} \\
     &\quad\quad\quad + F_{1}\widetilde{\mathcal{H}} - \left. \xi\left(\frac{\xi}{1+\xi}F_{1} + \frac{t}{4M^{2}}F_{2}\right)\widetilde{\mathcal{E}}  \right], \nonumber\\
    A_{LL}^{\cos(\phi)}
 \propto& \mathfrak{Re}\left[ \xi(F_{1} + F_{2})\left(\mathcal{H} + \frac{\xi}{1+\xi}\mathcal{E} \right)  \right. \label{ALLeq}\\
    &\quad\quad\left.  + \frac{\xi'}{\xi}F_{1}\widetilde{\mathcal{H}} - \xi'\left(\frac{\xi}{1+\xi}F_{1} + \frac{t}{4M^{2}}F_{2}\right)\widetilde{\mathcal{E}}\right].  \nonumber  
\end{align}

Eqs. \eqref{BSA}-\eqref{ALLeq} indicate that the target polarization state determines the vector and axial contributions to the asymmetry amplitude, \textit{}{e.g.} unpolarized target asymmetries are sensitive to the $F_{1}\mathcal{H} - (t/4M^{2})F_{2}\mathcal{E}$ and $(F_{1} + F_{2})\widetilde{\mathcal{H}}$ combinations. Moreover, the $\xi'/\xi$ dependence might enhance or suppress the contributions. For instance, at fixed $\xi'$, BCA measurements at small $\xi$ suppress the $\widetilde{H}$ contribution while enhancing it at large $\xi$. Finally, the experimentally accessed CFF combination is affected by the target type. For a proton target experiment, $F_{1}$ is sizable so $\mathcal{H}$ and $\widetilde{\mathcal{H}}$ dominate the asymmetries. Whereas for a neutron target, $F_{1}\approx 0$, so unpolarized target asymmetries access $\mathcal{E}$ and $\widetilde{\mathcal{H}}$ is suppressed in longitudinally polarized target asymmetries.

\section{Quark GPD models}\label{models}
GPD models are constrained by Form Factors (FFs), Parton Distribution Functions (PDFs) and polynomiality (see \cite{GPDproperties} for a review). A commonly used approach to satisfy most of the properties is the Double Distribution (DD) modeling of GPDs \cite{DD1,DD2}, where GPDs are written as:

\begin{align}
    H&(x,\xi,t)= \label{DD} \\
    &\int_{0}^{1}dy\int_{-1+y}^{1-y}dz \; \delta(x-y-\xi z) h\left(y,z,t\right). \nonumber \end{align}

\noindent
while GPD $\widetilde{E}$ is parameterized by the pion exchange in the $t$-channel \cite{PiPol1,PiPol2}. The main difference among models comes from the chosen $t$-dependence which cannot be fully restricted by theoretical consideration. Some models are currently available in open-source projects such as PARTONS \cite{PARTONS} and Gepard \cite{gepard}. Models currently do not have support for both real and imaginary parts of all four GPDs. The current GPD support of the considered models is summarized in table \ref{modelsuppor} while a brief description of the used quark GPD models is given in the following subsections. Once we select a particular model, nucleon GPDs are constructed for VCS processes as
\begin{align}
    F^{\text{p}}&=\frac{4}{9}F^{u} + \frac{1}{9}F^{d}, \\
    F^{\text{n}}&=\frac{1}{9}F^{u} + \frac{4}{9}F^{d}, 
\end{align}
\noindent
where in general each GPD $F^{q}$ has a valence and sea quark component.

\begin{table}[ht]
    \centering
    \begin{tabular}{|p{0.08\textwidth}>{\centering\arraybackslash}p{0.08\textwidth}>{\centering}p{0.08\textwidth}>{\centering}p{0.08\textwidth}>{\centering\arraybackslash}p{0.1\textwidth}|} \hline
        CFF & VGG & GK & EKM & EAFKM12 \strut \\ \hline \hline
        $\mathfrak{Re}(\mathcal{H})$ & \textcolor{green}{\checkmark}& \textcolor{green}{\checkmark}& \textcolor{green}{\checkmark}& \textcolor{green}{\checkmark}\strut \\
        $\mathfrak{Im}(\mathcal{H})$ & \textcolor{green}{\checkmark}& \textcolor{green}{\checkmark}& \textcolor{green}{\checkmark}& \textcolor{green}{\checkmark}\strut \\ \hline
        $\mathfrak{Re}(\mathcal{E})$ & \textcolor{green}{\checkmark}& \textcolor{green}{\checkmark}& \textcolor{red}{\xmark} & \textcolor{green}{\checkmark}\strut \\
        $\mathfrak{Im}(\mathcal{E})$ & \textcolor{green}{\checkmark}& \textcolor{green}{\checkmark} & \textcolor{red}{\xmark} & \textcolor{green}{\checkmark}\strut \\ \hline
        $\mathfrak{Re}(\mathcal{\widetilde{H}})$ & \textcolor{green}{\checkmark} & \textcolor{green}{\checkmark} & \textcolor{green}{\checkmark} & \textcolor{red}{\xmark} \strut \\ 
        $\mathfrak{Im}(\mathcal{\widetilde{H}})$ & \textcolor{green}{\checkmark} & \textcolor{green}{\checkmark} & \textcolor{green}{\checkmark}\ & \textcolor{red}{\xmark} \strut \\ \hline
        $\mathfrak{Re}(\mathcal{\widetilde{E}})$ & \textcolor{green}{\checkmark} & \textcolor{green}{\checkmark} & \textcolor{green}{\checkmark}\ & \textcolor{red}{\xmark} \strut \\
        $\mathfrak{Im}(\mathcal{\widetilde{E}})$ & \textcolor{red}{\xmark} & \textcolor{green}{\checkmark} & \textcolor{red}{\xmark}& \textcolor{red}{\xmark} \strut \\ \hline 
    \end{tabular}
    \caption{Model support for the different GPDs.}
    \label{modelsuppor}
\end{table}

\subsection{VGG model}
The Vanderhaegen-Guidal-Guichon (VGG) model arose from a series of publications between 1999 and 2005 \cite{VGG1,VGG2,VGG3,VGG4} considering factorized and unfactorized $t$-dependencies in the DD. In particular, we consider the following parameterization for the DD of GPD $H$:

\begin{align}
h(x,y)&=g(x,y)q(x)x^{-\alpha'(1-x)t}    
\end{align}

\noindent 
where $q(x)$ is the corresponding PDF, $\alpha'$ drives the $t$-dependence and is chosen to be $\alpha'$=$1.098$ so the GPD $H$ provides the correct FF values in the forward limit, and $g(x,y)$ is a profile function given by

\begin{align}
    g(x,y)&=\frac{\Gamma(2b+2)}{2^{2b+1}\Gamma(2b+1)}\frac{[(1-|x|)^{2}-y^{2}]^{b}}{(1-|x|)^{2b+1}},
\end{align}

\noindent
$b$ being a free parameter driving the $\xi$-dependence, chosen to be $b_{val}$=$1$ for the valence quark contribution and $b_{sea}$=$5$ for the sea contribution. Then, the only free parameter on the VGG model comes from the choice of the PDF parameterization, we consider the MRST02@NNLO model \cite{MRST} and the CTEQ18 global analysis \cite{CTEQ18}. 

Likewise, GPD $\widetilde{H}$ follows the same parameterization as $H$ with the replacement of the unpolarized PDF $q(x)$ by the polarized PDF $\Delta q(x)$. In the case of $E$, a factorized $t$-dependence is chosen but no DIS constraint exists for the $x$-dependence. The only restriction comes from the Pauli FF $F_{2}^{q}(0)$ which has to be recovered with the first Mellin moment (see \cite{GPDrev, SYthese} for more details). 

Given by the pion pole contribution, $\widetilde{E}$ has a constant real part while its imaginary part is set to zero. It can be written as:

\begin{align}
    \widetilde{E}^{u/p} &= -\widetilde{E}^{d/p} = \frac{1}{2\xi}\theta(\xi - |x|)\;h_{A}(t)\;\phi_{as}\left(\frac{x}{\xi}\right)
\end{align}

\noindent
with $\phi_{as}(z)$=$3/4\; (1-z^{2})$ and $h_{A}(x)$ being the asymptotic distribution amplitude  
 and the induced pseudo scalar FF of the nucleon respectively. Thanks to its simple form, CFF $\mathcal{\widetilde{E}}$ can be integrated directly, giving as a result:

\begin{align}
    \widetilde{\mathcal{E}}(\xi',\xi, t)& \label{CFFEt}\\
    =&\int_{-1}^{1}dx\left[\frac{1}{x-\xi'+i\epsilon} - \frac{1}{x+\xi' - i\epsilon}\right]\tilde{E}(x,\xi,t) \nonumber\\
    =&-\frac{h_{A}(t)}{8\xi^{3}}\left[4\xi\xi' + 2(\xi^{2} - \xi^{\prime 2})\log\left(\frac{\xi + \xi'}{\xi - \xi'}\right)\right]. \nonumber
\end{align}

\subsection{GK model}
The Goloskokov-Kroll (GK) model is also based on DDs but was developed by fitting high-energy DVMP data \cite{GK1, GK2} as it depends on the same GPDs entering the DVCS/TCS/DDVCS process. For GPD $H$ a different $t$-dependence is proposed,  given at $\xi$=$0$ by:

\begin{align}
    H(x,0,t)&=H(x,0,0)x^{-\alpha't}e^{bt}.
\end{align}
However, the full $H$ dependence is described through a series expansion of half-integer powers of $x$:

\begin{align}
    H(x, \xi, t)&=e^{bt}\sum_{j=0}^{3}c_{ij}H_{ij}(x,\xi,t)
\end{align}
\noindent
where the integrals $H_{ij}$ can be read from \cite{GK2} and the slope $b$ is set to zero for valence quarks. For GPD $\widetilde{H}$, the same DD modeling as GPD $H$ is chosen. Its forward limit is described by the Blümlein-Böttcher results \cite{BB}, the sea contribution is set to zero, and the valence contribution is constrained by HERMES data \cite{HERMES1, HERMES2}. It uses the following profile function:

\begin{align}
    \widetilde{H}_{val}(x,0,0)&=\xi A x^{-\alpha_{\widetilde{H}}(0)}(1-x)^{3}\sum_{j=0}^{2}\widetilde{c}_{ij}x^{j},
\end{align}

\noindent
where $\xi$ guarantees the correct normalization of its first moment, and the coefficients $A$ and $\widetilde{c}_{ij}$ can be found in \cite{GKcoeff} for each quark. For the GPD $E$, a DD ansatz is also used and $E(x,0,0)$ is parameterized with a classical PDF functional form
\begin{align}
    E_{val}(x,0,0)&=\frac{\Gamma(2-\alpha_{val}+\beta_{val})}{\Gamma(1-\alpha_{val})\Gamma(1+\beta_{val})}\kappa x^{-\alpha_{val}}(1-x)^{\beta_{val}},
\end{align}
\noindent
where the $b$ parameters in the DD are assumed to be equal to those in GPD $H$, $\beta_{val}$=$4$ for the up quark and $\beta_{val}$=$5.6$ for the down quark. Finally, GPD $\widetilde{E}$ is also given by the pion pole contribution, similar to the VGG case, but with a different ($x$, $\xi$)-dependence given by:

\begin{align}
    \widetilde{E}^{u/p} &= -\widetilde{E}^{d/p} = \frac{1}{4\xi}\;\theta(\xi - |x|)\;h_{A}(t)\;\Phi_{\pi}\left(\frac{x+\xi}{2\xi}\right)
\end{align}
\noindent
with 
\begin{align}
    \Phi_{\pi}(\tau)&=6\tau(1-\tau)[1+0.22C_{2}^{3/2}(2\tau -1 )]
\end{align}
\noindent
$C_{2}^{3/2}$ being a Gegenbauer polynomial. In particular, we consider the parameters from the GK19 model of PARTONS \cite{PARTONS} as it contains the most recent fits and supports DDVCS computations.

\subsection{Extended KM model}
The Kumericki-Müller (KM) model \cite{KM1} has a hybrid implementation of the Mellin-Barnes parameterization of GPDs for the sea quarks contribution and a dispersion relation approach for valence quarks. In general, the parameters of the KM model come from fits to H1/ZEUS, HERMES and JLab data \cite{KM1}. Several models are implemented in Gepard \cite{gepard} for DVCS computations only. To perform DDVCS computations, we have implemented a toy-model generalization where the sea quark contribution is given by the Mellin-Barnes (MB) representation in \cite{MellinBarnes} at a fixed renormalization scale $\mu=Q^{2} + Q^{\prime 2}$. The $\xi'$ dependence is thus introduced by restoring the formula for the Wilson coefficient in the CFF Taylor expansion (see Eqs. (29), (34) and (35) of \cite{MellinBarnes}, $\vartheta$=$1$ corresponds to the DVCS case implemented in Gepard). Furthermore, a simple generalization of the valence quark contribution is obtained from an ansatz inspired by the Dispersion Relation (DR) approach in the $\xi'>\xi$ region of \cite{MellinBarnes,DR_DDVCS}
\begin{align}
    \mathfrak{Re}[&\mathcal{H}(\xi',\vartheta,t)]= \\
    &\frac{1}{\pi}\int_{0}^{1}dx\; \mathfrak{Im}[\mathcal{H}(x,\vartheta,t)]\left[\frac{1}{x-\xi'}+\frac{1}{x+\xi'}\right] + C_{F}(\vartheta,t) \label{DR_DDVCS}
\end{align}

\noindent
where $\vartheta=\xi'/\xi$ and $C_{F}(\vartheta,t)$ is the subtraction constant related to the $D$-term, and the real part integral is computed over the trajectory $\xi=x/\vartheta$ \textit{\textit{i.e.}} at constant $\vartheta$=$\xi'/\xi$. 

To model the imaginary part of the valence contribution we start from the results of the EKM model. For DVCS, the imaginary part is modeled with a DD whose $t$-dependence is inspired by the spectator model \cite{spectator}. At $t=0$ it reads
\begin{align}
    h(y,z,0)=& \\
    &\frac{\Gamma(3/2 + b)}{\Gamma(1/2)\Gamma(1+b)}\frac{q(y)}{1-y}\left(1-\frac{z^{2}}{(1-y)^{2}}\right)^{b}, \nonumber
\end{align}
\noindent
leading to the GPD $H$ on the cross-over line:
\begin{align}
    &H^{+}(x,x,t)= \label{Hxx}\\
    &\frac{nr}{1+x}\left(\frac{2x}{1+x}\right)^{-\alpha(t)}\left(\frac{1-x}{1+x}\right)^{b}\frac{1}{\left(1 - \frac{1-x}{1+x}\frac{t}{M^{2}}\right)^{p}} \; . \nonumber
\end{align}
To compute the imaginary part of CFFs for DDVCS, we made a simple generalization of the GPD for the $x\neq\xi$ case by constructing a $t$-dependent DD that can reproduce the DVCS limit of Eq. (\ref{Hxx}) (see appendix \ref{app2}). After integrating it in the general $x\neq \xi$ case, we find the following functional form for the GPD $H$.
\begin{align}
     H^{+}&(x>0,\xi,t)=nr\frac{1+\vartheta}{2}\frac{\vartheta}{\vartheta+x} \\
     &\left(\frac{x+ \frac{x}{\vartheta}}{1+\frac{x}{\vartheta}}\right)^{-\alpha(t)} \left(\frac{1-\vartheta^{2}}{4} + \frac{\vartheta^{2}(1+\vartheta)}{2}\frac{(1-x)}{\vartheta + x} \right)^{b} \nonumber \\
    &\left(\frac{1}{1-\frac{t}{4M^{2}}(1-\vartheta^{2}) - \frac{t}{M^{2}}\frac{\vartheta^{2}(1+\vartheta)}{2}\frac{(1-x)}{x + \vartheta}}\right)^{p}\nonumber
\end{align} 
 \noindent
where $\vartheta=x/\xi$. This function is defined for $x>0$. We define the GPD in the $x<0$ region by $H^{+}(-x,\xi,t) = -H^{+}(|x|,\xi,t)$, as it is required to be an odd function of $x$. For GPD $\widetilde{H}$, a similar result is obtained:

\begin{align}
    \widetilde{H}^{+}&(x,\xi,t) = \pi \left(2\frac{4}{9} + \frac{1}{9}\right)
        \frac{\tilde{n}\vartheta}{\vartheta+x}\frac{1+\vartheta}{2} \nonumber\\
     &\left(\frac{x+ \frac{x}{\vartheta}}{1+\frac{x}{\vartheta}}\right)^{-\alpha(t)} \left(\frac{1-\vartheta^{2}}{4} + \frac{\vartheta^{2}(1+\vartheta)}{2}\frac{(1-x)}{\vartheta + x} \right)^{3/2} \nonumber \\
    &\left(\frac{1}{1-\frac{t}{4M^{2}}(1-\vartheta^{2}) - \frac{t}{M^{2}}\frac{\vartheta^{2}(1+\vartheta)}{2}\frac{(1-x)}{x + \vartheta}}\right).
    \end{align}

\begin{figure}[H]
    \centering
    \includegraphics[width=\linewidth]{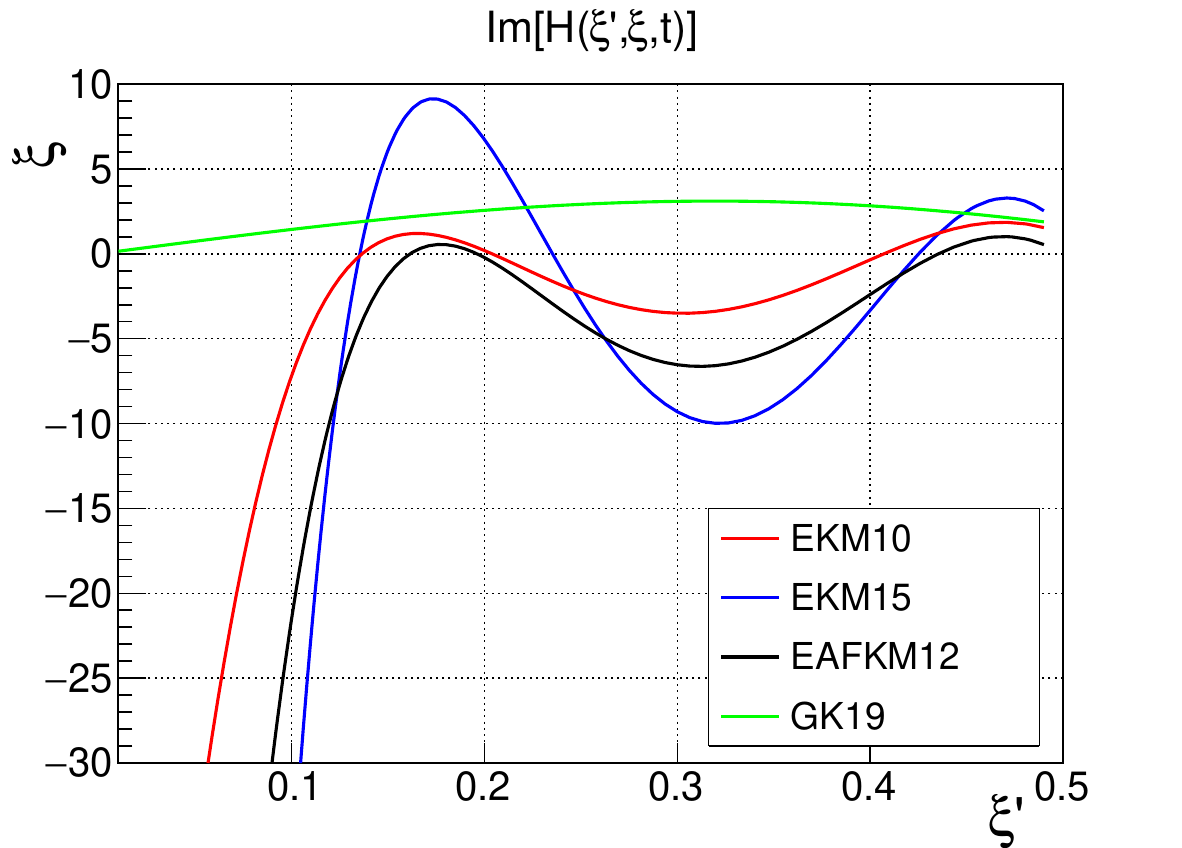}
    \caption{$\mathfrak{Im}[H(\xi',\xi,t)]$ as a function of $\xi'$ at fixed $\xi=0.5$ and $t$=$-0.2$ GeV$^{2}$ for the EKM15 and GK19 models.}
    \label{ImH_EKM}
\end{figure}

As there is no candidate function for $\mathfrak{Im}(\mathcal{E}(x,x,t))$, we do not implement a dispersion relation contribution for CFF $\mathcal{E}$. CFF $\widetilde{\mathcal{E}}$ is also computed from the pion pole contribution of Eq. (\ref{CFFEt}). Finally, we consider the EKM10, EKM15 and EAFKM12 \cite{AFKM12} (Aschenauer-Fazio EKM model adaptation for EIC kinematics) models to compute CFFs with Gepard. Then, CFFs are given as input to a local routine that computes general DDVCS cross-sections.  Notice from table \ref{modelsuppor} that the EAFKM12 model has support for $\mathfrak{Im}(\mathcal{E})$ but that is only from the Mellin-Barnes approach since at EIC kinematics the sea quark contribution is expected to have a major role in the cross-section. Finally, it is shown in Fig. \ref{ImH_EKM} the evolution of $\mathfrak{Im}[\mathcal{H}(\xi',\xi=0.5,t=-0.2)]$ as a function of $\xi'$ for the EKM10, EAFKM12, EKM15 and GK19 models. As a consequence of the $(\xi/\xi')^{n}$ dependence of the Mellin-Barnes representation, for $n$ even, the GPD diverges as $\xi'\rightarrow 0$ and presents two sign changes due to the effect of the evolution operators of the Wilson coefficients. Unlike the KM models, the GK19 prediction vanishes at $\xi'$=$0$ as expected from the leading valence-quark contribution. Besides, one can observe that the location of the maximum $\mathfrak{Im}[\mathcal{H}(\xi',\xi,t)]$ value at fixed $\xi$ and the GPD sign changes are discriminating features for GPD models. Thus, additional phenomenological studies for GPD models in the $\xi\neq \xi'$ region are needed.

\section{Experimental projections}\label{exps}
%\textit{i.e.} the GK19 model from PARTONS and the EKM10, EKM15 \& EAFKM12 models and the VGG model using the MRST\cite{MRST} and CTEQ18\cite{CTEQ18} parameterization for quark PDFs
In the following, we present the BSA, TSA, DSA and BCA predictions from the set of models introduced in the previous section. Motivated by CFF extraction, these predictions are computed over bins in the $(\xi', \xi)$-phase space and integrated over $x_{B}$ and $t$. On top of model predictions, we present experimental projections with statistical errors computed from simulations re-scaled to the foreseen integrated luminosity, including detector acceptance and geometrical and reconstruction efficiency effects.

Considering the experimental configuration at JLab, we explore the DDVCS measuring capabilities of the $\mu$CLAS detector in Hall-B and the SoLID$\mu$ spectrometer. Both spectrometers are foreseen to take data with a luminosity of $10^{37}\text{cm}^{-2}\cdot\text{s}^{-1}$ and beam energies up to 22 GeV. Furthermore, we study the EIC support for DDVCS measurements considering $5$ GeV electron and $41$ GeV proton beams with a luminosity of $10\; \mathrm{fb}^{-1}\cdot\mathrm{y}^{-1}$, as expected for the initial years of operations. In all cases, event generation with EpIC\cite{EpIC} is subject to the  $W>2$ GeV, $t>-1$ GeV$^{2}$, $Q^{2}+Q^{\prime 2}>1$ GeV$^{2}$ and $Q^{\prime 2}>2$ GeV$^{2}$ restrictions, where the latter excludes the vector meson resonance region. The studied configurations are: 
\begin{itemize}
    \item 11 GeV at JLab, 100 days and $\mathcal{L}=10^{37}\; \mathrm{cm}^{-2}\cdot\mathrm{s}^{-1}$;
    \item 22 GeV at JLab, 200 days  and $\mathcal{L}=10^{37}\; \mathrm{cm}^{-2}\cdot\mathrm{s}^{-1}$;
    \item EIC case, 1 year and $\mathcal{L}=10\; \mathrm{fb}^{-1}\cdot\mathrm{y}^{-1}$.\\
\end{itemize}
In the polarized target case, error bars were also scaled by $1/\sqrt{D}$, being $D$ a dilution factor depending on the target type. For instance, a polarized NH$_{3}$ target features $D$=$3/17$ corresponding to only 3 polarized nucleons out of 17. Finally, we include the error estimation from the beam and target polarization by quadratically adding them to the statistical error, being $P_{b}$=$0.86$ and $P_{t}$=$0.80$ typical polarization values obtained in $\mu$CLAS experiments with uncertainties of $\delta P_{b}$=$0.05$ and $\delta P_{t}$=$0.02$ respectively. For simplicity, we assume the same target polarization for both spin directions. To make the projection more realistic, the observable prediction for each bin, $A(\phi_{i})$ is replaced by a random number from a Gaussian distribution with mean $A(\phi_{i})$ and standard deviation $\delta A(\phi_{i})$.
\vspace{-0.5cm}
\subsection{Jefferson lab experimental configuration}\label{JLabkin}
Considering the set of generated DDVCS events, we apply geometrical and reconstruction efficiency effects to obtain a more realistic view of the phase space and number of reconstructed events in each kinematic bin for statistical error computations. In the $\mu$CLAS scenario, we use Geant4 Monte Carlo (GEMC) \cite{gemc} to simulate the response to the generated events. Then, particles were reconstructed using the latest software for real data processing in the $\mu$CLAS collaboration. In this case, acceptance and reconstruction efficiencies are automatically included in the simulation and reconstruction chain. In the SoLID$\mu$ scenario, geometry and decay effects were included by weighting the events by their acceptance efficiency retrieved from acceptance maps. To account for the reconstruction efficiency, the total number of events is re-scaled by 0.7 which quantifies PID and tracking efficiency \cite{solid2}.
% Both spectrometers are designed to reconstruct electrons in a similar range ($7\degree-35\degree$) but with different muon acceptance. Namely $7\degree - 18\degree$ for SoLID$\mu$ and $7\degree - 35\degree$ for $\mu$CLAS.

\begin{figure}[H]
    \centering
    \includegraphics[width=\linewidth]{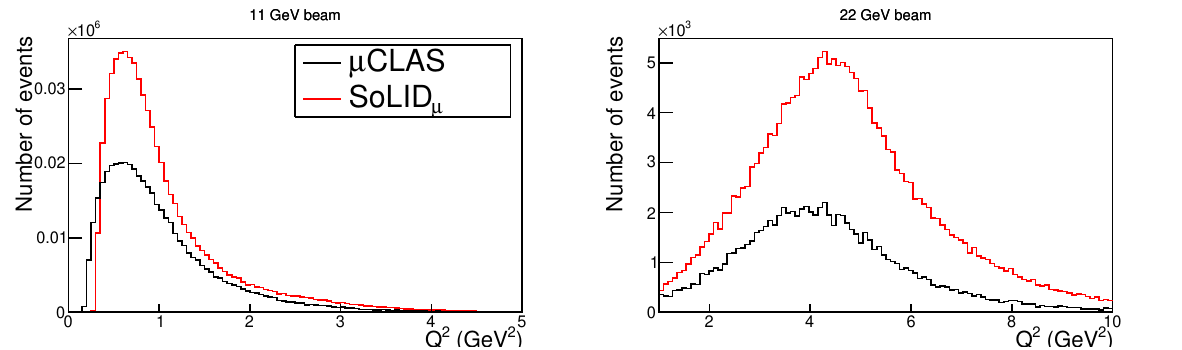}
    \caption{$Q^{2}$ distributions with an $11$ GeV (left) and a $22$ GeV beam (right).}
    \label{Q2dists}
\end{figure}

\begin{figure}[H]
    \centering
    \begin{subfigure}[b]{0.45\textwidth}
        \centering
        \includegraphics[width=0.86\textwidth]{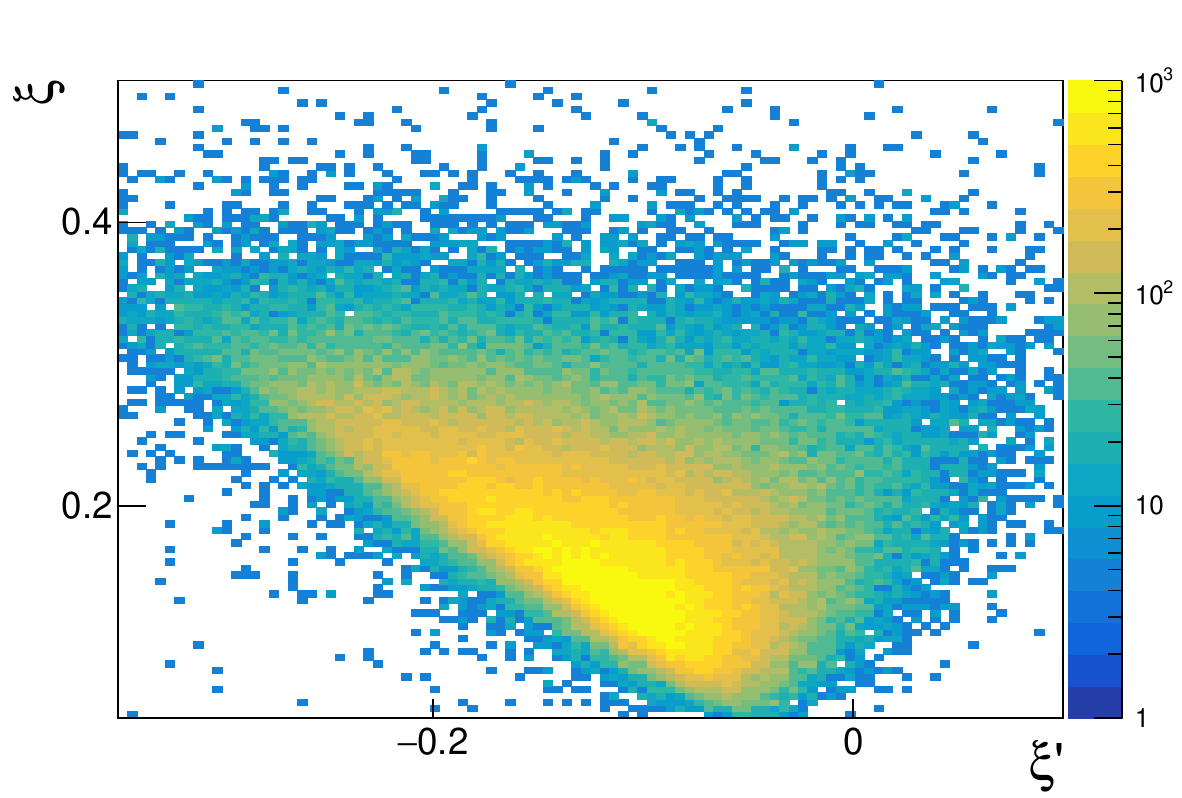}
        \caption{$\mu$CLAS $\xi'$ vs $\xi$ acceptance.}
    \end{subfigure}
    \begin{subfigure}[b]{0.45\textwidth}
        \centering
        \includegraphics[width=0.86\textwidth]{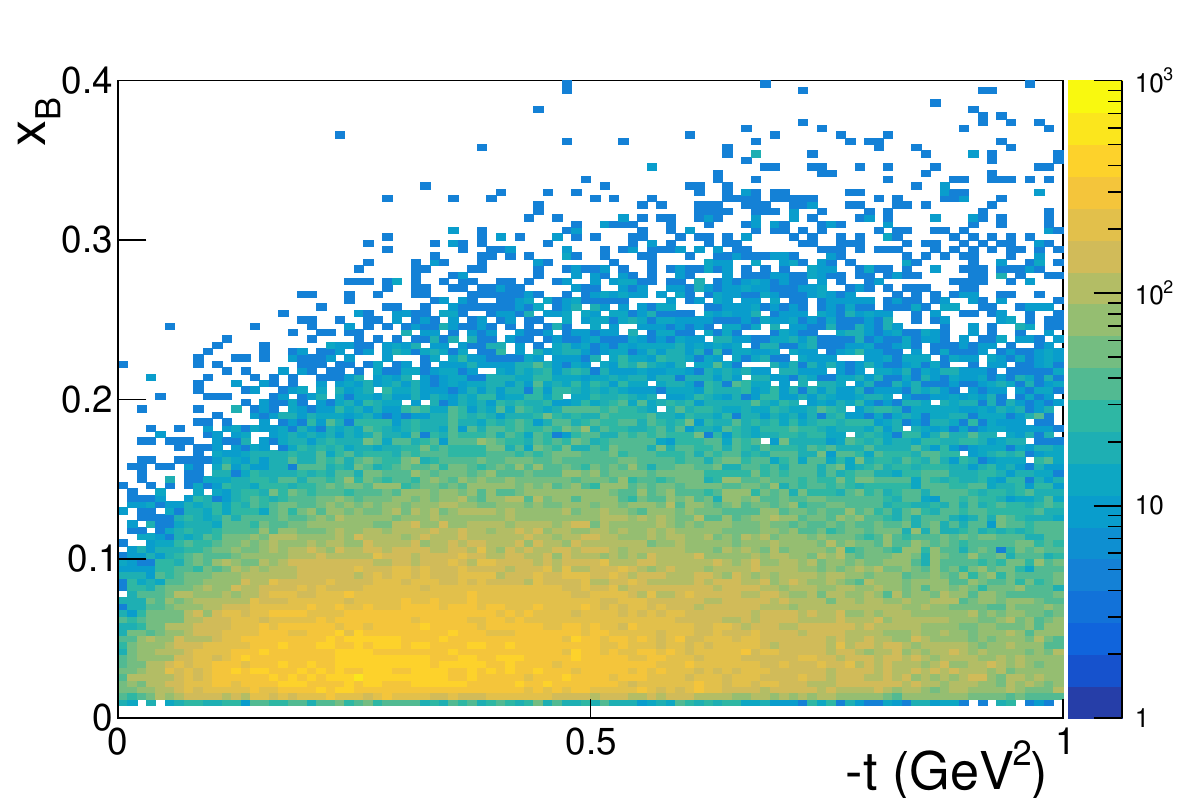}
        \caption{$\mu$CLAS $-t$ vs $x_{B}$ acceptance.}
    \end{subfigure}
    \begin{subfigure}[b]{0.45\textwidth}
        \centering
        \includegraphics[width=0.86\textwidth]{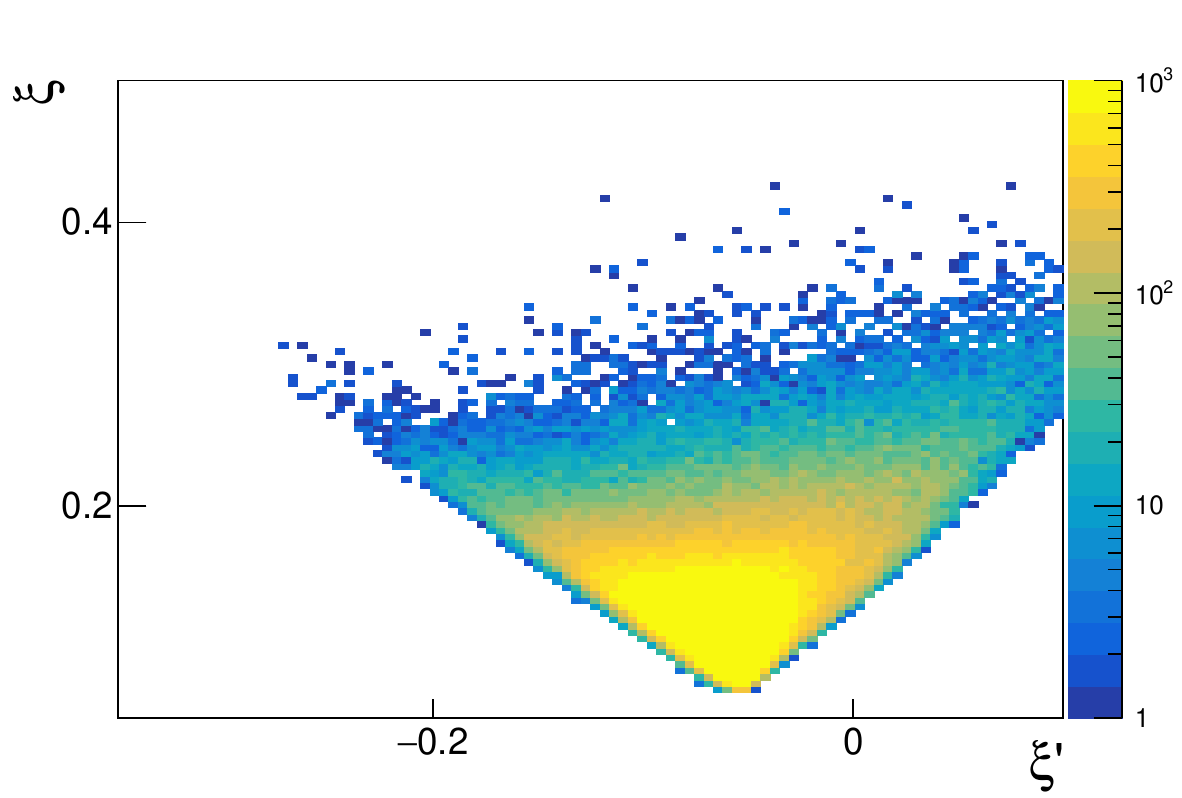}
        \caption{SoLID$\mu$ $\xi'$ vs $\xi$ acceptance.}
    \end{subfigure}
    \begin{subfigure}[b]{0.45\textwidth}
        \centering
        \includegraphics[width=0.86\textwidth]{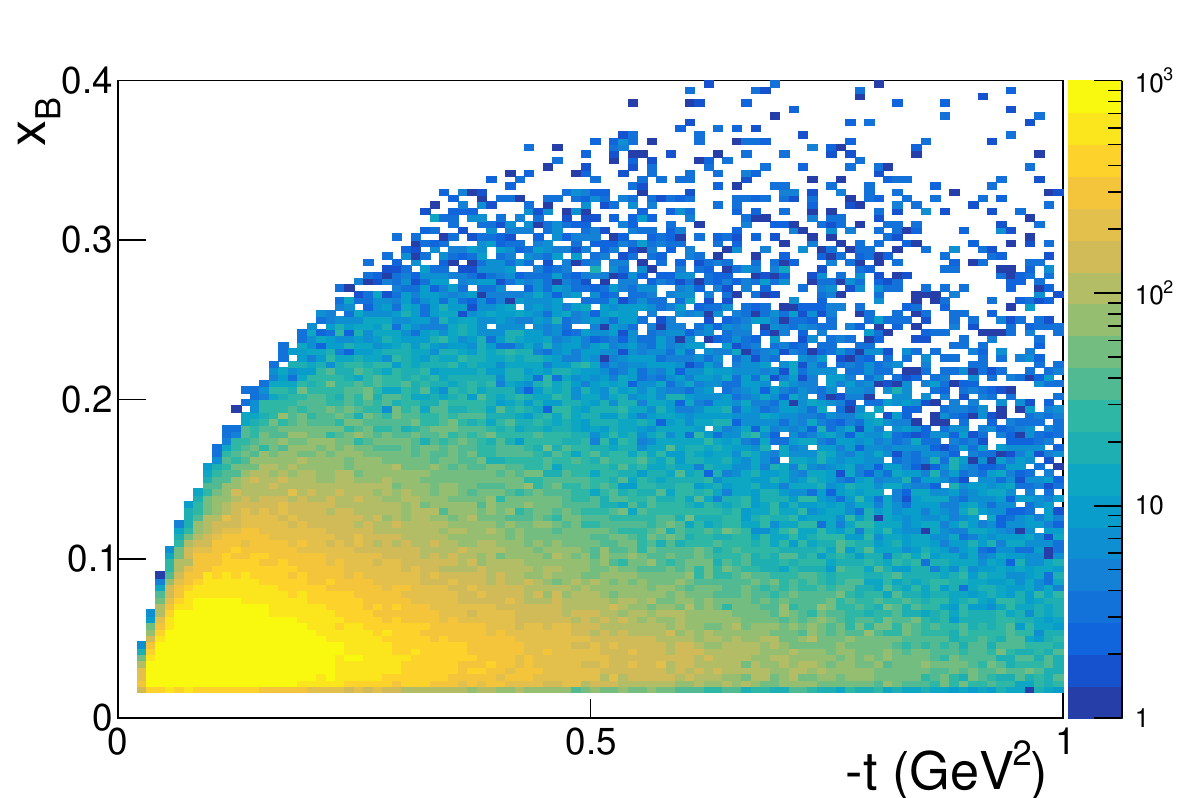}
        \caption{SoLID$\mu$ $-t$ vs $x_{B}$ acceptance.}
    \end{subfigure}
    \caption{Phase space coverage at JLab with an 11 GeV beam.}
    \label{JLabPS11}
\end{figure}

\begin{figure}[H]
    \centering
    \begin{subfigure}[b]{0.45\textwidth}
        \centering
        \includegraphics[width=0.86\textwidth]{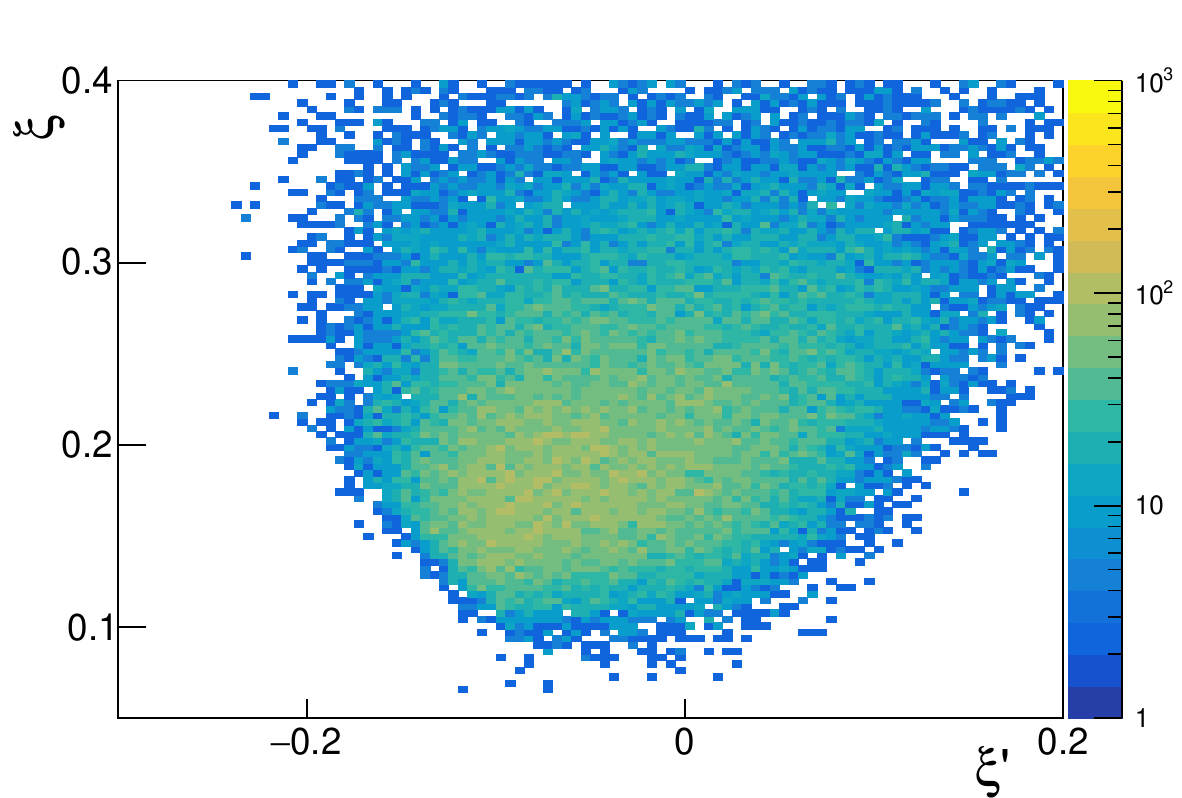}
        \caption{$\mu$CLAS $\xi'$ vs $\xi$ acceptance.}
    \end{subfigure}
    \begin{subfigure}[b]{0.45\textwidth}
        \centering
        \includegraphics[width=0.86\textwidth]{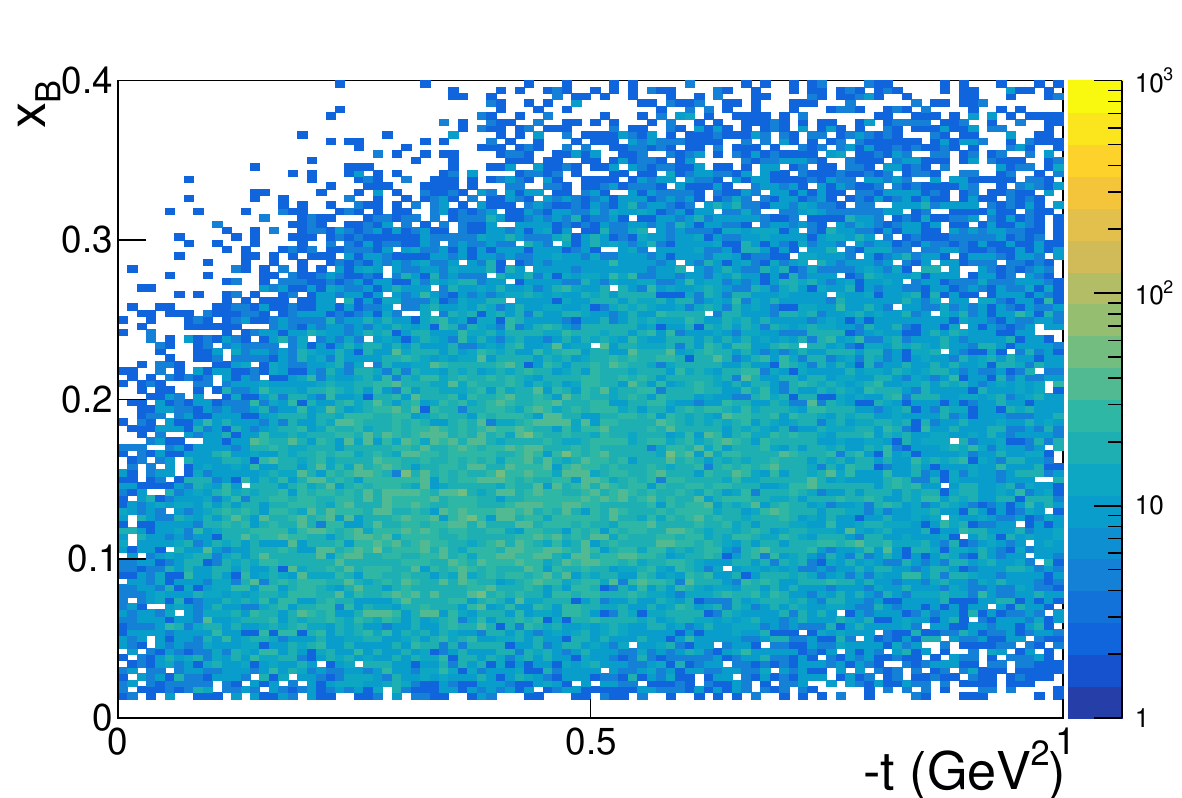}
        \caption{$\mu$CLAS $-t$ vs $x_{B}$ acceptance.}
    \end{subfigure}
    \begin{subfigure}[b]{0.45\textwidth}
        \centering
        \includegraphics[width=0.86\textwidth]{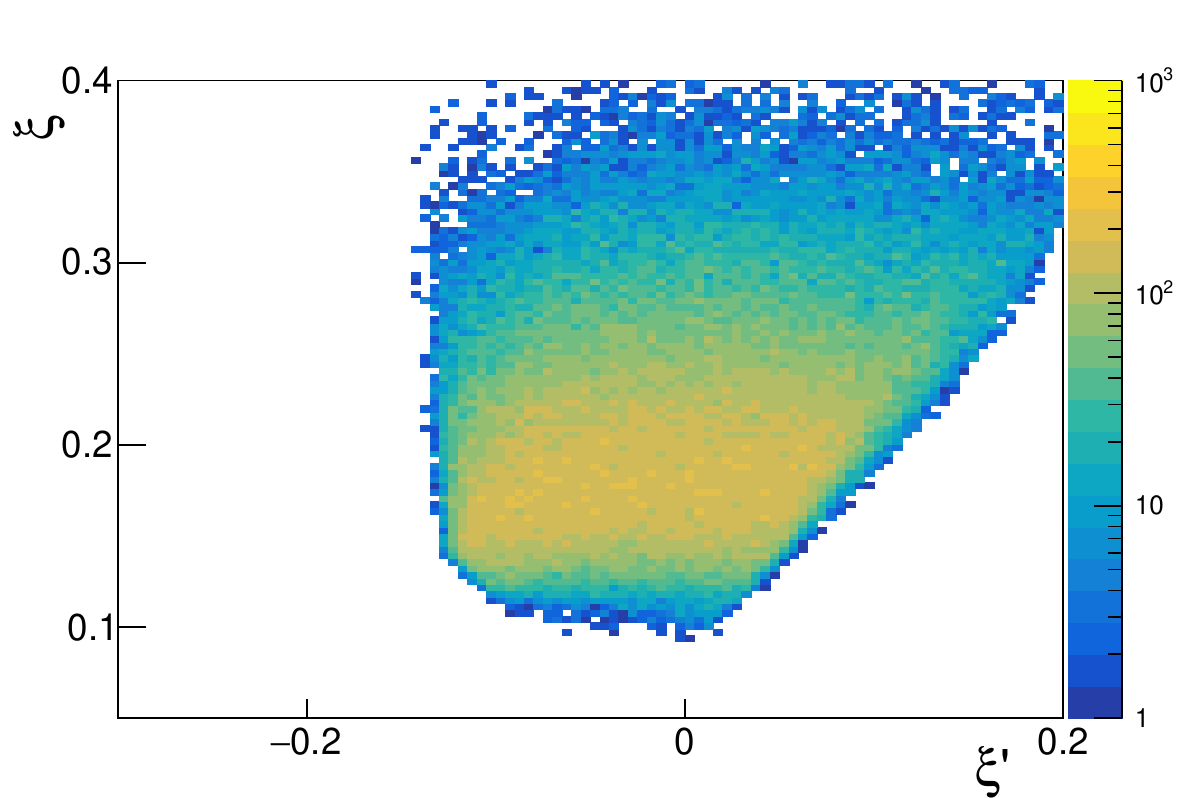}
        \caption{SoLID$\mu$ $\xi'$ vs $\xi$ acceptance.}
    \end{subfigure}
    \begin{subfigure}[b]{0.45\textwidth}
        \centering
        \includegraphics[width=0.86\textwidth]{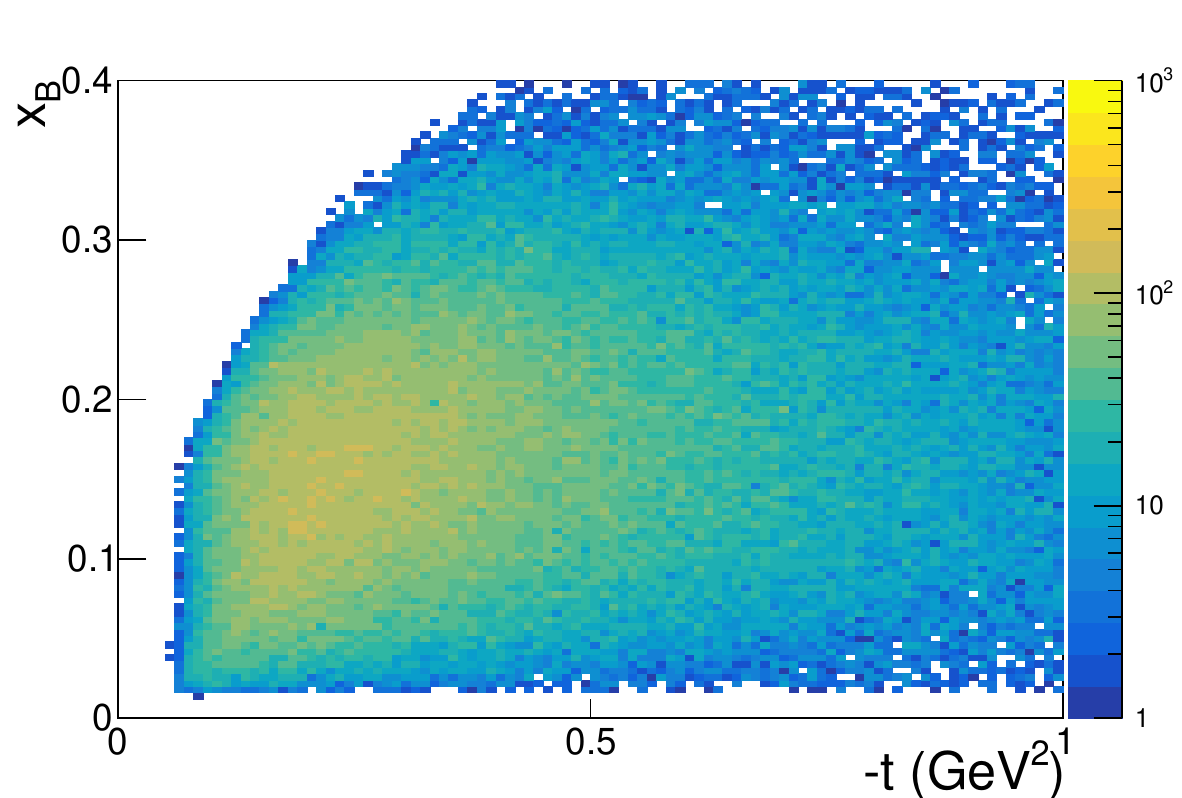}
        \caption{SoLID$\mu$ $-t$ vs $x_{B}$ acceptance.}
    \end{subfigure}
    \caption{Phase space coverage at JLab with a 22 GeV beam.}
    \label{JLabPS22}
\end{figure}

The reconstructed phase space in the relevant kinematic variables is illustrated in Figs. \ref{Q2dists}, \ref{JLabPS11} and \ref{JLabPS22} considering charged particles with energy above 1 GeV and beam energies of 11-22 GeV. Focusing on the $(\xi', \xi)$-phase space, we observe that the distribution of events reconstructed with the $\mu$CLAS detector has a larger density in the TCS-like region while SoLID$\mu$ provides a more uniform distribution of events.  Regarding the $x_{B}$ coverage, SoLID$\mu$ and $\mu$CLAS acceptances complement each other as the $\mu$CLAS detector targets smaller values of $x_{B}$. Finally, the $Q^{2}$ distributions in Fig. \ref{Q2dists} show that both spectrometers have an almost identical coverage for both beam energies. As TCS-like observables provide larger amplitudes in the phase space region accessed by the spectrometers, the DDVCS beam and target asymmetries are studied as a function of $\varphi_{\ell}$.

\onecolumngrid

\twocolumngrid

\begin{figure}[H]
    \centering
    \begin{subfigure}[b]{0.48\textwidth}
        \centering
        \includegraphics[width=\textwidth]{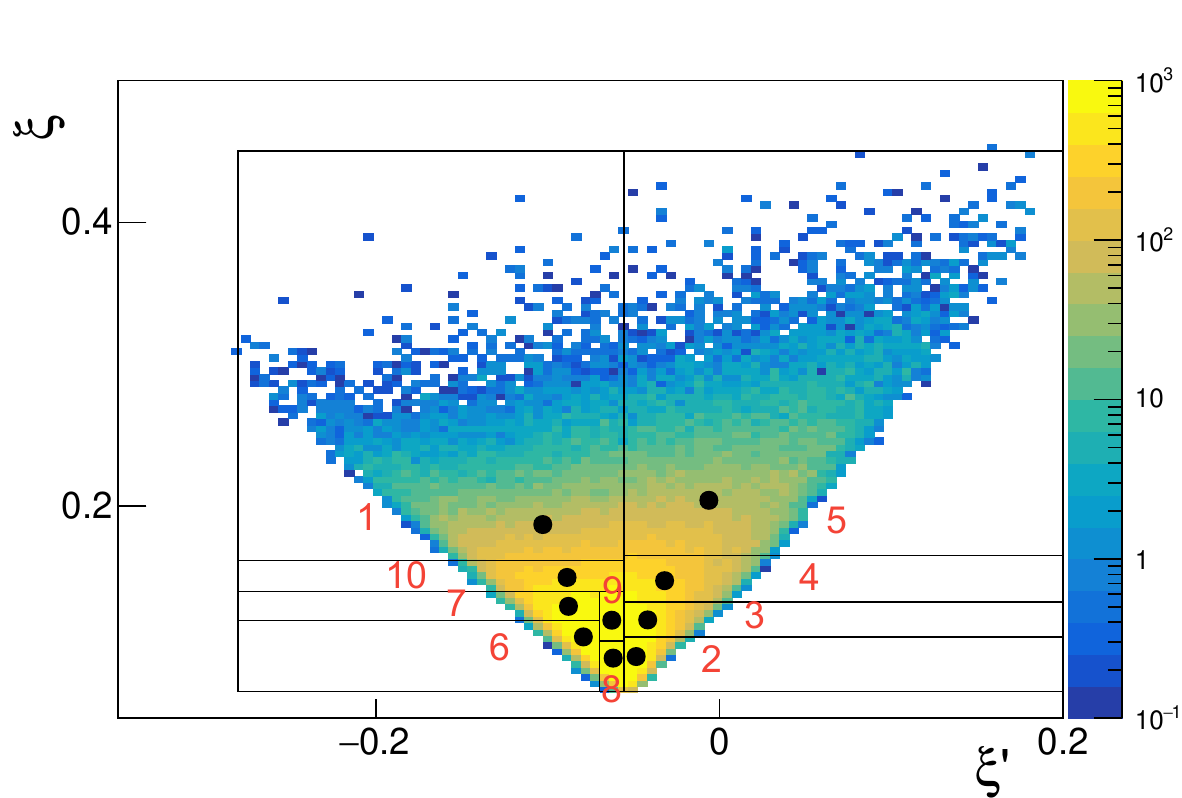}
        \caption{11 GeV beam.}
    \end{subfigure}
    \begin{subfigure}[b]{0.48\textwidth}
        \centering
        \includegraphics[width=\textwidth]{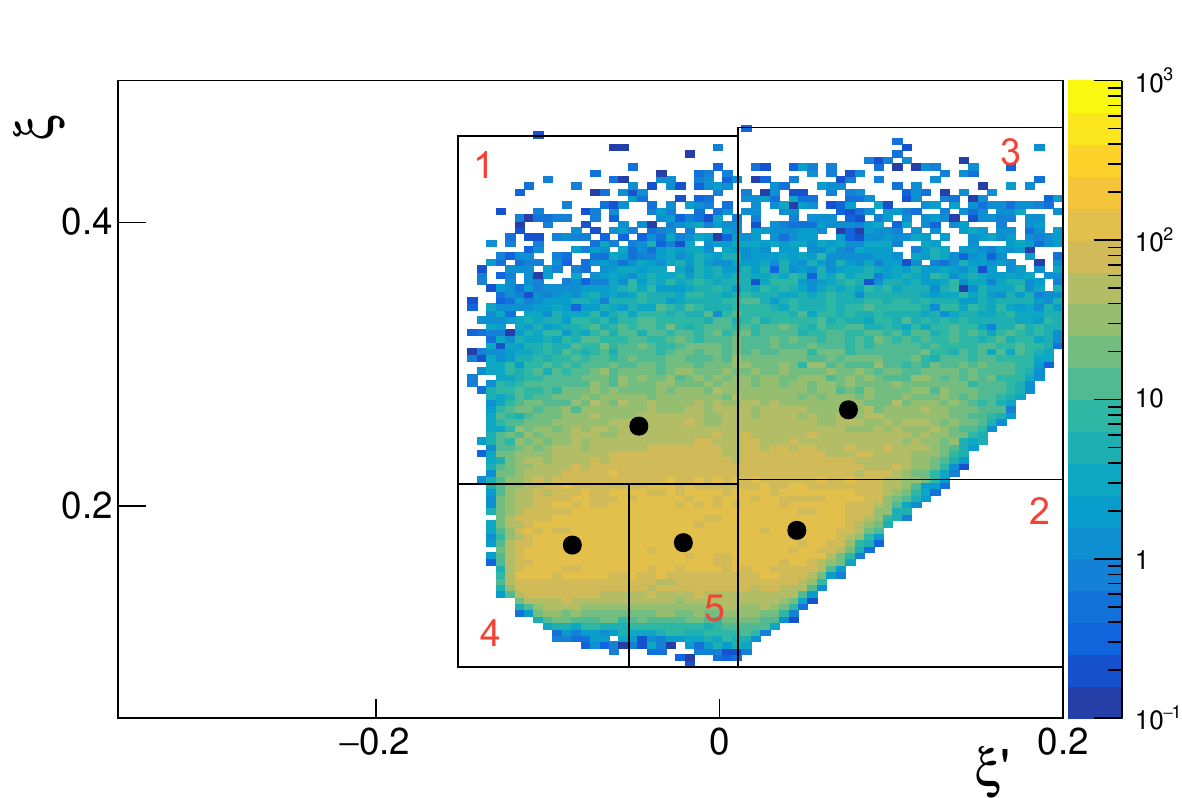}
        \caption{22 GeV beam.}
    \end{subfigure}
    \caption{Binning scheme for the JLab configurations with dots representing the mean value of the bin.}
    \label{binning}
\end{figure}

To compute experimental observables we defined a binning of a similar number of events in the $(\xi',\xi)$-phase space as shown in Fig. \ref{binning}. The use of the $(\xi',\xi)$-phase space instead of $(Q^{2}, Q^{\prime 2})$ is motivated by CFF extraction although $t$ and $x_{B}$ remain unconstrained and can be used for further binning depending on the available experimental statistics. The number of bins was chosen so that the error bars of the asymmetry projections show feasible measurements. .

Following the chosen binning, we can observe in Fig. \ref{binning} that the projected number of measurements for a 22 GeV beam is about half the number for the 11 GeV case. Moreover, the mean kinematics (represented with dots in Fig. \ref{binning}) indicates that the 11 GeV configuration would provide measurements mostly in the TCS-like region ($\xi'<0$) while a 22 GeV beam access similarly the TCS-like and DVCS-like ($\xi'>0$) regions. Therefore, we can conclude from this interplay between the beam energy and access to the GPD phase space that both configurations are complementary for GPD studies as 11/22 GeV beams would provide unique measurements in the TCS/DVCS regions, providing experimental evidence of the expected asymmetry sign change around $\xi'=0$.

\begin{figure}[ht]
    \centering
    \begin{subfigure}[b]{0.45\textwidth}
        \centering
        \includegraphics[width=0.95\textwidth]{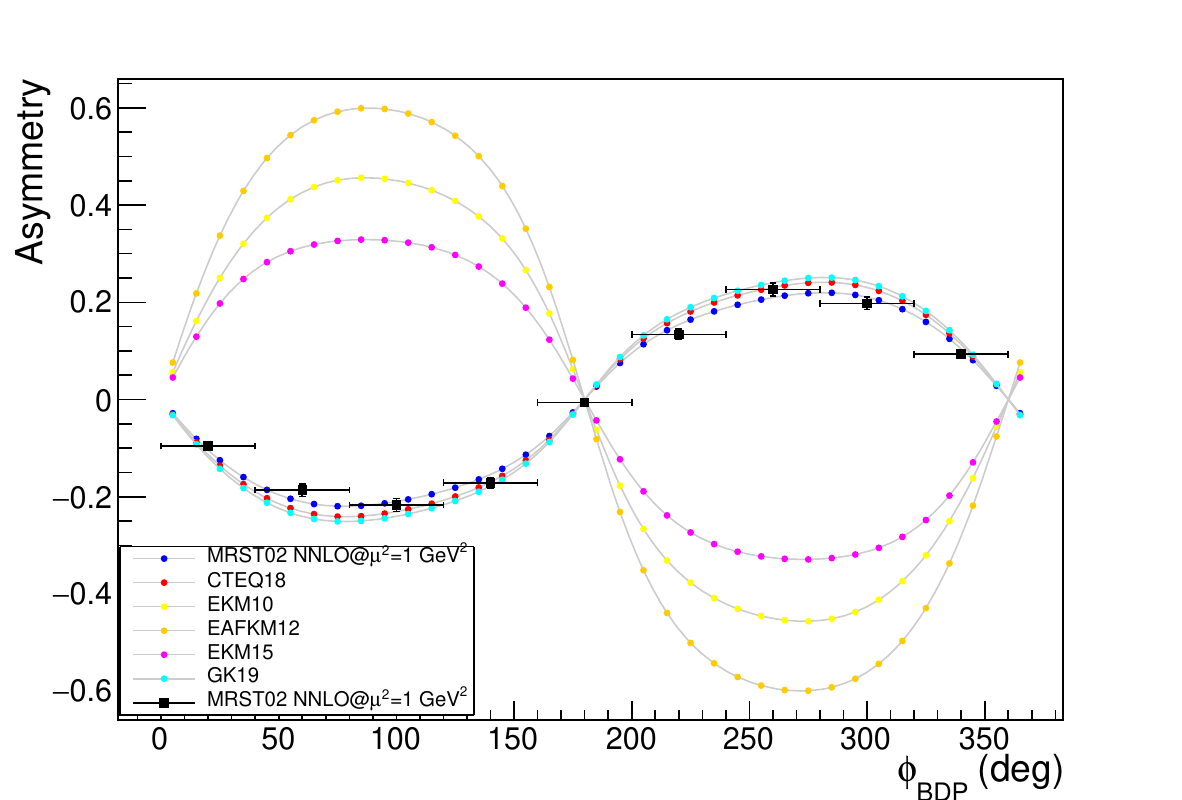}
        \caption{$A_{LU}$ in bin 1: $\xi'<-0.056$ and $\xi>0.45$.}
    \end{subfigure}
    \hspace{1.0cm}
    \begin{subfigure}[b]{0.45\textwidth}
        \centering
        \includegraphics[width=0.95\textwidth]{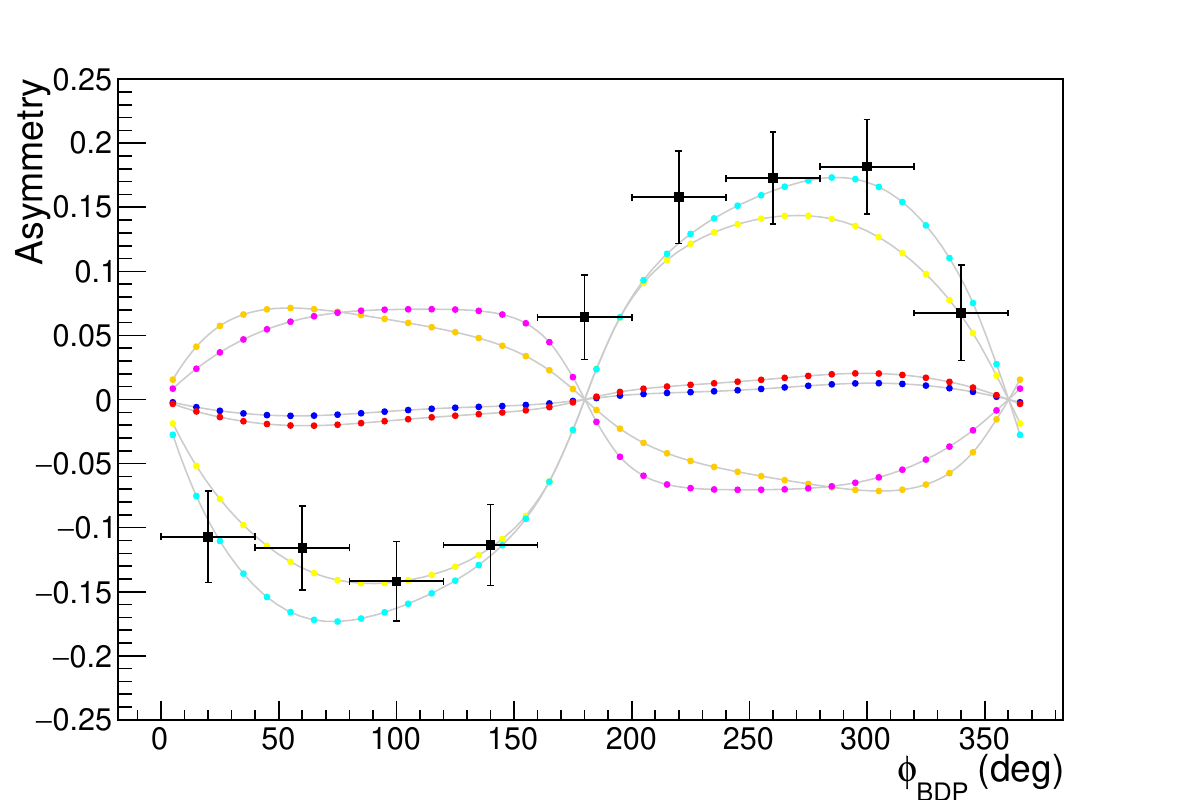}
        \caption{$A_{UL}$ in bin 6: $\xi'<-0.069$ and $\xi<0.107$.}
    \end{subfigure}
    \begin{subfigure}[b]{0.45\textwidth}
        \centering
        \includegraphics[width=0.95\textwidth]{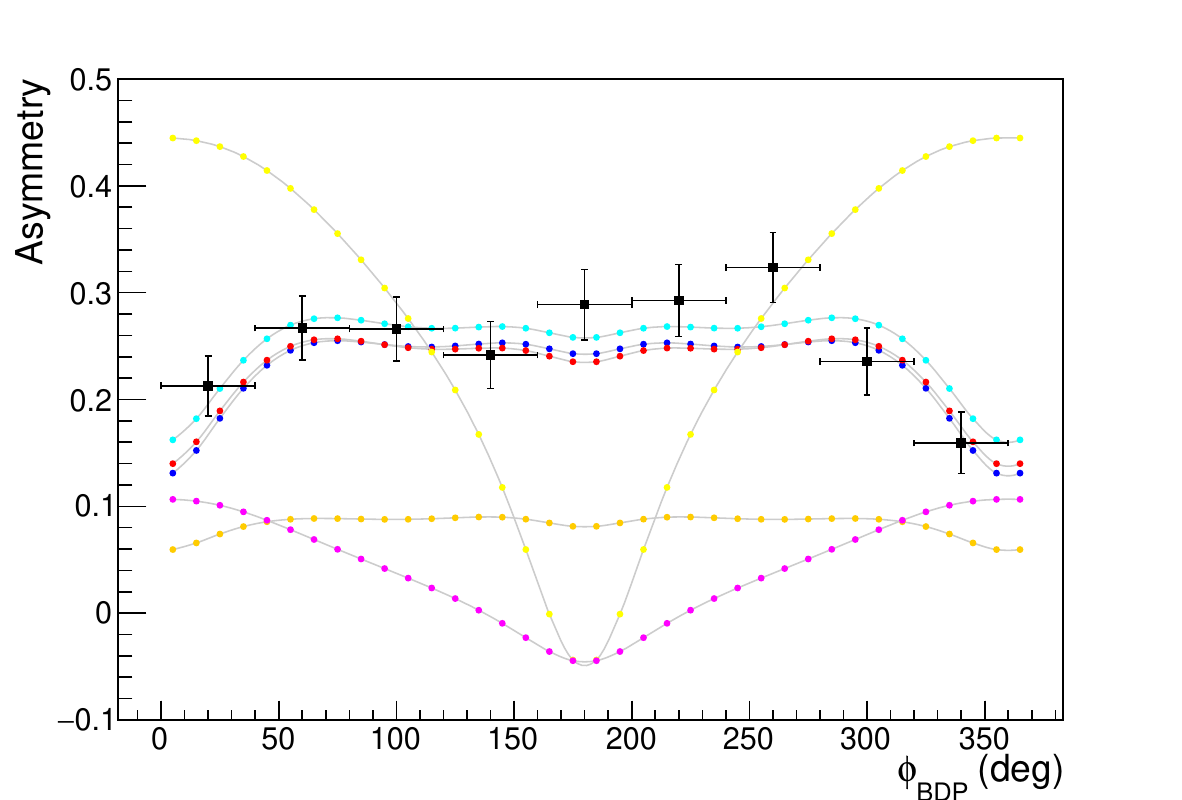}
        \caption{$A_{LL}$ in bin 10: $\xi'<-0.069$ and $0.139<\xi<0.45$.}
    \end{subfigure}
    \hspace{1.0cm}
    \begin{subfigure}[b]{0.45\textwidth}
        \centering
        \includegraphics[width=0.95\textwidth]{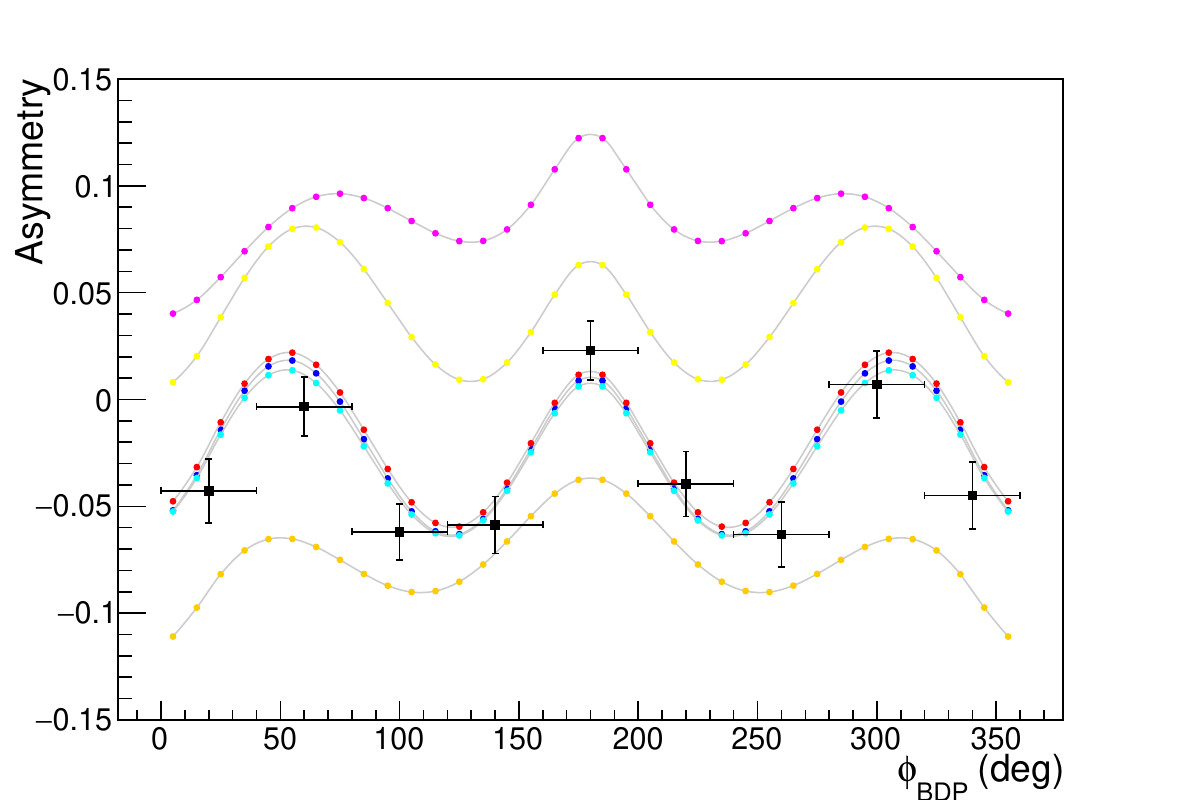}
        \caption{$A_{UU}^{C}$ in bin 6: $\xi'>-0.056$ and $\xi>0.165$.}
    \end{subfigure}
    \caption{\justifying{Asymmetry projections for DDVCS measurements with the $\mu$CLAS spectrometer and an 11 GeV beam. Error bars represent statistical errors.}}
    \label{JLabMeas}
\end{figure}

\begin{figure}[ht]
    \centering
    \begin{subfigure}[b]{0.45\textwidth}
        \centering
        \includegraphics[width=0.95\textwidth]{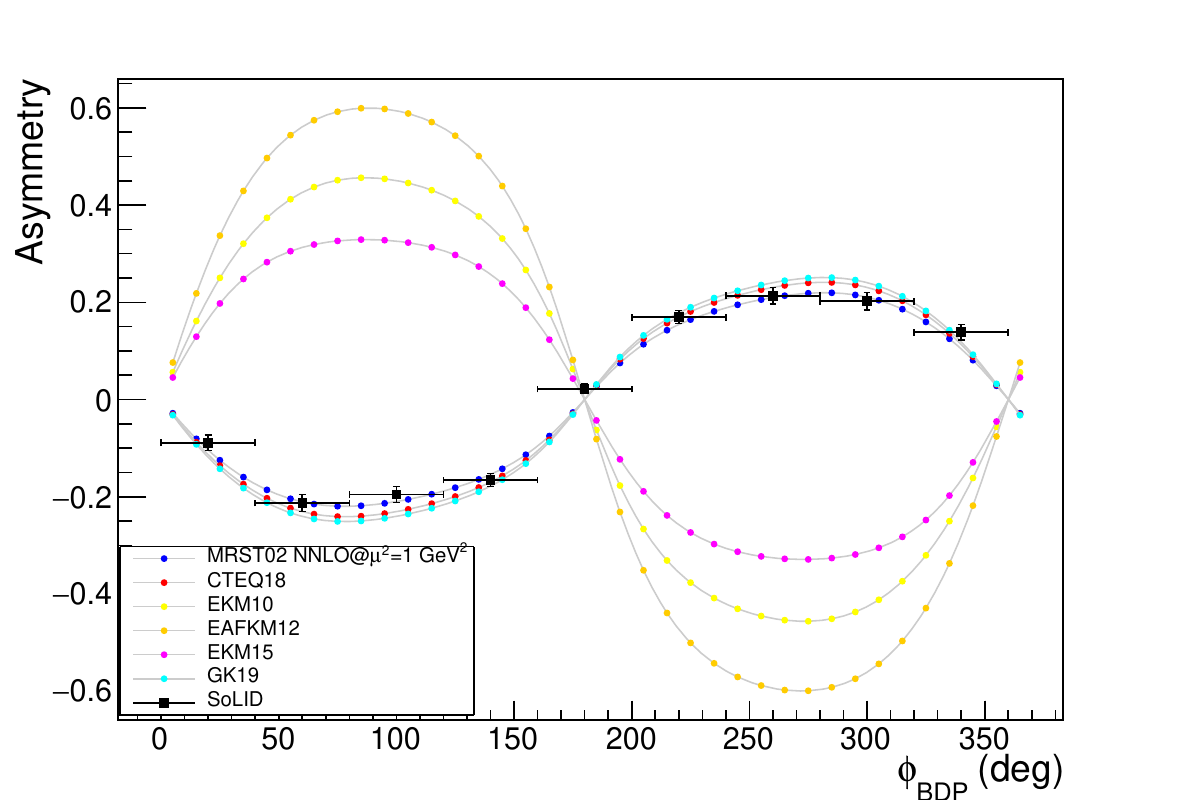}
        \caption{$A_{LU}$ in bin 1: $\xi'<-0.056$ and $\xi>0.45$.}
    \end{subfigure}
    \hspace{1.0cm}
    \begin{subfigure}[b]{0.45\textwidth}
        \centering
        \includegraphics[width=0.95\textwidth]{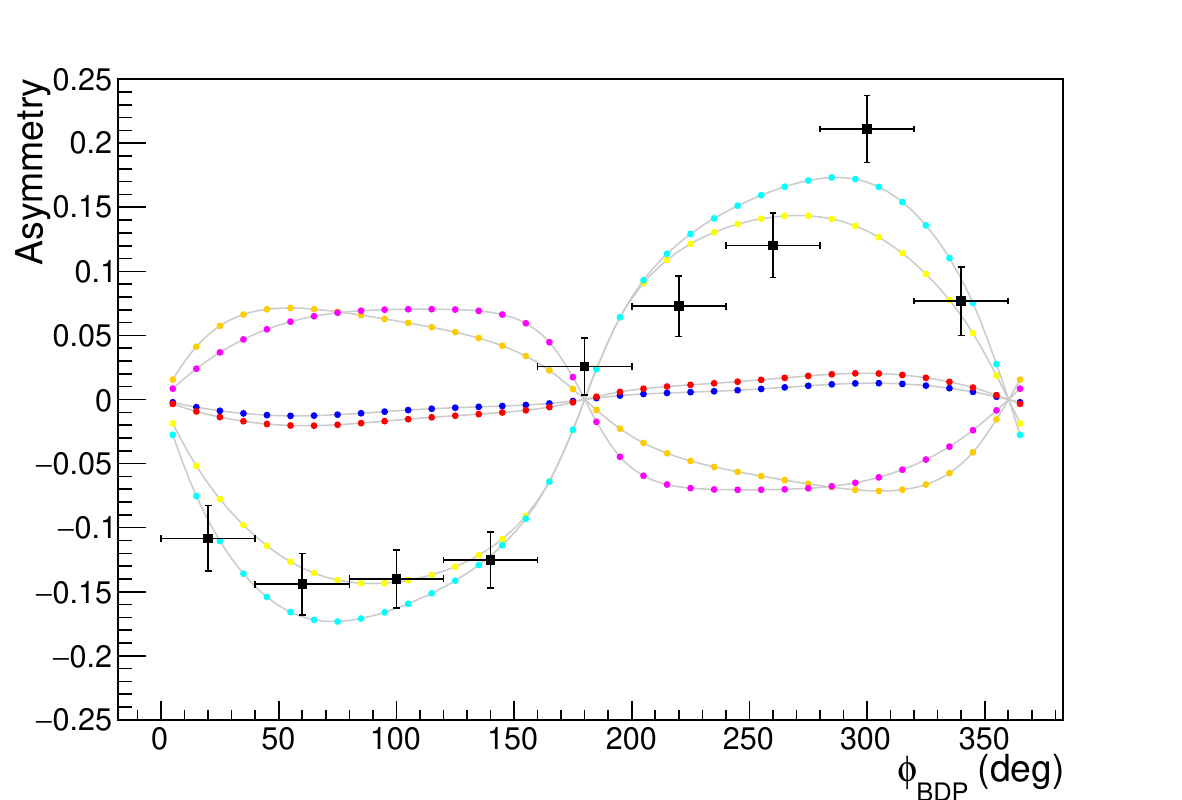}
        \caption{$A_{UL}$ in bin 6: $\xi'<-0.069$ and $\xi<0.107$.}
    \end{subfigure}
    \begin{subfigure}[b]{0.45\textwidth}
        \centering
        \includegraphics[width=0.95\textwidth]{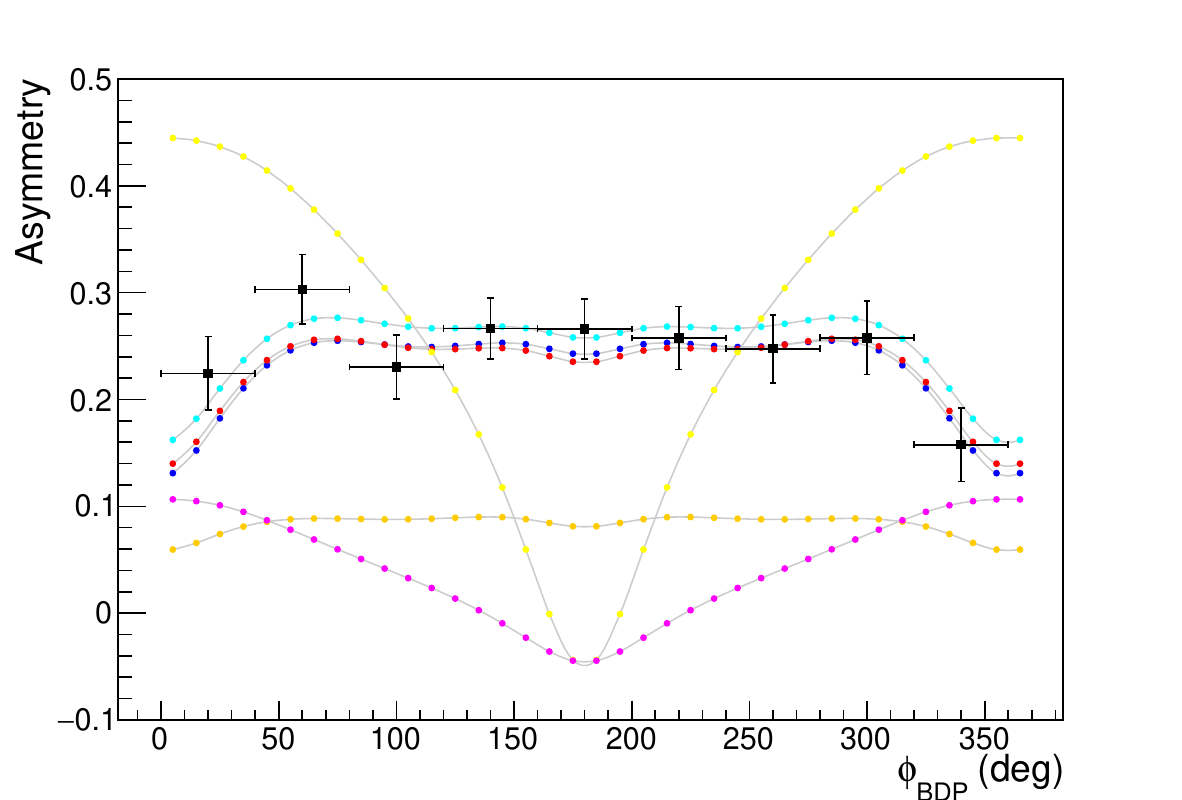}
        \caption{$A_{LL}$ in bin 10: $\xi'<-0.069$ and $0.139<\xi<0.45$.}
    \end{subfigure}
    \hspace{1.0cm}
    \begin{subfigure}[b]{0.45\textwidth}
        \centering
        \includegraphics[width=0.95\textwidth]{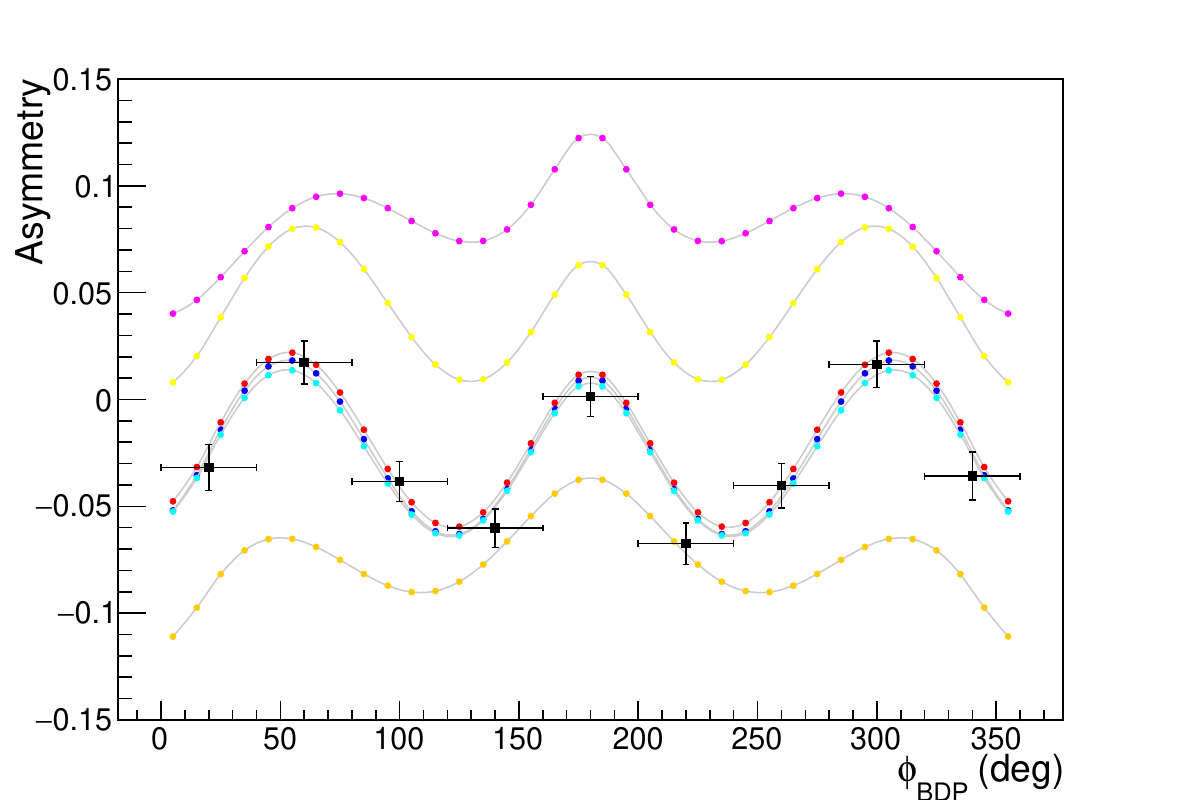}
        \caption{$A_{UU}^{C}$ in bin 6: $\xi'<-0.069$ and $\xi<0.107$.}
    \end{subfigure}
    \caption{\justifying{Asymmetry projections for DDVCS measurements with the SoLID$\mu$ spectrometer and an 11 GeV beam. Error bars represent statistical errors.}}
    \label{JLabMeas2}
\end{figure}

It is shown in Figs. \ref{JLabMeas} and \ref{JLabMeas2} sample experimental projections for the different DDVCS observables for each experimental set-up. We placed the BSA error bars on the smallest amplitude prediction to consider the most pessimistic scenario. In contrast, TSA and DSA error bars are placed on the largest predicted amplitude to illustrate the size of the asymmetries that can be measured with the same event count. In most cases, we observe similar VGG and GK19 predictions with large amplitudes facilitating their experimental measurements. An exception to such observation is the TSA predictions shown in Figs. \ref{JLabMeas} and \ref{JLabMeas2} where the VGG amplitude prediction is about $2\%$ and turns out experimentally compatible with a null asymmetry. The latter is an example of the model and GPD sensitivity of DDVCS observables providing, under certain conditions, null asymmetries due to a cancellation of the CFF combination of Eq. (\ref{AULeq}).

The observed cancellation in the shown TSA means that the observable is no longer $\widetilde{\mathcal{H}}$ dominated at that kinematic point, according to the VGG model. Thus, $\mathcal{H}$ acquires values that overcome the $\xi'$ suppression factor. Moreover, the EKM projections for the TSA shows a sign change resulting from an $ H$-dominated scenario. Such model predictions indicate either the need for a Mellin-Barnes representation for axial GPDs or the non-validity of the formalism for $\xi\neq \xi'$ values, thus requiring further phenomenological GPD studies. The same $H$ dominance in the EKM models is observed in the BSA predictions which, in addition to large amplitudes, present a sign change as expected from the model behavior on Fig. \ref{ImH_EKM}. Finally, the observed differences in the EKM prediction of the DSA and BCA result from $\mathfrak{Re}[\widetilde{\mathcal{H}}]$ values much larger than the VGG/GK prediction, thus overriding the $\mathcal{H}$ contribution in Eq. (\ref{ALLeq}). 

\begin{figure}[H]
    \centering
    \begin{subfigure}[b]{0.45\textwidth}
        \centering
        \includegraphics[width=0.95\textwidth]{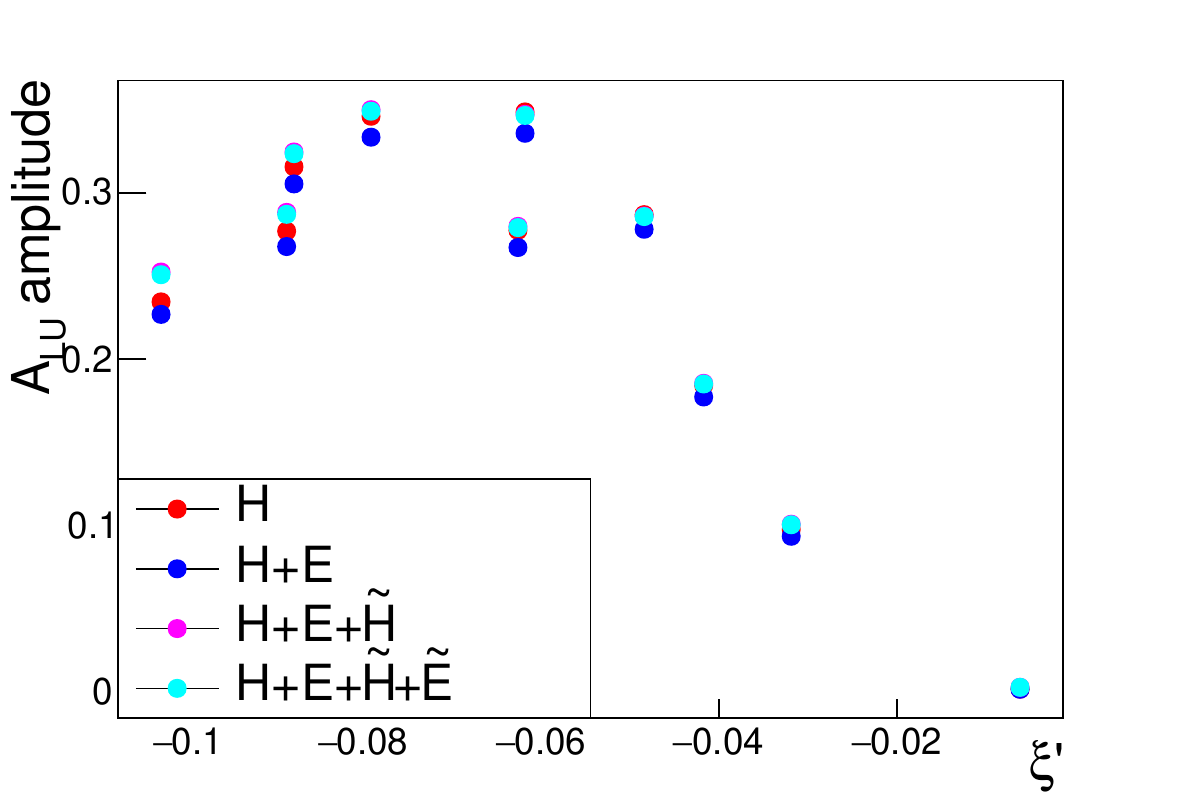}
        \caption{$A_{LU}$}
    \end{subfigure}
    \hspace{1.0cm}
    \begin{subfigure}[b]{0.45\textwidth}
        \centering
        \includegraphics[width=0.95\textwidth]{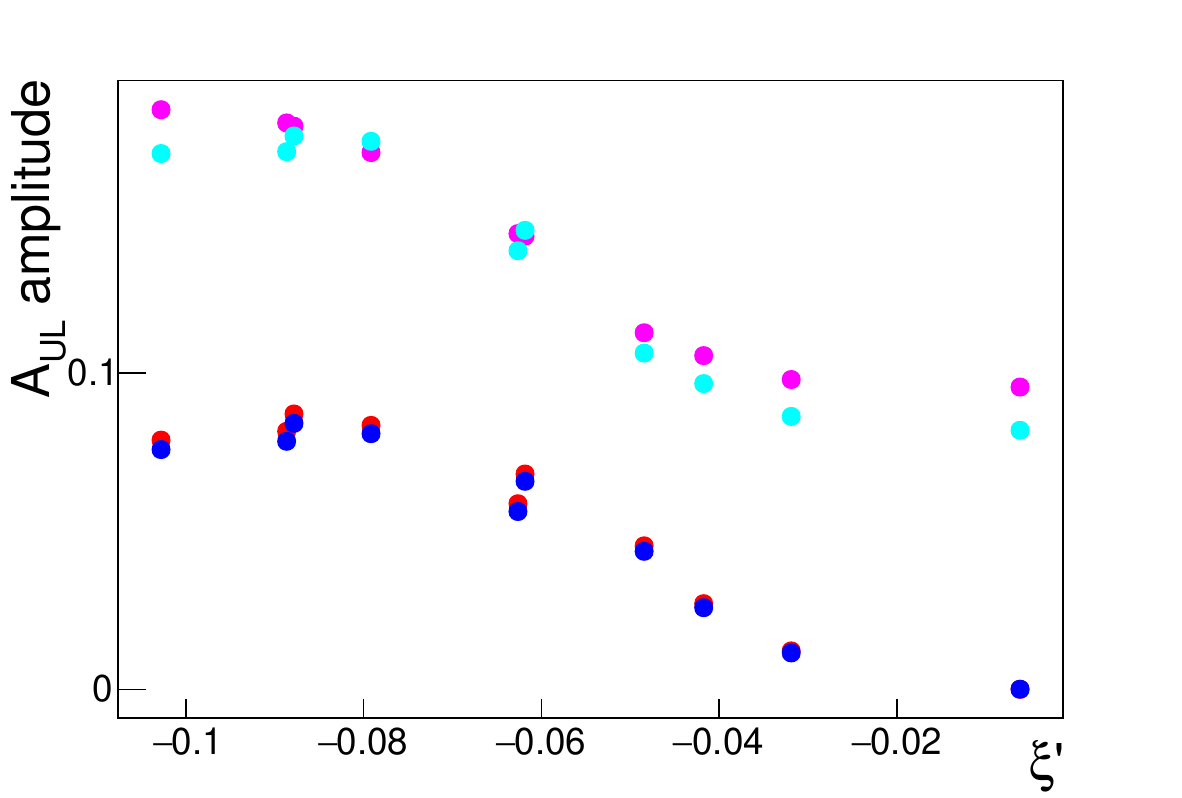}
        \caption{$A_{UL}$}
    \end{subfigure}
    \begin{subfigure}[b]{0.45\textwidth}
        \centering
        \includegraphics[width=0.95\textwidth]{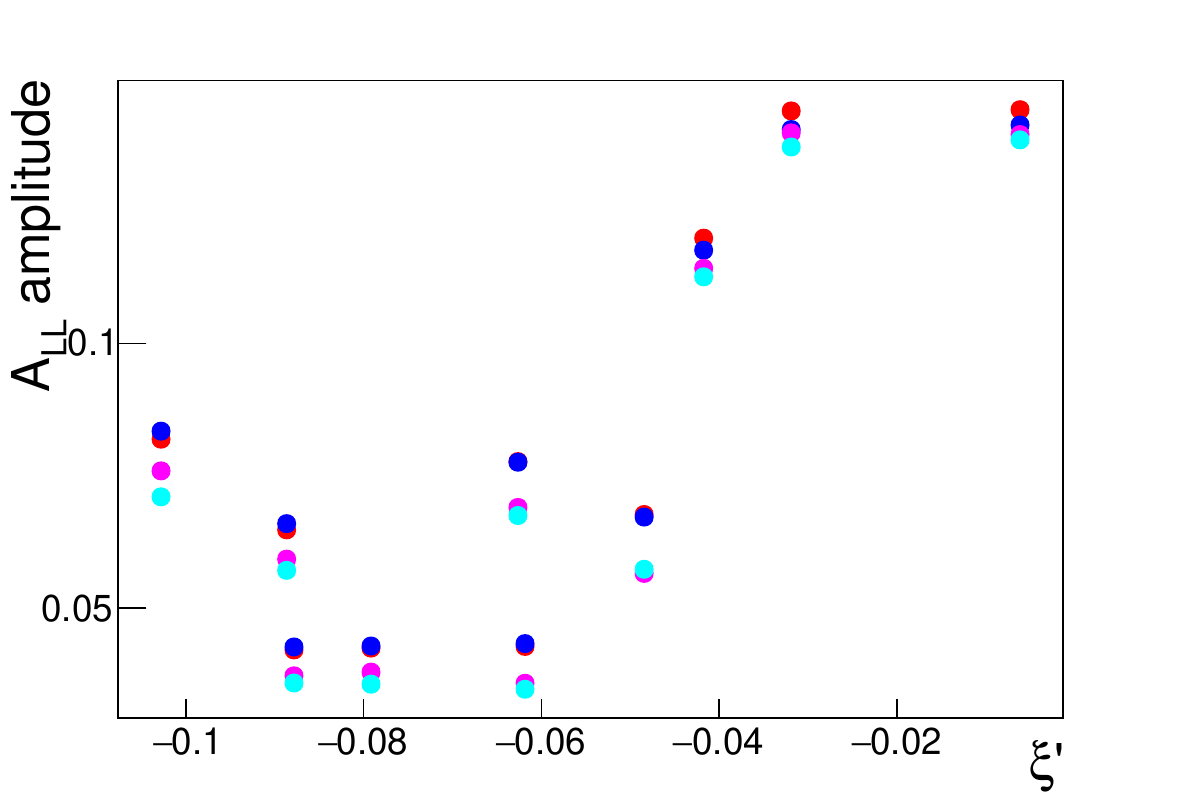}
        \caption{$A_{LL}$}
    \end{subfigure}
    \hspace{1.0cm}
    \begin{subfigure}[b]{0.45\textwidth}
        \centering
        \includegraphics[width=0.95\textwidth]{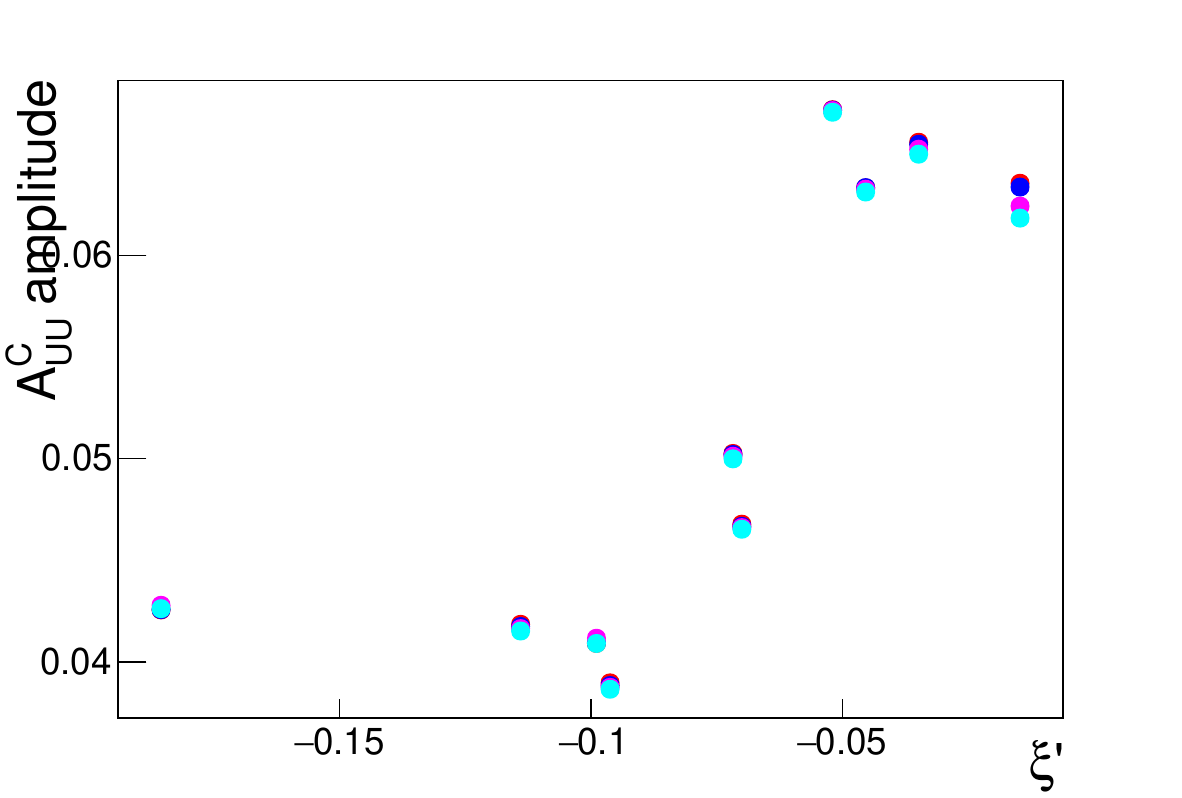}
        \caption{$A_{UU}^{C}$}
    \end{subfigure}
    \caption{GPD contribution to the DDVCS asymmetry amplitude according to the GK19 model. Case of an 11 GeV beam.}
    \label{JLabSen}
\end{figure}

\begin{figure}[H]
    \centering
    \begin{subfigure}[b]{0.45\textwidth}
        \centering
        \includegraphics[width=0.95\textwidth]{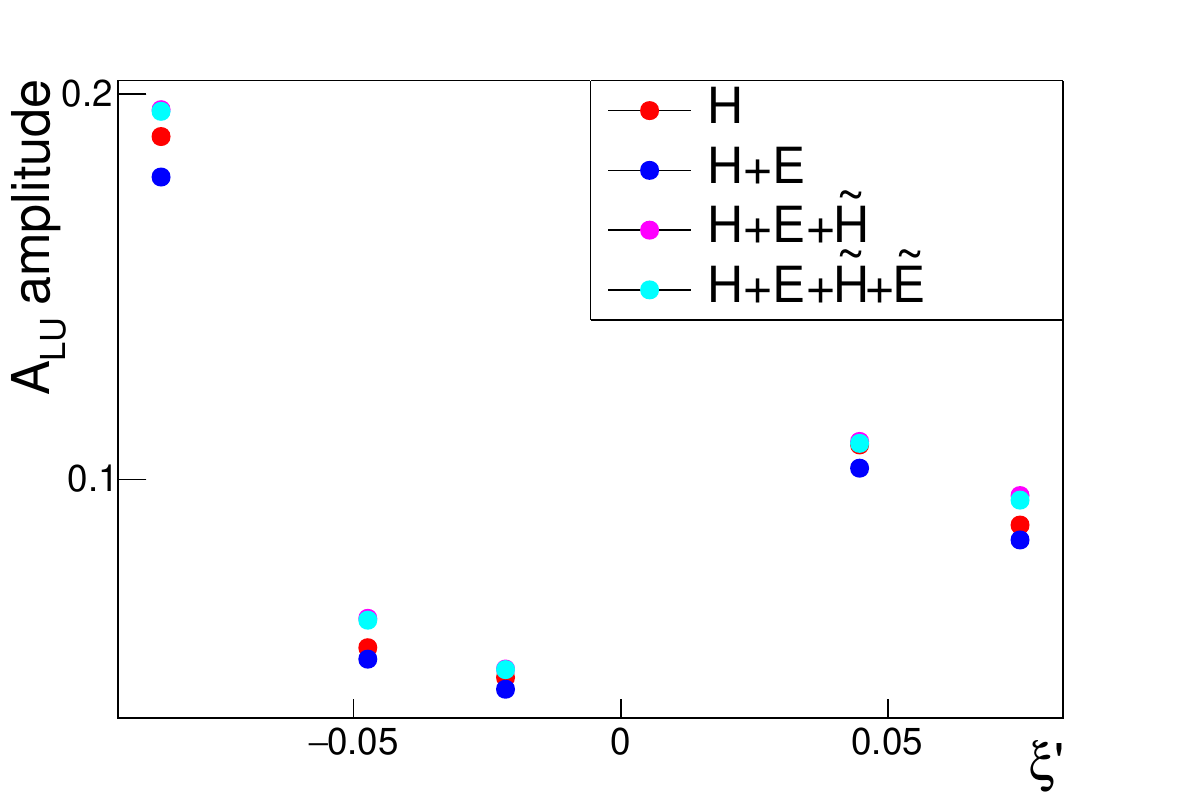}
        \caption{$A_{LU}$}
    \end{subfigure}
    \hspace{1.0cm}
    \begin{subfigure}[b]{0.45\textwidth}
        \centering
        \includegraphics[width=0.95\textwidth]{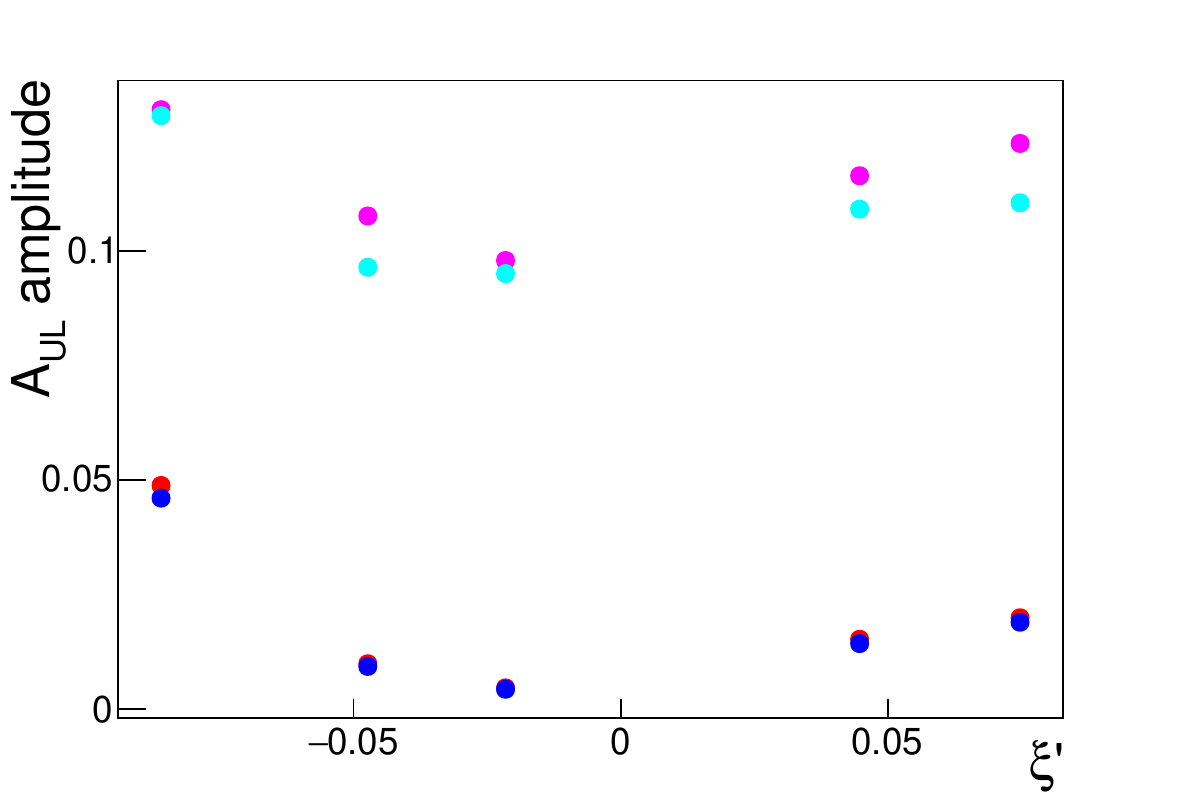}
        \caption{$A_{UL}$}
    \end{subfigure}
    \begin{subfigure}[b]{0.45\textwidth}
        \centering
        \includegraphics[width=0.95\textwidth]{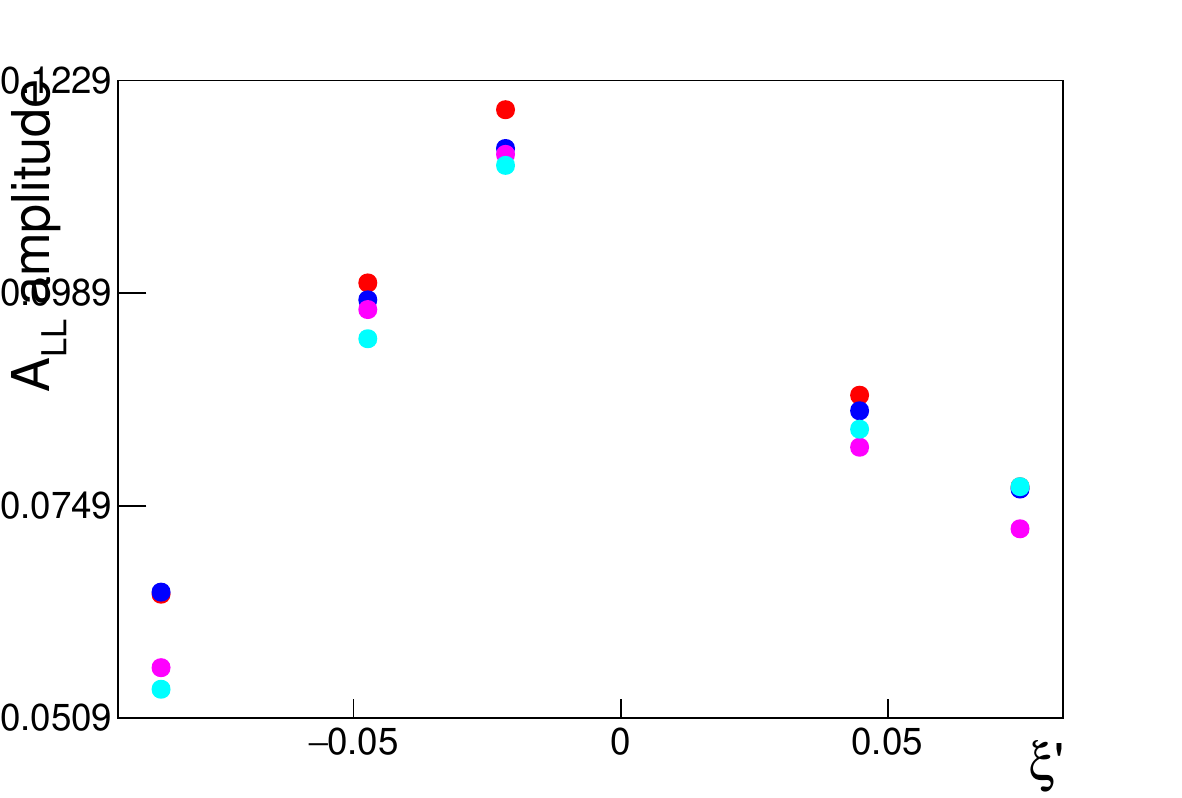}
        \caption{$A_{LL}$}
    \end{subfigure}
    \hspace{1.0cm}
    \begin{subfigure}[b]{0.45\textwidth}
        \centering
        \includegraphics[width=0.95\textwidth]{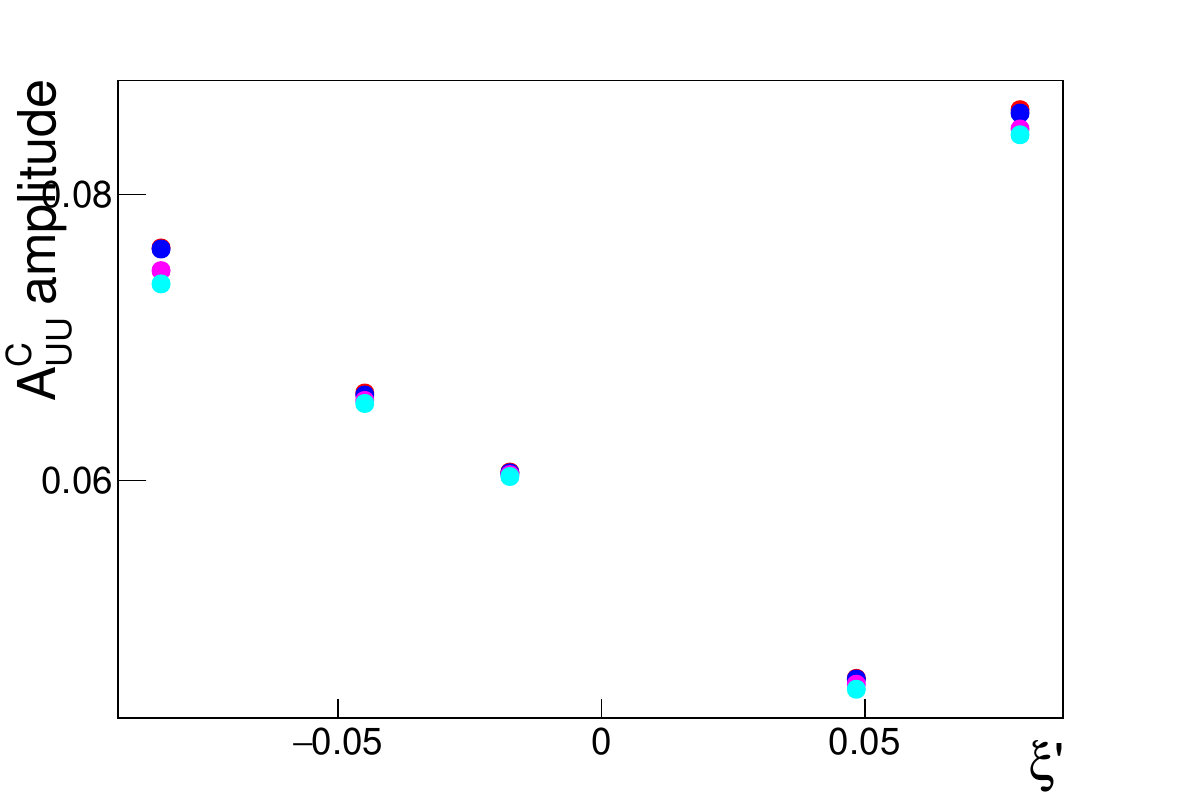}
        \caption{$A_{UU}^{C}$}
    \end{subfigure}
    \caption{GPD contribution to the DDVCS asymmetry amplitude according to the GK19 model. Case of a 22 GeV beam.}
    \label{JLabSen22}
\end{figure}

Regarding the overall GPD sensitivity of the observables, we show in Figs. \ref{JLabSen} and \ref{JLabSen22} the GK19 predictions as a function of the mean $\xi'$ value over bins, when the different GPD contributions are included in the computation. BSA, DSA, and BCA amplitudes are $H$ dominated while the TSA is dominated by $\widetilde{H}$ with a sizable contribution from $H$. Therefore, the GK19 predictions indicate that DDVCS measurements would mostly improve our knowledge of $H$ over the whole GPD phase space, and $\widetilde{H}$ in the target-polarized case, as other GPDs contribute by a few percent. This highlights the need for dedicated experiments to measure other GPDs. For instance, $E$ measurements through transversely-polarized target experiments \cite{transverse} or precise DDVCS $A_{UL}$ measurements at small $\xi'$ (see Eq. \ref{AULeq}) for $\widetilde{E}$ extraction.

\begin{figure}[ht]
    \centering
    \begin{subfigure}[b]{0.45\textwidth}
        \centering
        \includegraphics[width=0.95\textwidth]{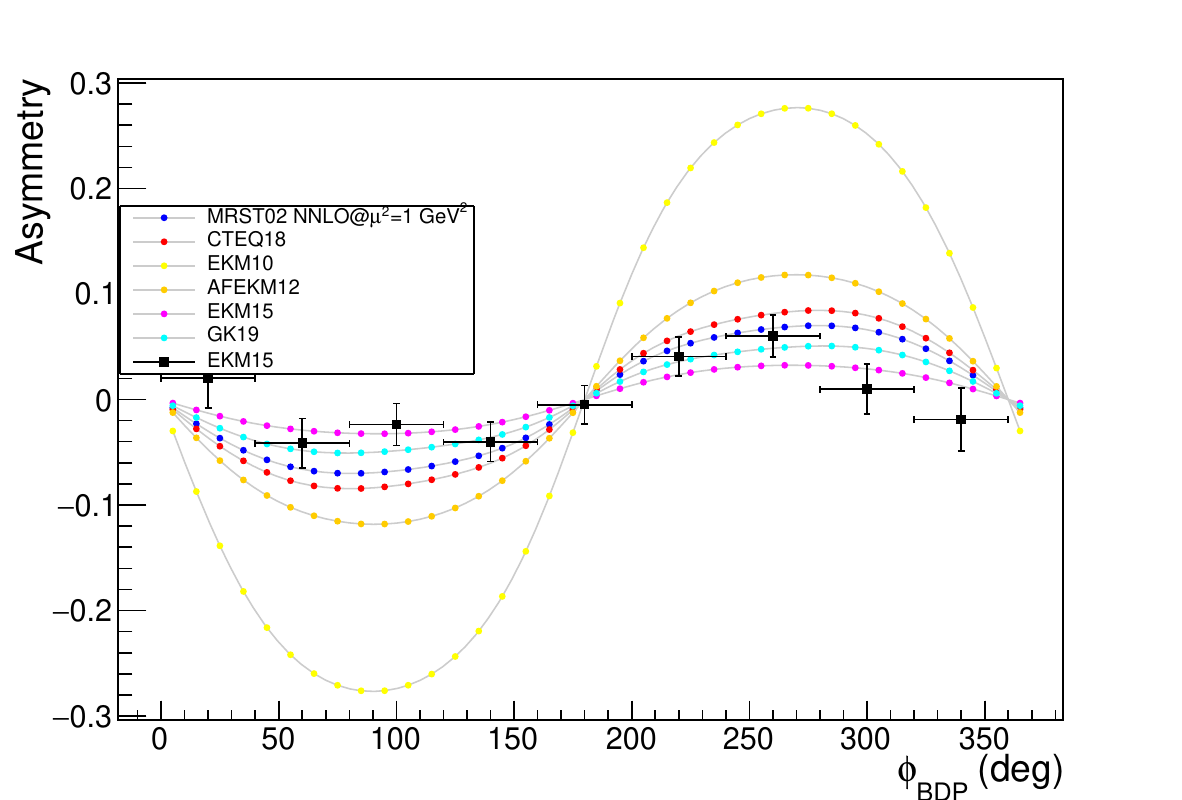}
        \caption{$A_{LU}$ in bin 5: $-0.053<\xi'<0.01$ and $\xi<0.214$.}
    \end{subfigure}
    \hspace{1.0cm}
    \begin{subfigure}[b]{0.45\textwidth}
        \centering
        \includegraphics[width=0.95\textwidth]{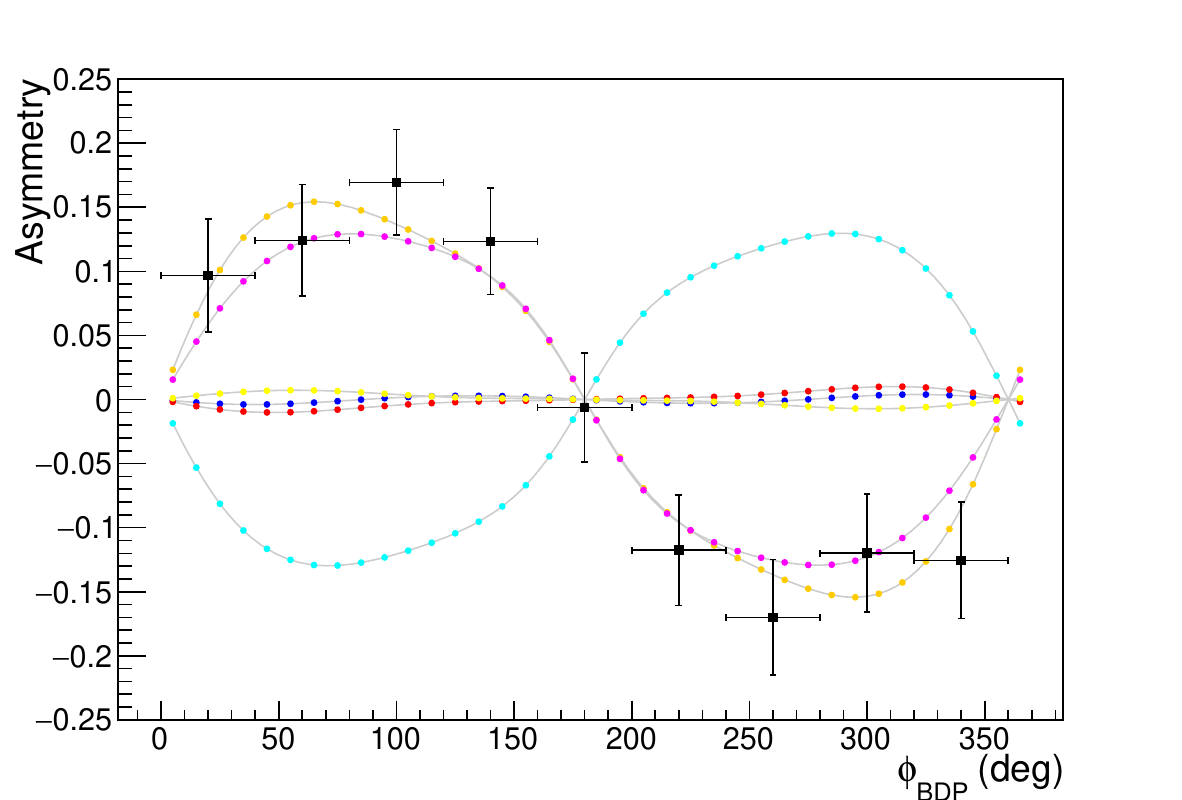}
        \caption{$A_{UL}$ in bin 4: $\xi'<-0.053$ and $\xi<0.214$.}
    \end{subfigure}
    \begin{subfigure}[b]{0.45\textwidth}
        \centering
        \includegraphics[width=0.95\textwidth]{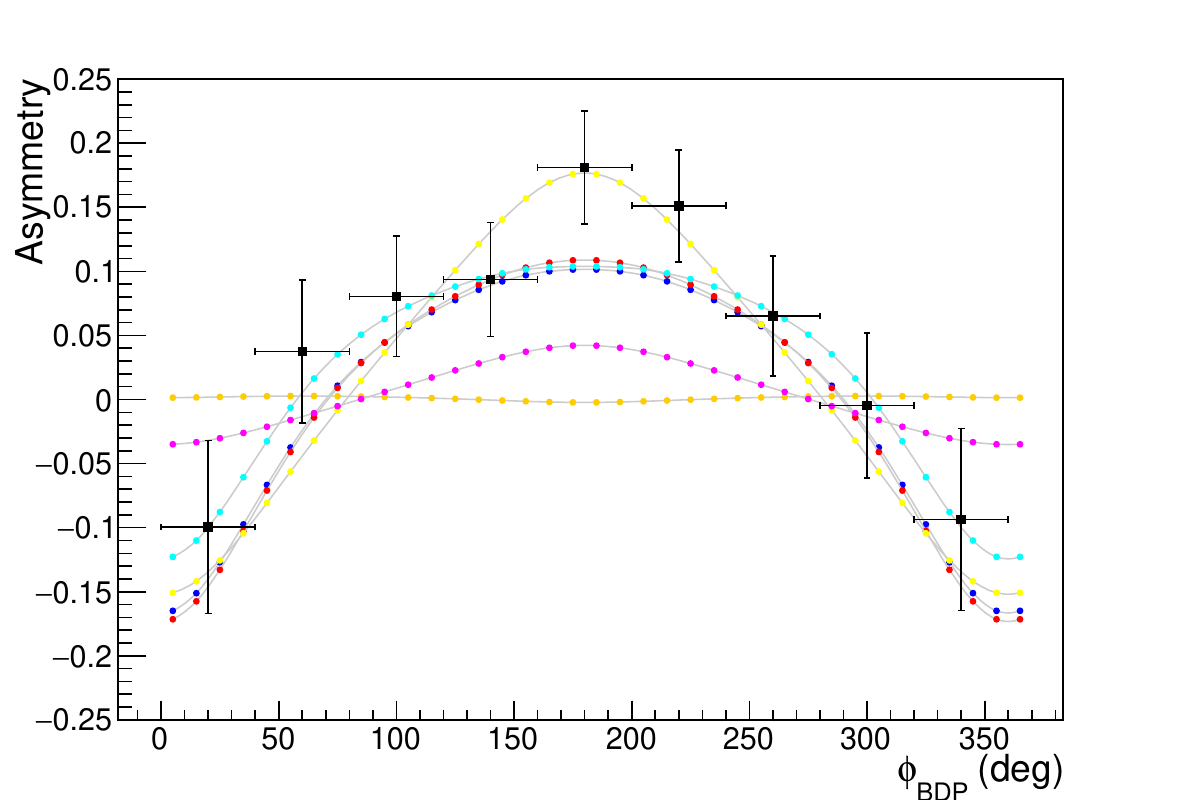}
        \caption{$A_{LL}$ in bin 5: $-0.053<\xi'<0.01$ and $\xi<0.214$.}
    \end{subfigure}
    \hspace{1.0cm}
    \begin{subfigure}[b]{0.45\textwidth}
        \centering
        \includegraphics[width=0.95\textwidth]{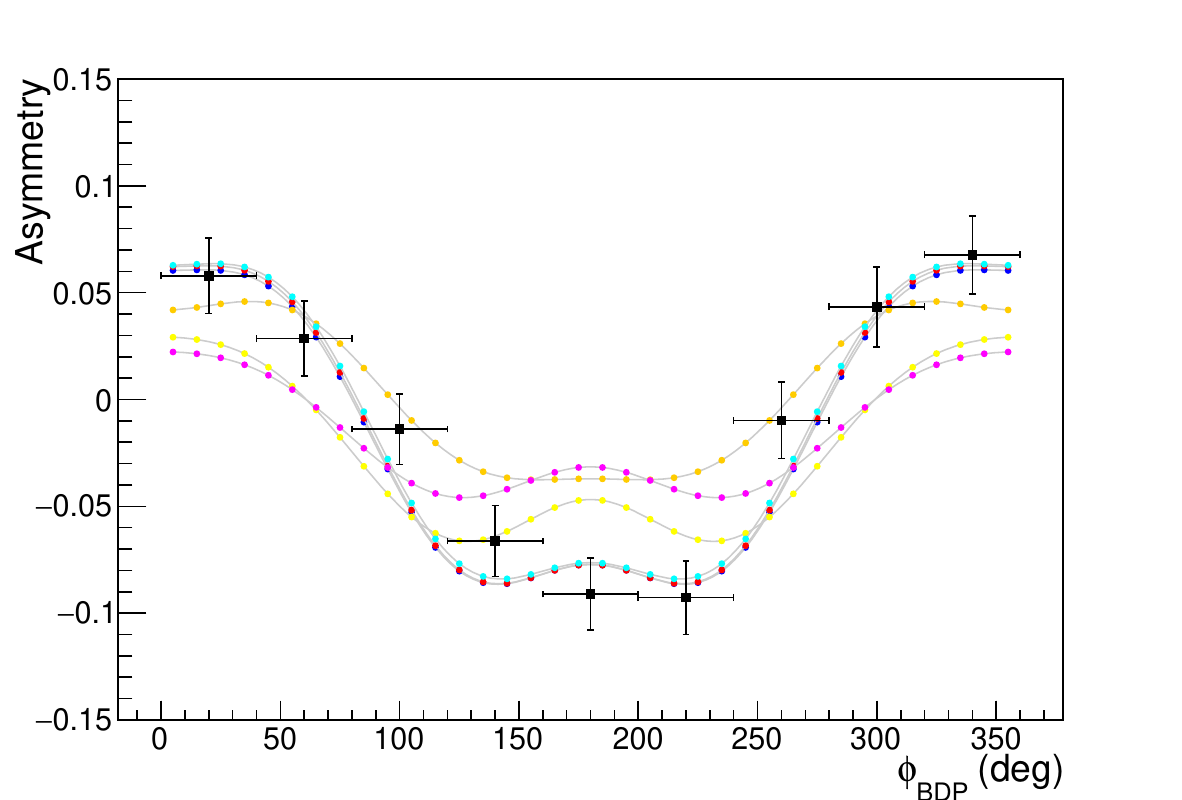}
        \caption{$A_{UU}^{C}$ in bin 3: $\xi'>0.01$ and $\xi>0.218$.}
    \end{subfigure}
    \caption{\justifying{Asymmetry projections for DDVCS measurements with the $\mu$CLAS spectrometer and a 22 GeV beam. Error bars represent statistical errors.}}
    \label{JLab22Meas}
\end{figure}

\begin{figure}[ht]
    \centering
    \begin{subfigure}[b]{0.45\textwidth}
        \centering
        \includegraphics[width=0.95\textwidth]{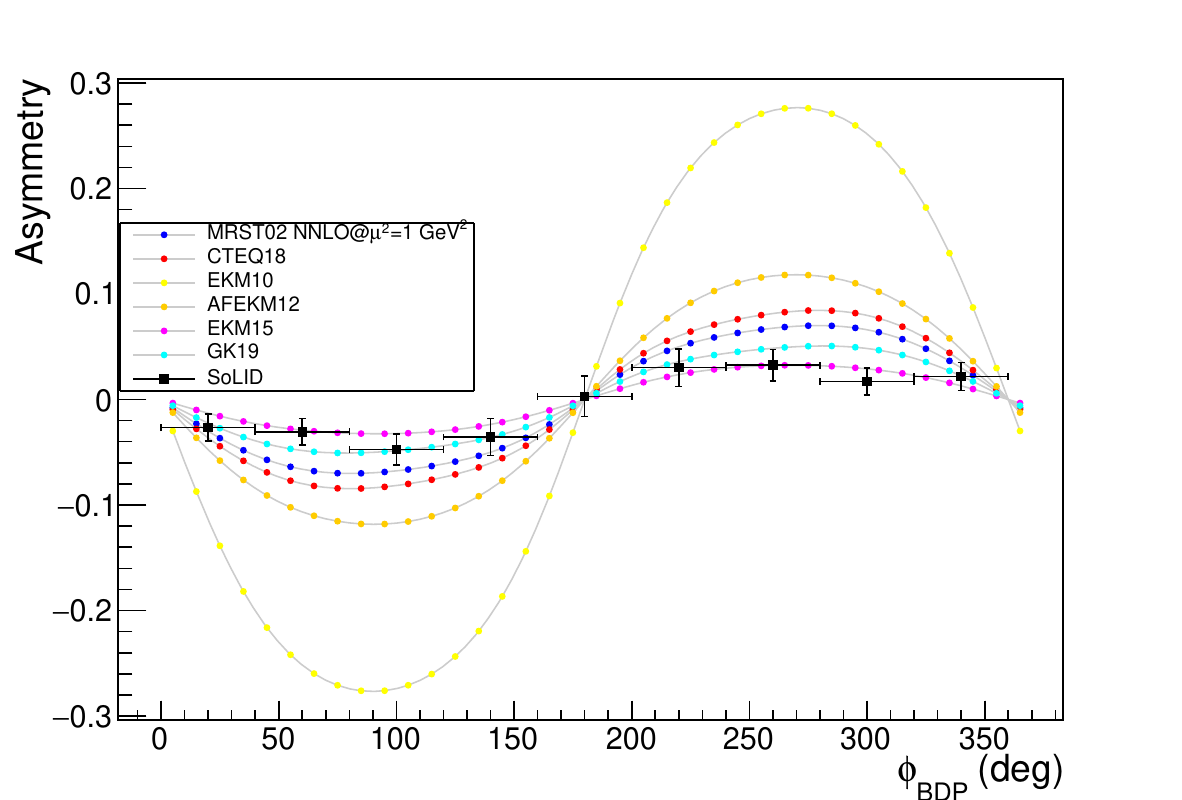}
        \caption{$A_{LU}$ in bin 5: $-0.053<\xi'<0.01$ and $\xi<0.214$.}
    \end{subfigure}
    \hspace{1.0cm}
    \begin{subfigure}[b]{0.45\textwidth}
        \centering
        \includegraphics[width=0.95\textwidth]{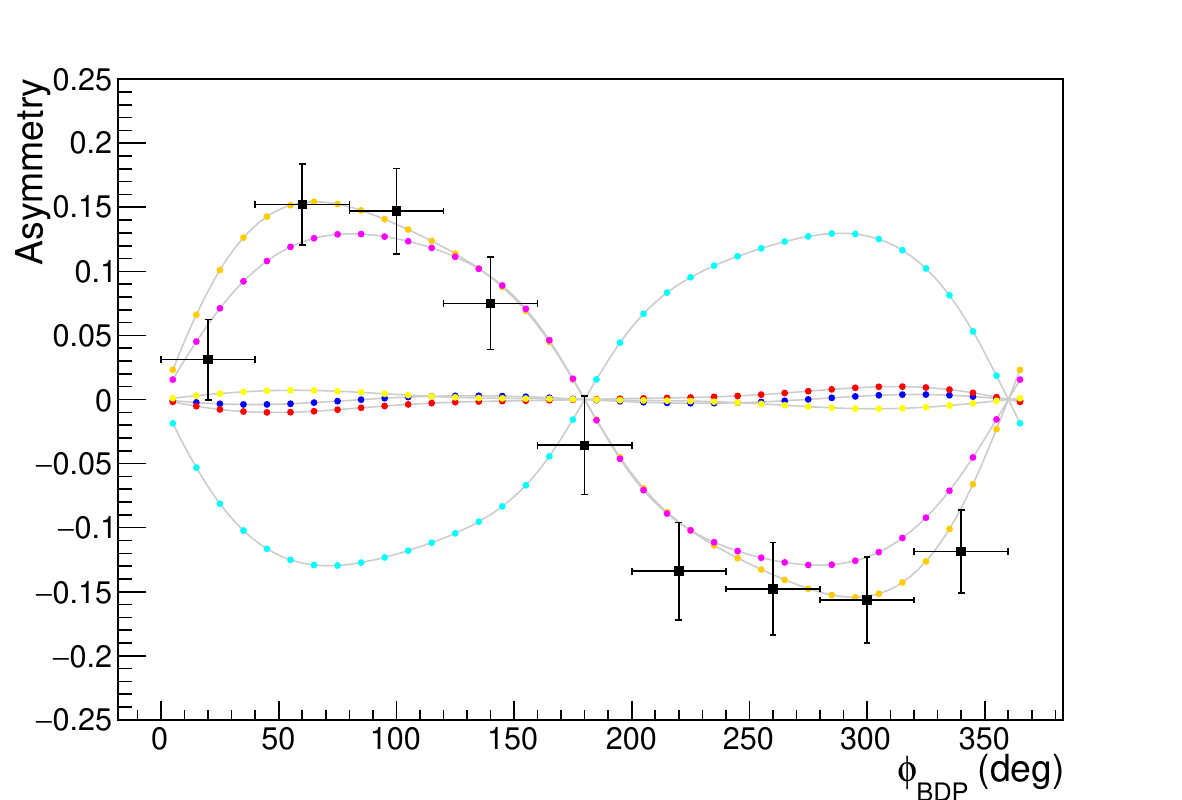}
        \caption{$A_{UL}$ in bin 4: $\xi'<-0.053$ and $\xi<0.214$.}
    \end{subfigure}
    \begin{subfigure}[b]{0.45\textwidth}
        \centering
        \includegraphics[width=0.95\textwidth]{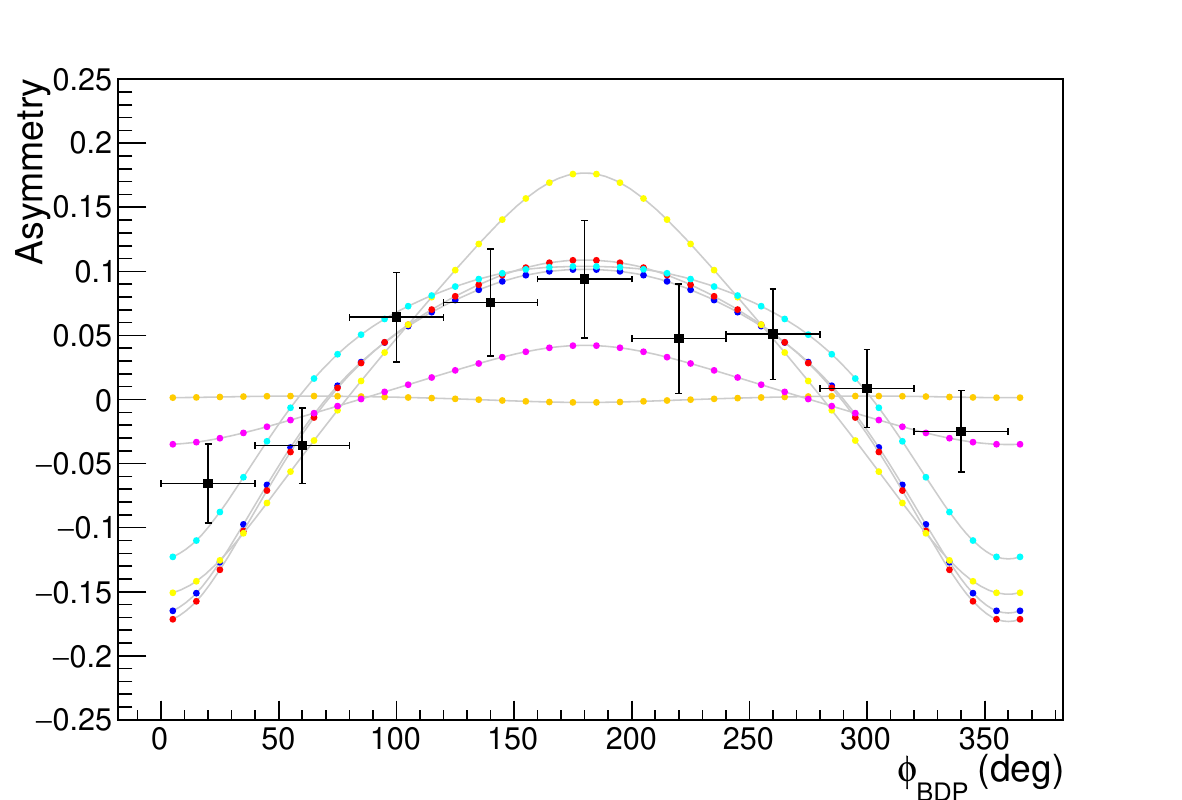}
        \caption{$A_{LL}$ in bin 5: $-0.053<\xi'<0.01$ and $\xi<0.214$.}
    \end{subfigure}
    \hspace{1.0cm}
    \begin{subfigure}[b]{0.45\textwidth}
        \centering
        \includegraphics[width=0.95\textwidth]{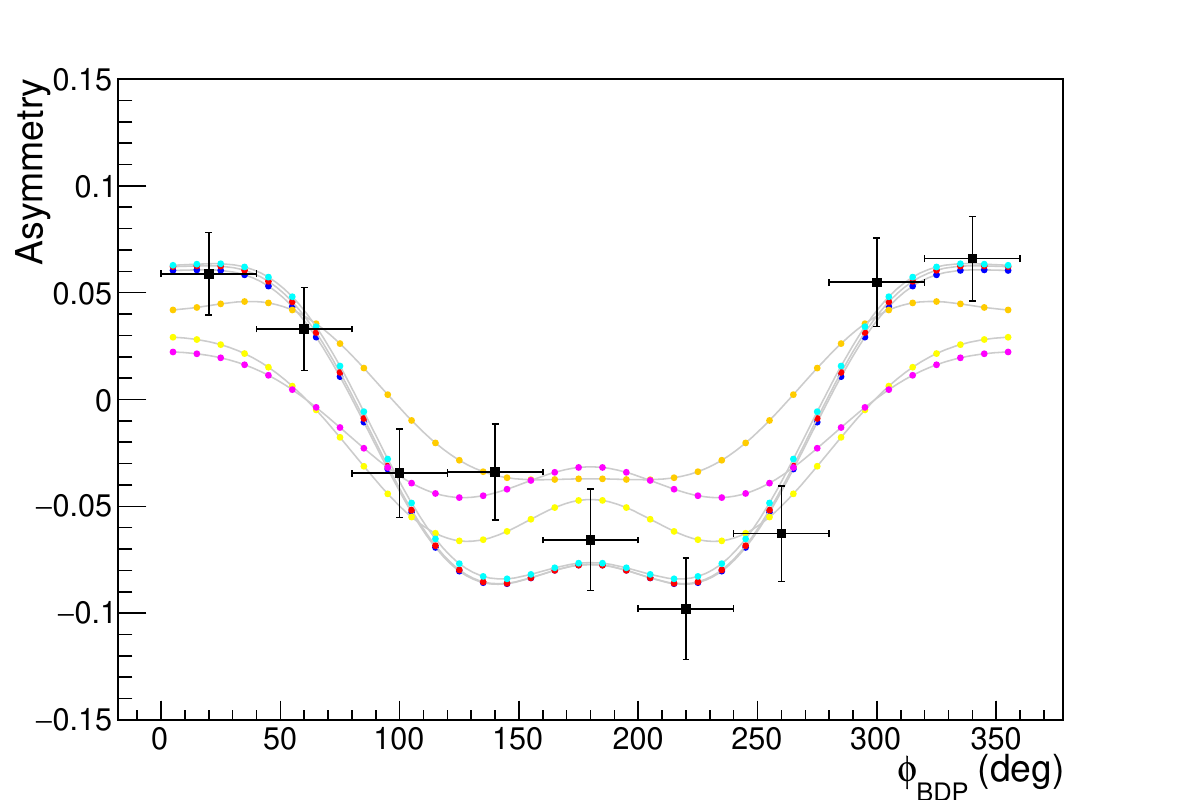}
        \caption{$A_{UU}^{C}$ in bin 3: $\xi'>0.01$ and $\xi>0.218$.}
    \end{subfigure}
    \caption{\justifying{Asymmetry projections for DDVCS measurements with the SoLID$\mu$ spectrometer and a 22 GeV beam. Error bars represent statistical errors.}}
    \label{JLab22Meas2}
\end{figure}

\clearpage
\twocolumngrid

%Unlike the spin-dependent observables, BCA model predictions are plotted as a function of $\phi_{\text{trento}}$ as larger amplitudes are obtained with such functional dependence.

%\textcolor{blue}{Finally, one might wonder about cross-section measurements at JLab. The cross-section can be computed as $\sigma=N/(\varepsilon\mathcal{L}T)$ with statistical error given by $\Delta\sigma=\sqrt{\sigma/(\varepsilon\mathcal{L}T)}$ where $N$ is the reconstructed number of events, $\mathcal{L}$ the luminosity, $T$ the data taking time and $\varepsilon$ an efficiency factor. As the DDVCS cross-section is at the pico-barn order of magnitude, we would be interested in cross-section measurements within $0.1-1 \%$ accuracy \textit{i.e.} $\Delta \sigma$ within $0.001\sigma-0.01 \sigma$. Therefore, the required integrated luminosity $\mathcal{LT}$ for such measurements is about $10^{40} - 10^{42}$ cm$^{-2}$ which is easily achievable within 100 days of beam time ($T=8.64\times10^{6}\;\mathrm{s}$).} and a $1\%$ detection efficiency $\varepsilon=0.01$

On the experimental side, we consider a nominal luminosity of $\mathcal{L}$=$10^{37}\;\mathrm{cm}^{-2}\cdot\mathrm{s}^{-1}$ to balance the small cross-section of the DDVCS process. For an 11 GeV beam, we consider 100 days of data taking while for a 22 GeV the value is doubled. We can observe from the experimental projections in Figs. \ref{JLab22Meas} and \ref{JLab22Meas2} that the detection capabilities of the $\mu$CLAS and SoLID$\mu$ detectors are similar for both beam energies. As a result, the expected accuracy would allow measurements of all the spin-dependent observables according to model predictions. In particular, BSA and DSA measurements are predicted to be accurately measured thanks to the large amplitude predicted by all models. In contrast, the experimental accuracy would allow TSA and BCA measurements if the observed amplitude is about $5\%$ or larger. Besides, it would discriminate the VGG and GK19 models.

\subsection{EIC experimental configuration}\label{EICkin}

In the following, we study the possible DDVCS measurements from electron-proton collisions at the Electron-Ion Collider (EIC). Unlike the JLab experimental configuration, the EIC will allow measurements of the sea-quark and gluon contributions to GPDs at $x_{B}$ values as small as $10^{-4}$. Beam and target-polarized configurations are part of the baseline capabilities of EIC allowing for BSA, TSA and DSA measurements. At a later stage, positron beam and muon detection capabilities may also be considered due to their physics impact \cite{EICpos1, EICpos2, EICpos3}. %, giving access to charge-sensitive observables like the BCA and detection of exclusive processes like DDVCS or $J/\psi$ photo-production. %In fact, preliminary TCS and $J/\psi$ capabilities have also been investigated \cite{carlos} with the ECCE detector \cite{ECCE1, ECCE2, ECCE3}. 

As the EIC will access $x_{B}$ values down to $10^{-4}$, DDVCS measurements will be in the $\xi, \xi' \ll t/4M^{2}$ regime. Therefore, the CFF dependence of the observables in Eqs. (\ref{BSA})-(\ref{ALLeq}) simplifies to:
\begin{align}
    A_{LU}^{\sin(\phi)} &\propto \mathfrak{Im}\left[F_{1}\mathcal{H} - \frac{t}{4M^{2}}F_{2}\mathcal{E}\right] , \\
    A_{UU}^{C\; \cos(\phi)} &\propto \frac{\xi'}{\xi}\mathfrak{Re}\left[F_{1}\mathcal{H} - \frac{t}{4M^{2}}F_{2}\mathcal{E}\right], \\
    A_{UL}^{\sin(\phi)} &\propto \mathfrak{Im}\left[F_{1}\tilde{\mathcal{H}}  \right],  \\   
    A_{LL}^{\cos(\phi)} &\propto \mathfrak{Re}[F_{1}\tilde{\mathcal{H}}].
\end{align}

Despite most of the CFF dependence of the experimental observables being lost, the EIC offers the opportunity to perform accurate measurements of GPD $H$ and $\widetilde{H}$ with corrections coming from GPD $E$. Recent measurements of neutron DVCS provide information of $E$ for the neutron \cite{Mostafa}. However, proton $E$ measurements are still required to restrict $E$ models of the proton and perform flavor separation. 

The expected luminosity of $10-100\;\mathrm{fb}^{-1}\cdot\mathrm{year}^{-1}$ translating into $10^{33}-10^{34}\;\mathrm{cm}^{-2}\cdot\mathrm{s}^{-1}$, three to four orders of magnitude smaller than the foreseen JLab luminosity. As the DDVCS cross-section increases with decreasing values of $x_{B}$, the kinematic reach of the EIC can compensate for the smaller luminosity. However, the amplitude of the spin asymmetries decreases as well. In the following, we will consider a nominal luminosity of $10\;\mathrm{fb}^{-1}\cdot\mathrm{y}^{-1}$, expected for the starting operation, electrons at 5 GeV and protons at 41 GeV. The beam energy configuration is such that sizable asymmetries are obtained according to the different model predictions.

\begin{figure}[H]
    \centering
    \begin{subfigure}[b]{0.48\textwidth}
        \centering
        \includegraphics[width=\textwidth]{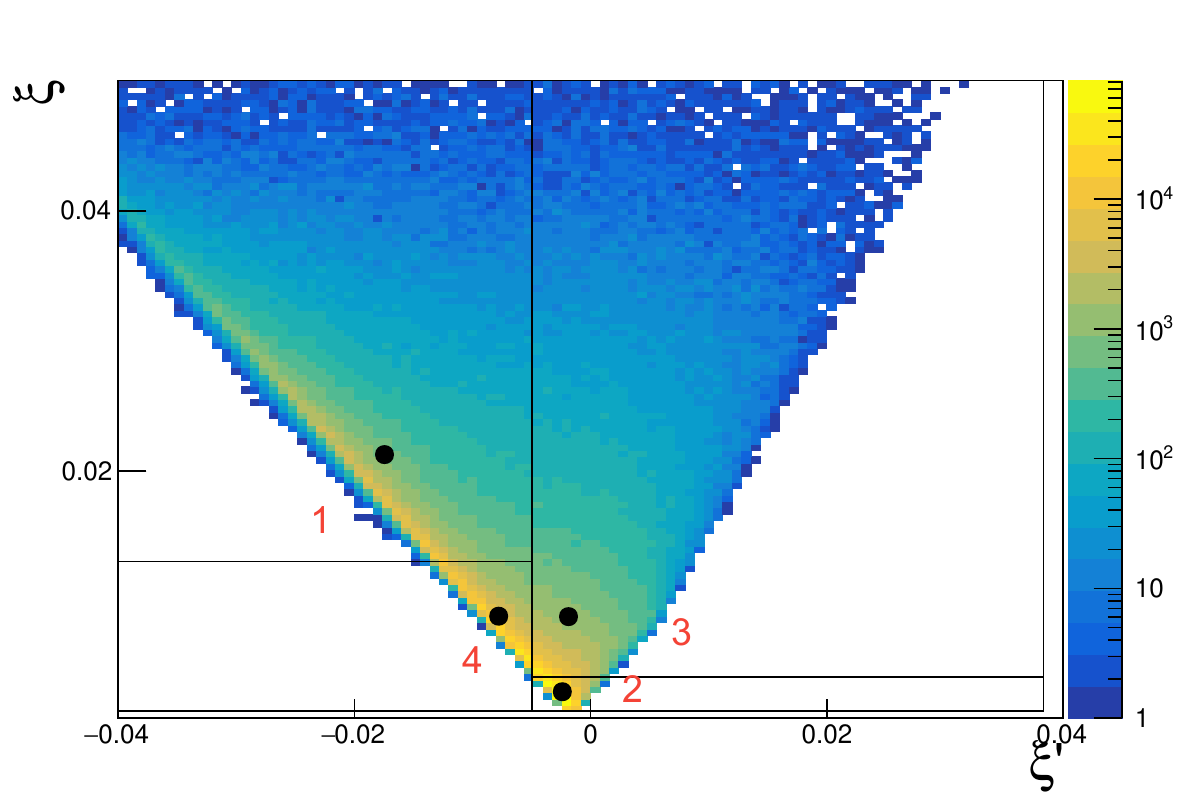}
        \caption{EIC-($5\times 41$) $\xi'$ vs $\xi$ acceptance with dots representing the mean value of the bin.}
    \end{subfigure}
    \begin{subfigure}[b]{0.48\textwidth}
        \centering
        \includegraphics[width=\textwidth]{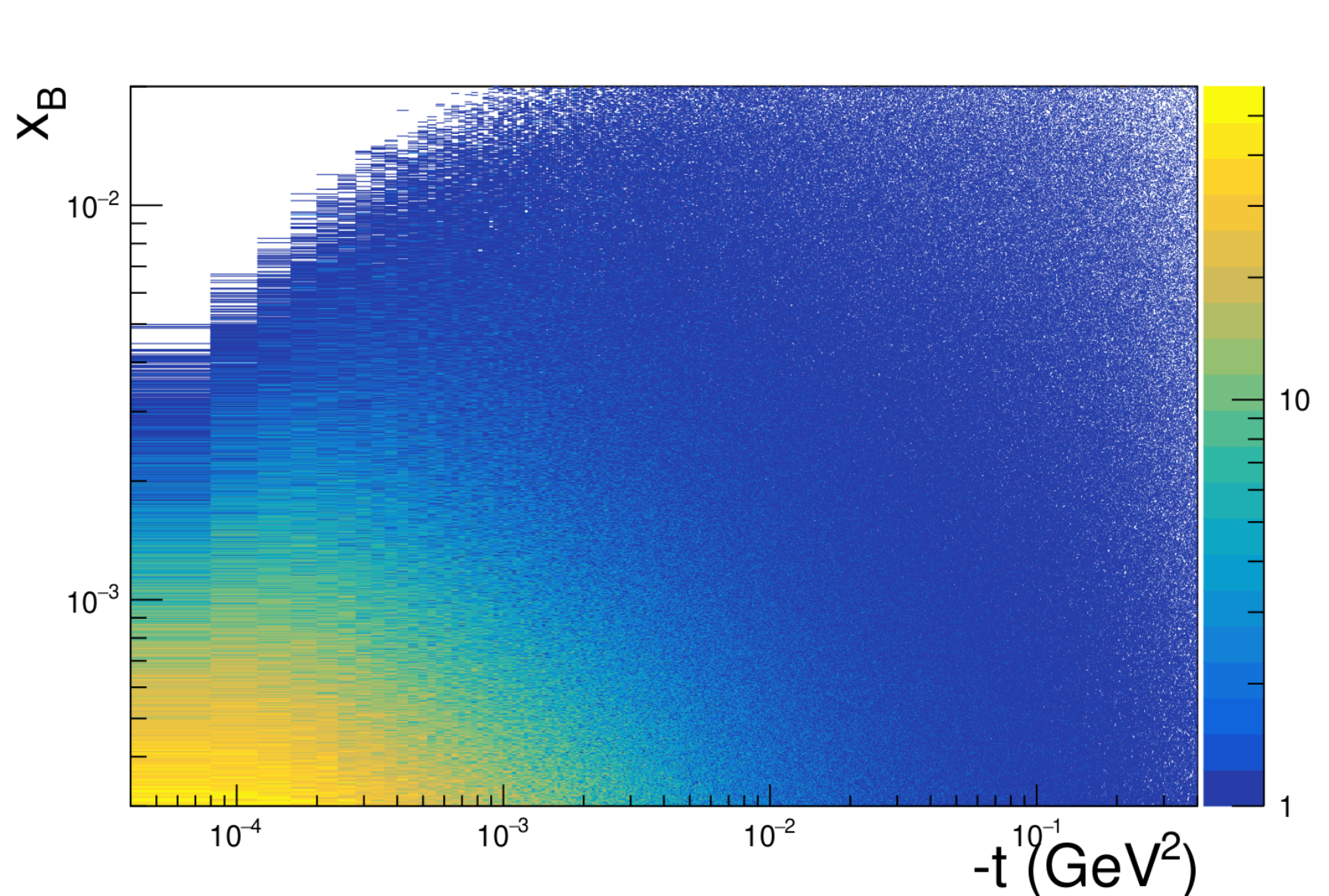}
        \caption{EIC-($5\times 41$) $-t$ vs $x_{B}$ acceptance.}
    \end{subfigure}
    \begin{subfigure}[b]{0.48\textwidth}
        \centering
        \includegraphics[width=\textwidth]{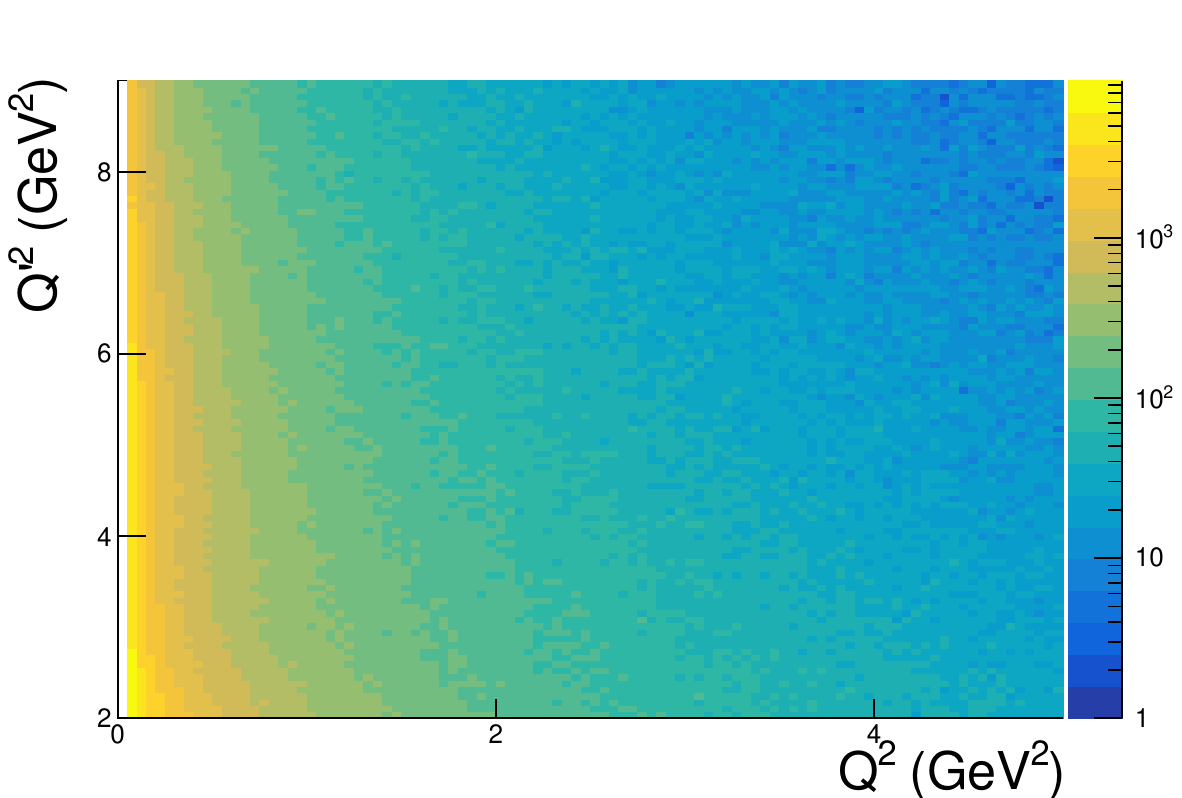}
        \caption{EIC-($5\times 41$) $Q^{2}$ vs $Q^{\prime 2}$ acceptance.}
    \end{subfigure}
    \caption{Binning scheme of the $(\xi',\xi)$-phase space for the EIC configuration.}
    \label{PSEIC}
\end{figure}

To compute statistical errors we generated $10^{7}$ events using EpIC \cite{EpIC} considering the ePIC detector acceptance for electrons \cite{ePIC_det}($2\degree <\theta < 178\degree$) and muons in the $2\degree<\theta<140\degree$ range as suggested in \cite{carlos}, and a global tracking efficiency of 0.7. Nucleon information is obtained from four-momentum conservation. The phase space associated with such conditions is shown in Fig. \ref{PSEIC} where a binning scheme was defined over the $(\xi', \xi)$ domain. Similar to the JLab case, the number of bins is determined by the size of the asymmetry amplitudes and the expected detector accuracy. In the EIC case, model predictions point to asymmetries suppressed by the relatively large unpolarized cross-section in the small $x_{B}$ limit. Therefore, bins enclosing a large amount of events are required.  

Despite the phase space covering the DVCS-like and TCS-like regions, the bulk of events is located close to the TCS limit, thus shifting the mean value of the bin to the $\xi'<0$ region. One can also notice that the events accumulate at small $Q^{2}$ values. As the maximum kinematically allowed value for $Q^{2}$ is given by $Q_{max}^{2}=2MEx_{B}$, $E=437$ GeV being the equivalent fixed-target beam energy in the $(5\times 41)$ configuration, the extent of the distribution is limited by the small accessed $x_{B}$ values. Likewise, the accumulation of events at small $t$ is fixed by the kinematic limits of the reaction.

Highly suppressed target-polarized and beam charge observables are found as the value of GPDs does not compensate for the kinematic factors. In such a case, they are not experimentally accessible as their amplitudes are smaller than $2\%$ for all bins and for both $\phi$ and $\varphi_{\ell}$ dependencies. Particularly for the EKM models, null target-polarized asymmetries are obtained as there is no Mellin-Barnes support for $\widetilde{H}$ and $\widetilde{E}$, only the valence-quark contribution. Considering that our current GPD knowledge mainly constrains the valence contribution to GPDs, one might question the validity of such models at EIC kinematics where the sea-quark contribution dominates. Therefore, new measurements are needed to resolve this issue.

\begin{figure}[H]
    \centering
    \begin{subfigure}[b]{0.45\textwidth}
        \centering
        \includegraphics[width=0.95\textwidth]{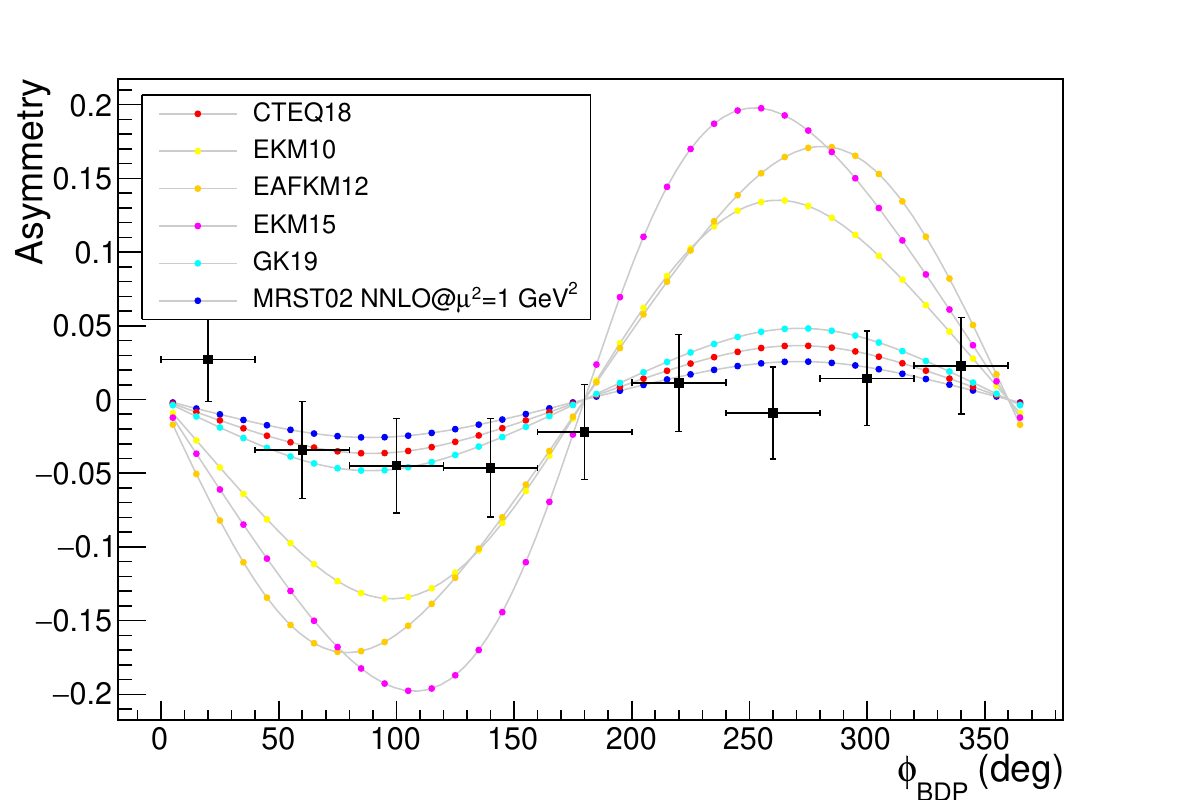}
        \caption{Bin 1: $\xi'<-0.005$ and $\xi>0.013$.}
    \end{subfigure}
    \hspace{1.0cm}
    \begin{subfigure}[b]{0.45\textwidth}
        \centering
        \includegraphics[width=0.95\textwidth]{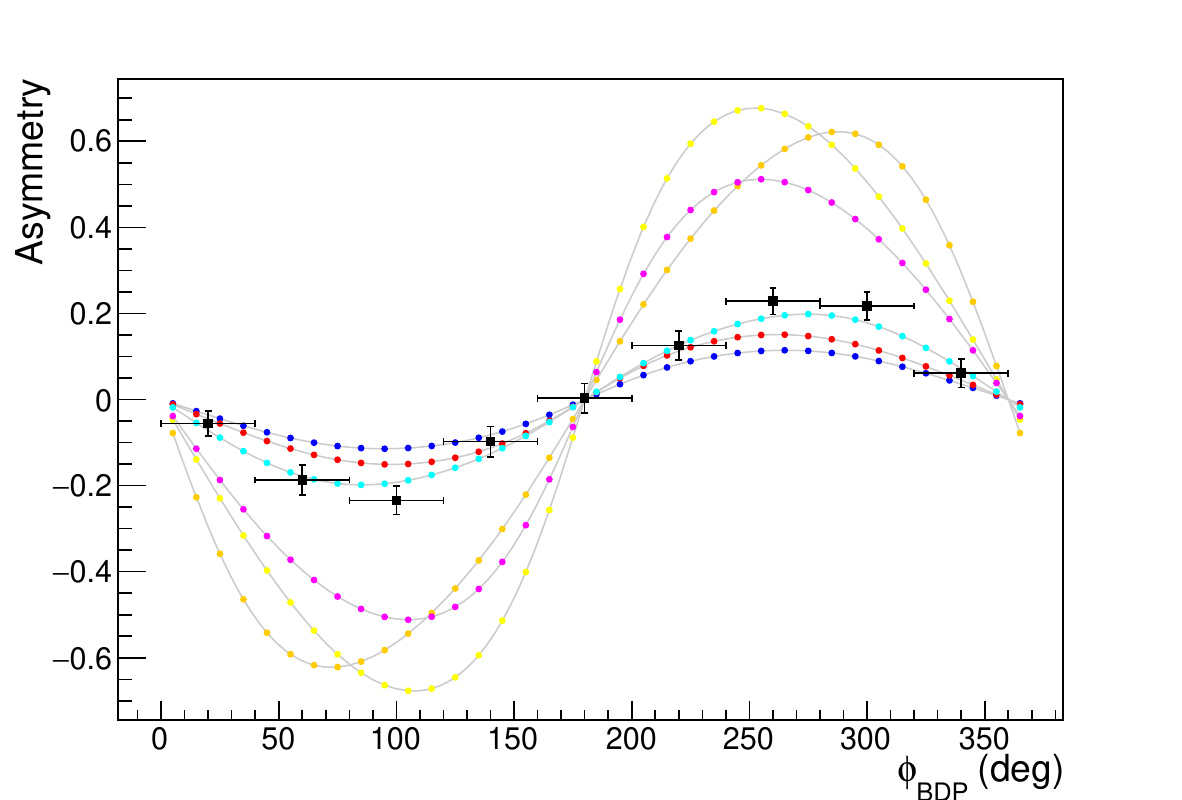}
        \caption{Bin 2: $\xi'>-0.005$ and $\xi<0.00417$.}
    \end{subfigure}
    \hspace{1.0cm}
    \begin{subfigure}[b]{0.45\textwidth}
        \centering
        \includegraphics[width=0.95\textwidth]{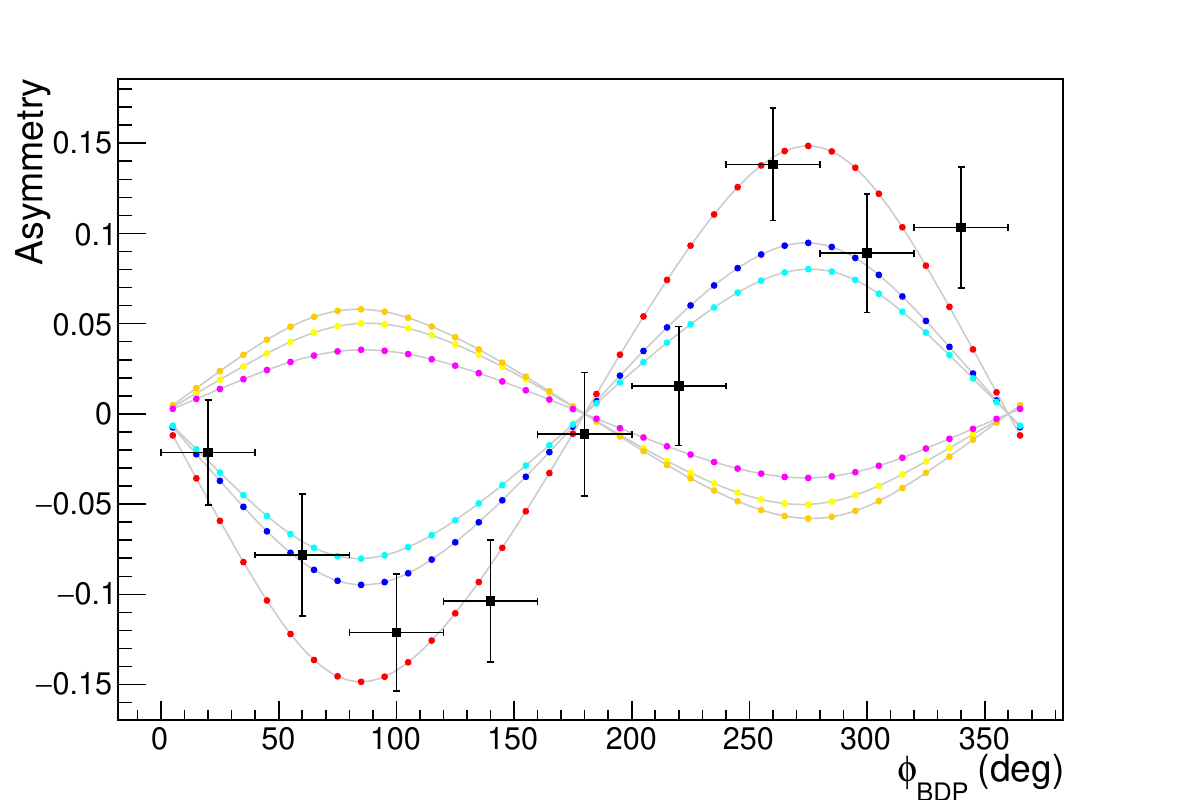}
        \caption{Bin 3: $\xi'>-0.005$ and $\xi>0.00417$.}
    \end{subfigure}
    \hspace{1.0cm}
    \begin{subfigure}[b]{0.45\textwidth}
        \centering
        \includegraphics[width=0.95\textwidth]{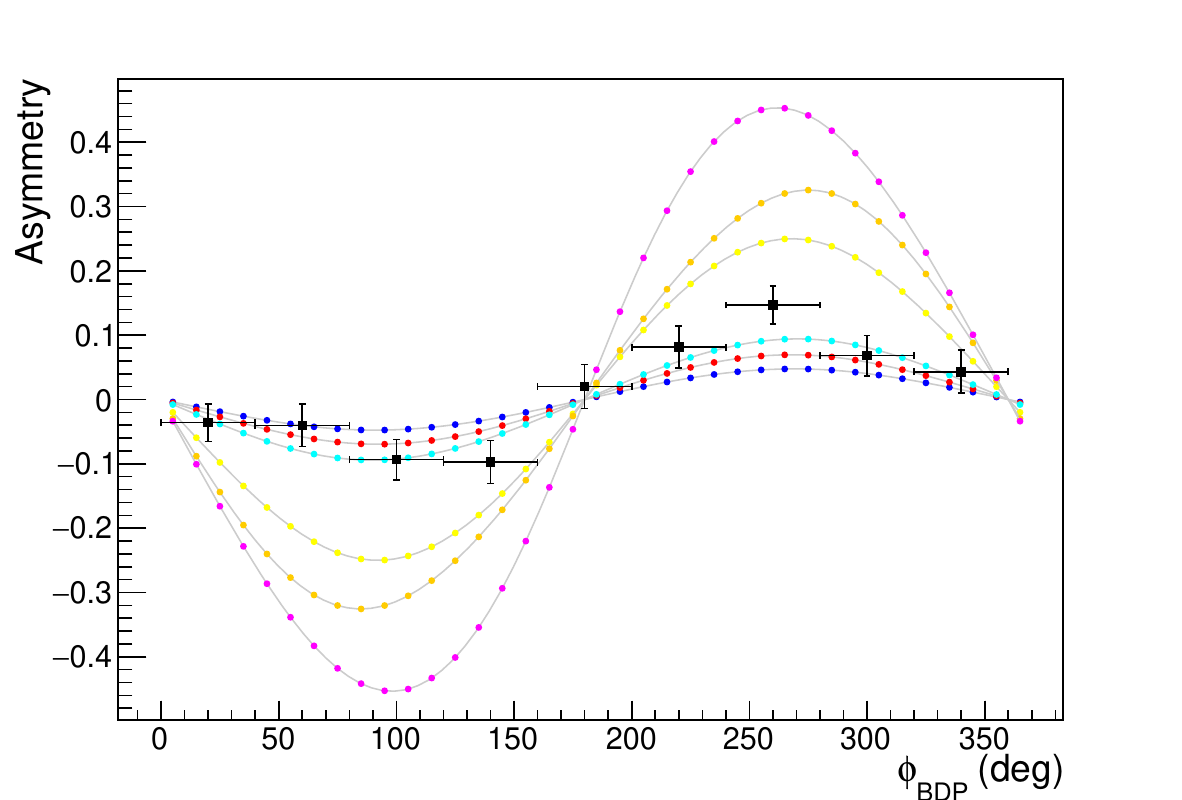}
        \caption{Bin 4: $\xi'<-0.005$ and $\xi<0.013$.}
    \end{subfigure}
    \caption{BSA experimental projections for DDVCS measurements at EIC. Error bars represent statistical errors.}
    \label{measEIC}
\end{figure}

On the contrary, model predictions provide large BSA amplitudes, as shown in Fig. \ref{measEIC}, confirming that the asymmetry is dominated by $\mathcal{H}$. Sizable BSA amplitudes are provided despite the small $\xi$ and $\xi'$ values as $\mathfrak{Im}[\mathcal{H}]$ depends on the $\xi'/\xi$ ratio. As shown in Fig. \ref{ImH_EKM} for a fixed $\xi=0.5$ value, $\mathfrak{Im}[\mathcal{H}]$ does not decrease monotonically to zero with decreasing $\xi'$ for the GK19 model. Instead, $\mathfrak{Im}[\mathcal{H}]$ grows to a maximum value before decreasing towards $\xi'=0$. Nevertheless, these model predictions provide rough estimations of the asymmetry size as further studies are needed to include the gluon contribution in models \cite{gluonic}.

%Error bars were computed as $\Delta \sigma = \sigma\sqrt{2/N_{\text{measured}}}$, where $N_{\text{measured}}$ is the number of measured events
Finally, one might wonder about the feasibility of unpolarized cross-section measurements at EIC. For an ideal experiment comprising full acceptance and 100\% detection efficiency, the experimental cross-section is computed as $\sigma$=$N/\mathcal{L}T$ with statistical error $\Delta \sigma$=$\sqrt{\sigma/\mathcal{L}T}$, $N$ being the reconstructed number of events, $\mathcal{L}$ the luminosity and $T$ the data-taking time. As the DDVCS cross-section is at the pico-barn level, we would be interested in cross-section measurements within 1-10\% accuracy. Therefore, the integrated luminosity $\mathcal{L}T$ required for such measurements is about $10^{38}$-$10^{40}$ cm$^{-2}$. Ideally, the integrated luminosity of the EIC after one year of data taking ($10\;\mathrm{fb}^{-1}$=$10^{40}\;\mathrm{cm}^{-2}$) would allow the measurement of the DDVCS cross-sections up to the $1\%$ accuracy. However, acceptance and reconstruction efficiency might point to higher luminosity requirements. 

\begin{figure}[ht]
    \centering
    \begin{subfigure}[b]{0.45\textwidth}
        \centering
        \includegraphics[width=0.95\textwidth]{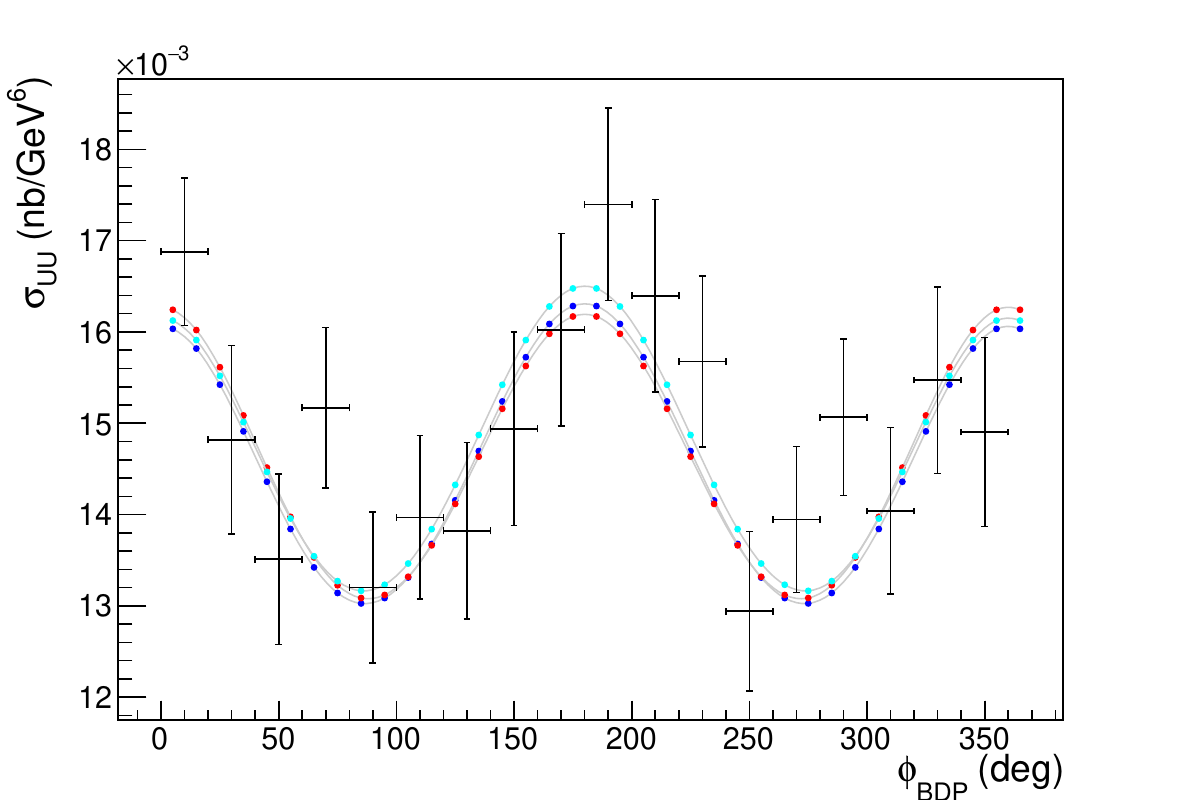}
        \caption{$\mathcal{L}=10$ fb$^{-1}$ and ePIC detector acceptance.}
        \label{UU1}
    \end{subfigure}
    \hspace{1.0cm}
    \begin{subfigure}[b]{0.45\textwidth}
        \centering
        \includegraphics[width=0.95\textwidth]{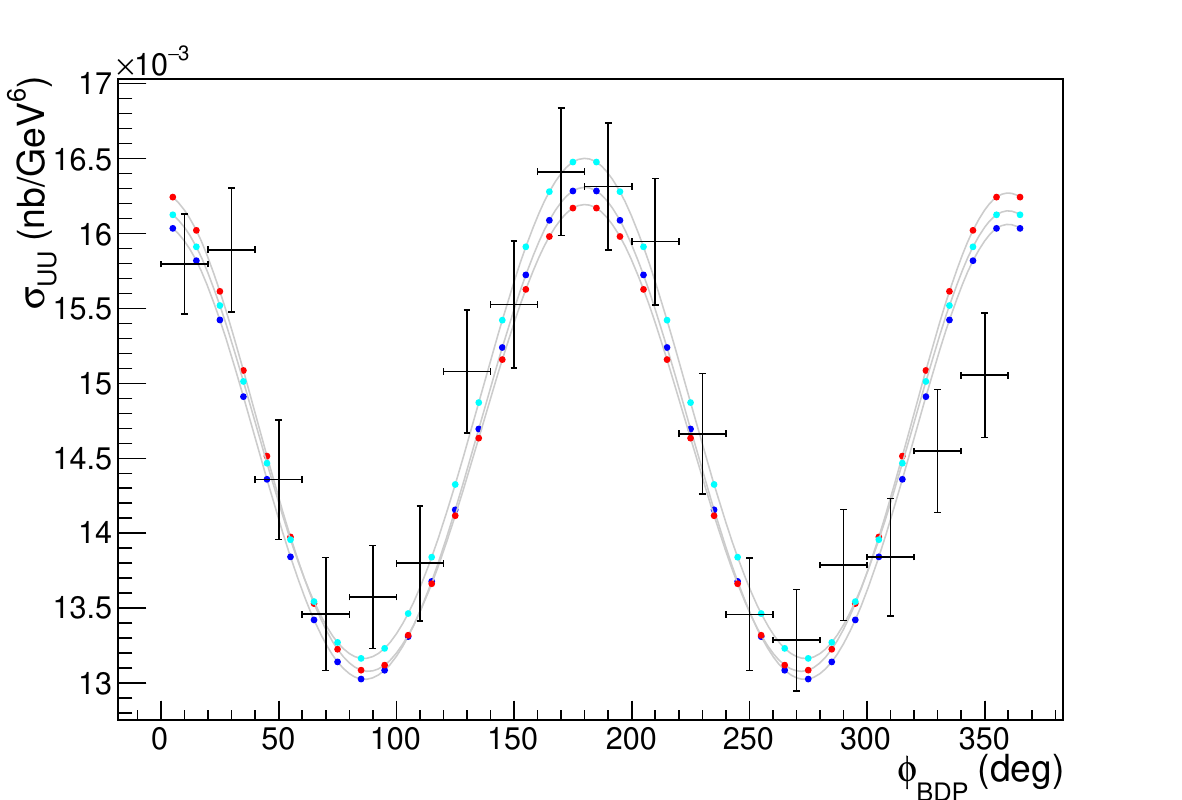}
        \caption{$\mathcal{L}=10$ fb$^{-1}$ and full acceptance.}
        \label{UU2}
    \end{subfigure}
    \hspace{1.0cm}
    \begin{subfigure}[b]{0.45\textwidth}
        \centering
        \includegraphics[width=0.95\textwidth]{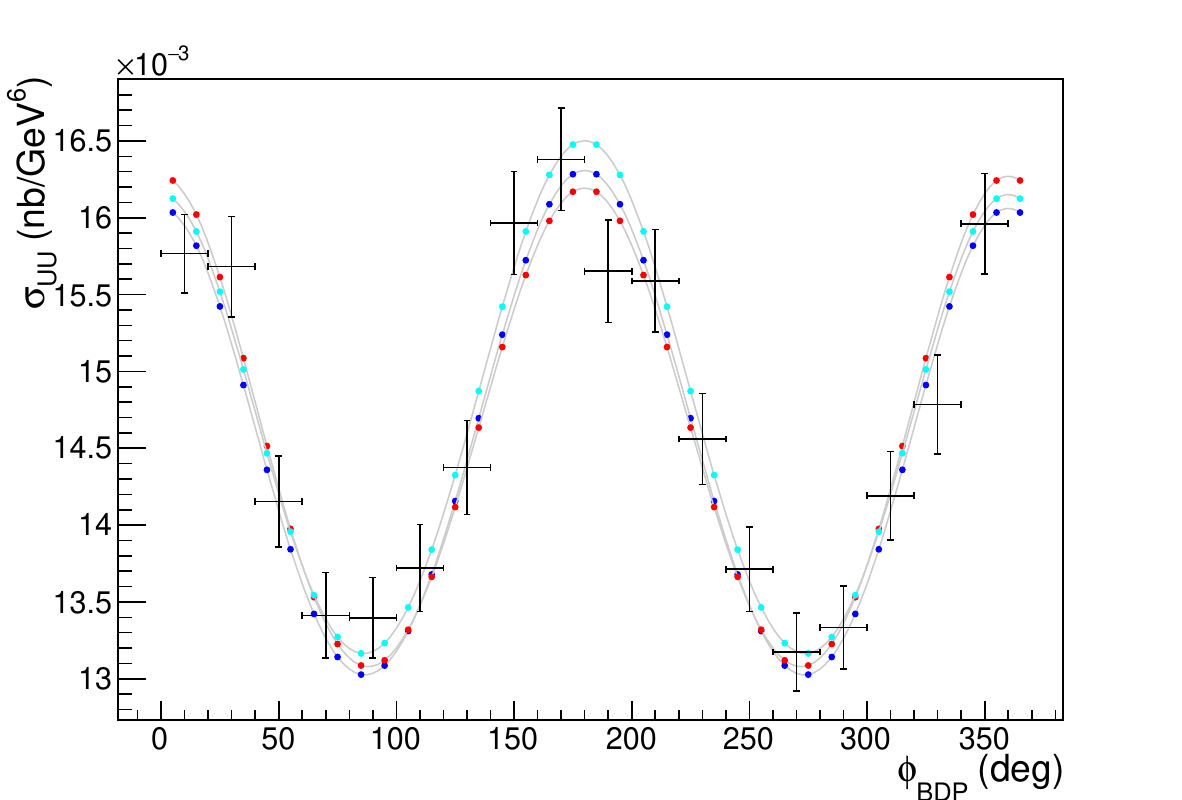}
        \caption{$\mathcal{L}=100$ fb$^{-1}$ and ePIC detector acceptance.}
        \label{UU3}
    \end{subfigure}
    \caption{\justifying{$\sigma_{UU}$ experimental projections for the EIC setup over bin 1: $\xi'<-0.005$ and $\xi>0.013$. Error bars account for statistical errors only.}}
    \label{UUmeasEIC}
\end{figure}

We show in Fig. \ref{UUmeasEIC} the cross-section projection over a sample bin (bin 1: $\xi'<-0.005$ and $\xi>0.013$) for three different scenarios. While Fig. \ref{UU1} shows the projected cross-section measurement subject to the $2\degree<\theta<178\degree$ electron acceptance, \ref{UU2} shows the projection assuming a $180\degree$ coverage as expected with the development of the ePIC far-forward detectors \cite{EICYellow}. Finally, Fig. \ref{UU3} shows the cross-section projection with electron detection within $2\degree<\theta<178\degree$ and $100\;\mathrm{fb}^{-1}$ luminosity. By comparing the cross-section projection of Figs. \ref{UU1} and \ref{UU2} we can observe the impact of small-angle electrons on the statistical error. Although preliminary cross-section measurements might be possible within one year of data taking, it is clear that better results could be obtained when considering a full acceptance scenario.  Furthermore, from Figs. \ref{UU2} and \ref{UU3} we can observe that one year of data taking at the maximum expected luminosity provides similar accuracy than the full acceptance scenario at minimum luminosity.

\begin{figure}[H]
    \centering
    % \begin{subfigure}[b]{0.45\textwidth}
    %     \centering
    %     \includegraphics[width=0.95\textwidth]{ALU.pdf}
    %     \caption{Amplitude as a function of the bin number.}
    % \end{subfigure}
    % \hspace{1.0cm}
    \begin{subfigure}[b]{0.45\textwidth}
        \centering
        \includegraphics[width=0.95\textwidth]{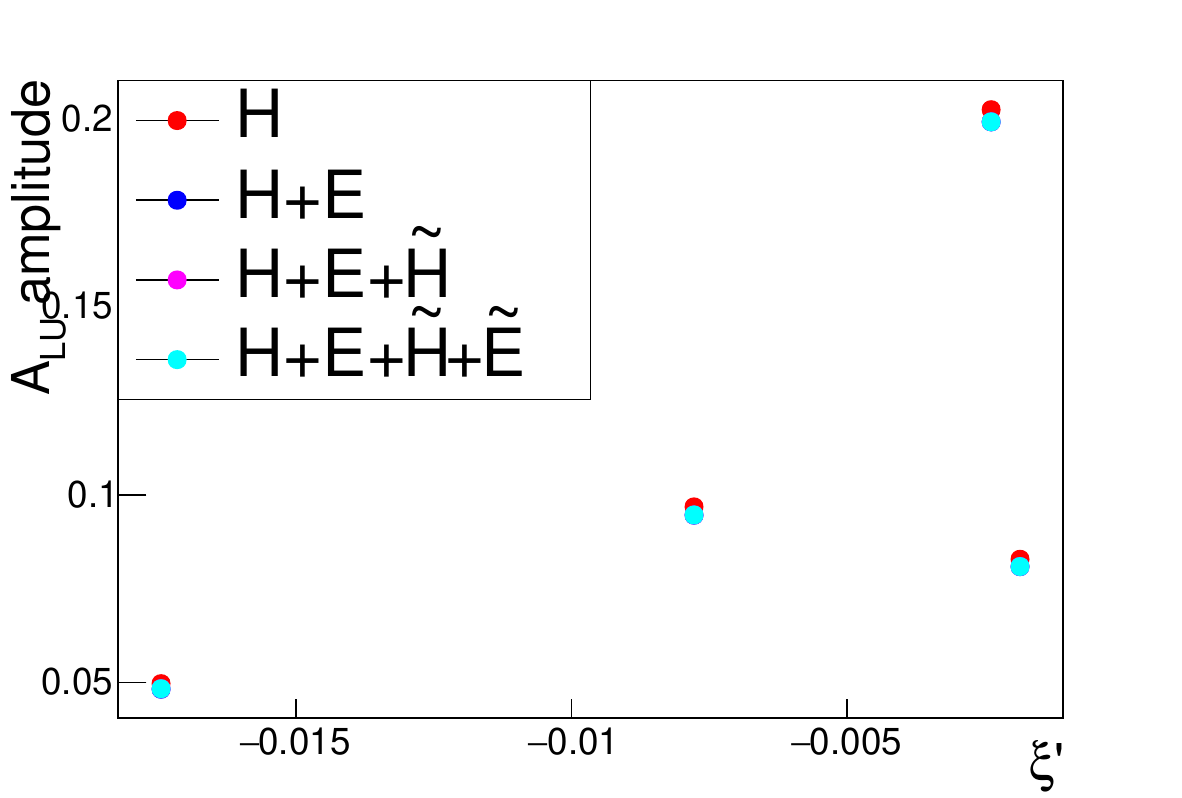}
    \end{subfigure}
        \caption{GPD contribution to the DDVCS asymmetry amplitude according to the GK19 model. EIC case.}
    \label{summEIC}
\end{figure}

\section{Conclusions}
DDVCS is a promising channel for GPD studies due to its unconstrained access to the ($\xi$, $\xi'$)-phase space of GPDs through polarized $e^{\mp}$ beam and polarized target observables. Assuming the foreseen capabilities of the $\mu$CLAS and SoLID$\mu$ spectrometers we presented experimental projections supporting the feasibility of DDVCS asymmetry measurements. The studied experiments access similar regions in the $(\xi',\xi)$-phase space and provide a complementary kinematical coverage, mostly in the time-like region of GPDs. Moreover, the 11 GeV and 22 GeV configurations would allow the observation of the asymmetries sign-change when transitioning between the space-like and time-like regions. Regarding the GPD sensitivity, models predict large asymmetries from the $\varphi_{\ell}$-dependent 5-fold cross-section and a dominance of $H$ in most amplitudes while $\widetilde{H}$ provides significant contributions in the target-polarized observables. In contrast, $E$ and $\widetilde{E}$ have negligible effects on the asymmetry amplitudes.

Based on our binning scheme, comprising three out of the five degrees of freedom of the 5-fold differential DDVCS cross-section, we have argued that a luminosity of $10^{37}\;\mathrm{cm}^{-2}\cdot\mathrm{s}^{-1}$ would allow exploratory three(four)-dimensional DDVCS measurements for $\widetilde{H}$($H$) extraction rather than a full characterization of the process. The 22 GeV configuration represents an additional challenge as the cross-section is about four times smaller as a consequence of the larger accessed values of $Q^{2}$ as shown in Fig. \ref{Q2dists}, thus requiring twice the beam time to access half the number of measurements compared to the 11 GeV case. Therefore, a full mapping of DDVCS at 11 GeV would require a minimum luminosity of $10^{38}\;\mathrm{cm}^{-2}\cdot\mathrm{s}^{-1}$, an analogous exploration of the GPD phase space with a 22 GeV beam would require four times larger luminosity. 

Exploring DDVCS at EIC kinematics demonstrated the possibility of accessing GPDs in the small $\xi$, $\xi'$ limit where the spin-dependent observables lose most of their GPD dependence. Thus, $H$ can be accessed through BSA measurements with corrections proportional to $E$. $\widetilde{H}$ extraction is not foreseen unless the target-polarized asymmetries are at least in the $5$-$10$\% range. Moreover, the expected accuracy would allow preliminary cross-section measurements within one year of data taking. 

Theoretical models have been constrained mainly by experimental data accessing the valence-quark contribution. This implies that predictions might not be accurate at EIC kinematics where the sea-quark and gluon contributions play a major role. The vanishing target-polarized asymmetries given by the EKM models result from a vanishing valence-quark contribution in the $\xi,\;\xi'\rightarrow 0$ limit and a non-existent sea-quark contribution. In contrast, models like the GK19 are expected to be valid at EIC kinematics from the DVCS perspective, however, the behavior of GPDs in the $\xi\neq \xi'$ region is unknown in the absence of DDVCS experimental data. Therefore, exploratory polarized beam and target DDVCS measurements at JLab and EIC are crucial to gain further insights into the sea-quark and gluon contribution to GPDs and their behavior for $\xi\neq \xi'$. For instance, while there is a CFF dispersion relation for DVCS an analogous relation, like that of Eq. \ref{DR_DDVCS} for Mellin-Barnes representation, is still under discussion. Moreover, the Mellin-Barnes approach needs further studies for $\xi\neq\xi'$. Although our current implementation restores the $\xi'$ dependence based on the original results on \cite{MellinBarnes}, conformal GPD moments are modeled by the $\xi,\xi'$-independent ansatz of DVCS and there is no generalization of the D-term.

\section{Acknowledgements}
The authors would like to thank Victor Martínez-Fernandez, Cédric Mezrag, Pawel Sznajder, Jakub Wagner and Zhiwen Zhao for fruitful discussions. This work was supported by the Programme blanc of the Physics Graduate School of the PHENIICS doctoral school [Contract No. D22-ET11] and the french Centre National de la Recherche Scientifique (CNRS). This work was benefited by Institut Pascal at Université Paris-Saclay with the support of the program “Investissements d’avenir” ANR-11-IDEX-0003-01.

% \nocite{*}

\bibliographystyle{apsrev4-2}
\bibliography{References.bib}% Produces the bibliography via BibTeX.

\begin{thebibliography}{67}%
\makeatletter
\providecommand \@ifxundefined [1]{%
 \@ifx{#1\undefined}
}%
\providecommand \@ifnum [1]{%
 \ifnum #1\expandafter \@firstoftwo
 \else \expandafter \@secondoftwo
 \fi
}%
\providecommand \@ifx [1]{%
 \ifx #1\expandafter \@firstoftwo
 \else \expandafter \@secondoftwo
 \fi
}%
\providecommand \natexlab [1]{#1}%
\providecommand \enquote  [1]{``#1''}%
\providecommand \bibnamefont  [1]{#1}%
\providecommand \bibfnamefont [1]{#1}%
\providecommand \citenamefont [1]{#1}%
\providecommand \href@noop [0]{\@secondoftwo}%
\providecommand \href [0]{\begingroup \@sanitize@url \@href}%
\providecommand \@href[1]{\@@startlink{#1}\@@href}%
\providecommand \@@href[1]{\endgroup#1\@@endlink}%
\providecommand \@sanitize@url [0]{\catcode `\\12\catcode `\$12\catcode `\&12\catcode `\#12\catcode `\^12\catcode `\_12\catcode `\%12\relax}%
\providecommand \@@startlink[1]{}%
\providecommand \@@endlink[0]{}%
\providecommand \url  [0]{\begingroup\@sanitize@url \@url }%
\providecommand \@url [1]{\endgroup\@href {#1}{\urlprefix }}%
\providecommand \urlprefix  [0]{URL }%
\providecommand \Eprint [0]{\href }%
\providecommand \doibase [0]{https://doi.org/}%
\providecommand \selectlanguage [0]{\@gobble}%
\providecommand \bibinfo  [0]{\@secondoftwo}%
\providecommand \bibfield  [0]{\@secondoftwo}%
\providecommand \translation [1]{[#1]}%
\providecommand \BibitemOpen [0]{}%
\providecommand \bibitemStop [0]{}%
\providecommand \bibitemNoStop [0]{.\EOS\space}%
\providecommand \EOS [0]{\spacefactor3000\relax}%
\providecommand \BibitemShut  [1]{\csname bibitem#1\endcsname}%
\let\auto@bib@innerbib\@empty
%</preamble>
\bibitem [{\citenamefont {{X. Ji}}(2003)}]{wigner1}%
  \BibitemOpen
  \bibfield  {author} {\bibinfo {author} {\bibnamefont {{X. Ji}}},\ }\href@noop {} {\bibfield  {journal} {\bibinfo  {journal} {Phys. Rev. Lett.}\ }\textbf {\bibinfo {volume} {91}},\ \bibinfo {pages} {062001} (\bibinfo {year} {2003})}\BibitemShut {NoStop}%
\bibitem [{\citenamefont {{A. V. Belitsky {\it et al}.}}(2004)}]{wigner2}%
  \BibitemOpen
  \bibfield  {author} {\bibinfo {author} {\bibnamefont {{A. V. Belitsky {\it et al}.}}},\ }\href@noop {} {\bibfield  {journal} {\bibinfo  {journal} {Phys. Rev. D}\ }\textbf {\bibinfo {volume} {69}},\ \bibinfo {pages} {074014} (\bibinfo {year} {2004})}\BibitemShut {NoStop}%
\bibitem [{\citenamefont {{D. Müller {\it et al}}}(1994)}]{GPD0}%
  \BibitemOpen
  \bibfield  {author} {\bibinfo {author} {\bibnamefont {{D. Müller {\it et al}}}},\ }\href@noop {} {\bibfield  {journal} {\bibinfo  {journal} {Fortschr. Phys.}\ }\textbf {\bibinfo {volume} {42}},\ \bibinfo {pages} {101} (\bibinfo {year} {1994})}\BibitemShut {NoStop}%
\bibitem [{\citenamefont {{M. Diehl}}(2003)}]{GPD1}%
  \BibitemOpen
  \bibfield  {author} {\bibinfo {author} {\bibnamefont {{M. Diehl}}},\ }\href@noop {} {\bibfield  {journal} {\bibinfo  {journal} {Phys. Rep.}\ }\textbf {\bibinfo {volume} {388}},\ \bibinfo {pages} {41} (\bibinfo {year} {2003})}\BibitemShut {NoStop}%
\bibitem [{\citenamefont {{A. V. Belitsky and A. V. Radyushkin}}(2005)}]{GPD2}%
  \BibitemOpen
  \bibfield  {author} {\bibinfo {author} {\bibnamefont {{A. V. Belitsky and A. V. Radyushkin}}},\ }\href@noop {} {\bibfield  {journal} {\bibinfo  {journal} {Phys. Rep.}\ }\textbf {\bibinfo {volume} {418}},\ \bibinfo {pages} {1} (\bibinfo {year} {2005})}\BibitemShut {NoStop}%
\bibitem [{\citenamefont {{M. Burkardt}}(2000)}]{GPD3}%
  \BibitemOpen
  \bibfield  {author} {\bibinfo {author} {\bibnamefont {{M. Burkardt}}},\ }\href@noop {} {\bibfield  {journal} {\bibinfo  {journal} {Phys. Rev. D}\ }\textbf {\bibinfo {volume} {62}},\ \bibinfo {pages} {071503} (\bibinfo {year} {2000})}\BibitemShut {NoStop}%
\bibitem [{\citenamefont {{M. V. Polyakov}}(2003{\natexlab{a}})}]{GPD4}%
  \BibitemOpen
  \bibfield  {author} {\bibinfo {author} {\bibnamefont {{M. V. Polyakov}}},\ }\href@noop {} {\bibfield  {journal} {\bibinfo  {journal} {Physics Lett. B}\ }\textbf {\bibinfo {volume} {555}},\ \bibinfo {pages} {57} (\bibinfo {year} {2003}{\natexlab{a}})}\BibitemShut {NoStop}%
\bibitem [{\citenamefont {{X. Ji}}(1997)}]{GPD5}%
  \BibitemOpen
  \bibfield  {author} {\bibinfo {author} {\bibnamefont {{X. Ji}}},\ }\href@noop {} {\bibfield  {journal} {\bibinfo  {journal} {Phys. Rev. Lett.}\ }\textbf {\bibinfo {volume} {78}},\ \bibinfo {pages} {610} (\bibinfo {year} {1997})}\BibitemShut {NoStop}%
\bibitem [{\citenamefont {{M. V. Polyakov}}(2003{\natexlab{b}})}]{GFF1}%
  \BibitemOpen
  \bibfield  {author} {\bibinfo {author} {\bibnamefont {{M. V. Polyakov}}},\ }\href@noop {} {\bibfield  {journal} {\bibinfo  {journal} {Phys. Lett. B}\ }\textbf {\bibinfo {volume} {555}},\ \bibinfo {pages} {57} (\bibinfo {year} {2003}{\natexlab{b}})}\BibitemShut {NoStop}%
\bibitem [{\citenamefont {{K. Kumeri{\v{c}}ki and D. Mueller}}(2010{\natexlab{a}})}]{GFF2}%
  \BibitemOpen
  \bibfield  {author} {\bibinfo {author} {\bibnamefont {{K. Kumeri{\v{c}}ki and D. Mueller}}},\ }\href@noop {} {\bibfield  {journal} {\bibinfo  {journal} {Nuc. Phys. B}\ }\textbf {\bibinfo {volume} {841}},\ \bibinfo {pages} {1} (\bibinfo {year} {2010}{\natexlab{a}})}\BibitemShut {NoStop}%
\bibitem [{\citenamefont {{ V. D. Burkert {\it et al}. }}(2018)}]{GFF3}%
  \BibitemOpen
  \bibfield  {author} {\bibinfo {author} {\bibnamefont {{ V. D. Burkert {\it et al}. }}},\ }\href@noop {} {\bibfield  {journal} {\bibinfo  {journal} {Nature}\ }\textbf {\bibinfo {volume} {557}},\ \bibinfo {pages} {396} (\bibinfo {year} {2018})}\BibitemShut {NoStop}%
\bibitem [{\citenamefont {{A. V. Belitsky {\it et al}.}}(2003)}]{belitsky}%
  \BibitemOpen
  \bibfield  {author} {\bibinfo {author} {\bibnamefont {{A. V. Belitsky {\it et al}.}}},\ }\href@noop {} {\bibfield  {journal} {\bibinfo  {journal} {Phys. Rev. D}\ }\textbf {\bibinfo {volume} {68}},\ \bibinfo {pages} {116005} (\bibinfo {year} {2003})}\BibitemShut {NoStop}%
\bibitem [{\citenamefont {{M. Guidal and M. Vanderhaeghen}}(2003)}]{guidal}%
  \BibitemOpen
  \bibfield  {author} {\bibinfo {author} {\bibnamefont {{M. Guidal and M. Vanderhaeghen}}},\ }\href@noop {} {\bibfield  {journal} {\bibinfo  {journal} {Phys. Rev. Lett.}\ }\textbf {\bibinfo {volume} {90}},\ \bibinfo {pages} {012001} (\bibinfo {year} {2003})}\BibitemShut {NoStop}%
\bibitem [{\citenamefont {{ K. Deja {\it et al}}}(2023)}]{victor}%
  \BibitemOpen
  \bibfield  {author} {\bibinfo {author} {\bibnamefont {{ K. Deja {\it et al}}}},\ }\href@noop {} {\bibfield  {journal} {\bibinfo  {journal} {Phys. Rev. D}\ }\textbf {\bibinfo {volume} {107}},\ \bibinfo {pages} {094035} (\bibinfo {year} {2023})}\BibitemShut {NoStop}%
\bibitem [{\citenamefont {{M. Murray and Hermes Collaboration}}(2011)}]{HERMES}%
  \BibitemOpen
  \bibfield  {author} {\bibinfo {author} {\bibnamefont {{M. Murray and Hermes Collaboration}}},\ }in\ \href@noop {} {\emph {\bibinfo {booktitle} {AIP Conf. Proc.}}},\ Vol.\ \bibinfo {volume} {1350}\ (\bibinfo {organization} {American Institute of Physics},\ \bibinfo {year} {2011})\ pp.\ \bibinfo {pages} {64--68}\BibitemShut {NoStop}%
\bibitem [{\citenamefont {{A. Ferrero {\it et al}.}}(2011)}]{COMPASS}%
  \BibitemOpen
  \bibfield  {author} {\bibinfo {author} {\bibnamefont {{A. Ferrero {\it et al}.}}},\ }in\ \href@noop {} {\emph {\bibinfo {booktitle} {J. Phys. Conf. Ser.}}},\ Vol.\ \bibinfo {volume} {295}\ (\bibinfo {organization} {IOP Publishing},\ \bibinfo {year} {2011})\ p.\ \bibinfo {pages} {012039}\BibitemShut {NoStop}%
\bibitem [{\citenamefont {{J. Arrington {\it et al}}}(2022)}]{CEBAF}%
  \BibitemOpen
  \bibfield  {author} {\bibinfo {author} {\bibnamefont {{J. Arrington {\it et al}}}},\ }\href@noop {} {\bibfield  {journal} {\bibinfo  {journal} {Prog. Part. Nucl. Phys.}\ }\textbf {\bibinfo {volume} {127}},\ \bibinfo {pages} {103985} (\bibinfo {year} {2022})}\BibitemShut {NoStop}%
\bibitem [{\citenamefont {{B. A. Mecking {\it et al}}}(2003)}]{CEBAFold}%
  \BibitemOpen
  \bibfield  {author} {\bibinfo {author} {\bibnamefont {{B. A. Mecking {\it et al}}}},\ }\href@noop {} {\bibfield  {journal} {\bibinfo  {journal} {Nucl. Instrum. Meth. A}\ }\textbf {\bibinfo {volume} {503}},\ \bibinfo {pages} {513} (\bibinfo {year} {2003})}\BibitemShut {NoStop}%
\bibitem [{\citenamefont {{R. Abdul Khalek {\it et al}.}}(2022)}]{EICYellow}%
  \BibitemOpen
  \bibfield  {author} {\bibinfo {author} {\bibnamefont {{R. Abdul Khalek {\it et al}.}}},\ }\href@noop {} {\bibfield  {journal} {\bibinfo  {journal} {Nucl. Phys. A}\ }\textbf {\bibinfo {volume} {1026}},\ \bibinfo {pages} {122447} (\bibinfo {year} {2022})}\BibitemShut {NoStop}%
\bibitem [{\citenamefont {{ I.V. Anikin {\it et al}}}(2018)}]{nearfut}%
  \BibitemOpen
  \bibfield  {author} {\bibinfo {author} {\bibnamefont {{ I.V. Anikin {\it et al}}}},\ }\href@noop {} {\bibfield  {journal} {\bibinfo  {journal} {Acta Phys. Polon. B}\ }\textbf {\bibinfo {volume} {49}} (\bibinfo {year} {2018})}\BibitemShut {NoStop}%
\bibitem [{\citenamefont {{V. D. Burkert {\it et al}}}(2020)}]{CLAS12}%
  \BibitemOpen
  \bibfield  {author} {\bibinfo {author} {\bibnamefont {{V. D. Burkert {\it et al}}}},\ }\href@noop {} {\bibfield  {journal} {\bibinfo  {journal} {Nucl. Instrum. Meth. A}\ }\textbf {\bibinfo {volume} {959}},\ \bibinfo {pages} {163419} (\bibinfo {year} {2020})}\BibitemShut {NoStop}%
\bibitem [{\citenamefont {{S. Stepanyan {\it et al}.}}(2016)}]{LOIDDVCS}%
  \BibitemOpen
  \bibfield  {author} {\bibinfo {author} {\bibnamefont {{S. Stepanyan {\it et al}.}}},\ }\href@noop {} {\bibfield  {journal} {\bibinfo  {journal} {Jefferson Lab Experiment LOI12-16-004}\ } (\bibinfo {year} {2016})}\BibitemShut {NoStop}%
\bibitem [{\citenamefont {{E. Voutier {\it et al}.}}(2015)}]{solid}%
  \BibitemOpen
  \bibfield  {author} {\bibinfo {author} {\bibnamefont {{E. Voutier {\it et al}.}}},\ }\href@noop {} {\bibfield  {journal} {\bibinfo  {journal} {Jefferson Lab Experiment LOI12-15-005}\ } (\bibinfo {year} {2015})}\BibitemShut {NoStop}%
\bibitem [{\citenamefont {{A. Camsonne {\it et al}.}}(2015)}]{solid2}%
  \BibitemOpen
  \bibfield  {author} {\bibinfo {author} {\bibnamefont {{A. Camsonne {\it et al}.}}},\ }\href@noop {} {\bibfield  {journal} {\bibinfo  {journal} {Jefferson Lab Experiment LOI12-23-012}\ } (\bibinfo {year} {2015})}\BibitemShut {NoStop}%
\bibitem [{\citenamefont {{A. Accardi {\it et al}.}}(2016{\natexlab{a}})}]{EIC}%
  \BibitemOpen
  \bibfield  {author} {\bibinfo {author} {\bibnamefont {{A. Accardi {\it et al}.}}},\ }\href@noop {} {\bibfield  {journal} {\bibinfo  {journal} {Eur. Phys. J. A}\ }\textbf {\bibinfo {volume} {52}},\ \bibinfo {pages} {1} (\bibinfo {year} {2016}{\natexlab{a}})}\BibitemShut {NoStop}%
\bibitem [{\citenamefont {{K. Deja {\it et al}}}(2023)}]{EIC1}%
  \BibitemOpen
  \bibfield  {author} {\bibinfo {author} {\bibnamefont {{K. Deja {\it et al}}}},\ }\href@noop {} {\bibfield  {journal} {\bibinfo  {journal} {arXiv:2401.13064}\ } (\bibinfo {year} {2023})}\BibitemShut {NoStop}%
\bibitem [{\citenamefont {{\it et al}.}(2021)}]{zhaoWP}%
  \BibitemOpen
  \bibfield  {author} {\bibinfo {author} {\bibfnamefont {V.~B.}\ \bibnamefont {{\it et al}.}},\ }\href@noop {} {\bibfield  {journal} {\bibinfo  {journal} {Eur. Phys. J. A}\ }\textbf {\bibinfo {volume} {57}},\ \bibinfo {pages} {186} (\bibinfo {year} {2021})}\BibitemShut {NoStop}%
\bibitem [{\citenamefont {{A. Accardi {\it et al}.}}(2021)}]{positron}%
  \BibitemOpen
  \bibfield  {author} {\bibinfo {author} {\bibnamefont {{A. Accardi {\it et al}.}}},\ }\href@noop {} {\bibfield  {journal} {\bibinfo  {journal} {Eur. Phys. J. A}\ }\textbf {\bibinfo {volume} {57}},\ \bibinfo {pages} {261} (\bibinfo {year} {2021})}\BibitemShut {NoStop}%
\bibitem [{\citenamefont {{A. Bacchetta {\it et al}.}}(2004)}]{trento}%
  \BibitemOpen
  \bibfield  {author} {\bibinfo {author} {\bibnamefont {{A. Bacchetta {\it et al}.}}},\ }\href@noop {} {\bibfield  {journal} {\bibinfo  {journal} {Phys. Rev. D}\ }\textbf {\bibinfo {volume} {70}},\ \bibinfo {pages} {117504} (\bibinfo {year} {2004})}\BibitemShut {NoStop}%
\bibitem [{\citenamefont {{E. Berger, M. Diehl and B. Pire}}(2002)}]{BDP}%
  \BibitemOpen
  \bibfield  {author} {\bibinfo {author} {\bibnamefont {{E. Berger, M. Diehl and B. Pire}}},\ }\href@noop {} {\bibfield  {journal} {\bibinfo  {journal} {Eur. Phys. J. C}\ }\textbf {\bibinfo {volume} {23}},\ \bibinfo {pages} {675} (\bibinfo {year} {2002})}\BibitemShut {NoStop}%
\bibitem [{\citenamefont {{S. Zhao}}(2020)}]{SYthese}%
  \BibitemOpen
  \bibfield  {author} {\bibinfo {author} {\bibnamefont {{S. Zhao}}},\ }\emph {\bibinfo {title} {{Studying the nucleon structure via Double Deeply Virtual Compton Scattering at the Jefferson Laboratory.}}},\ \href@noop {} {Ph.D. thesis},\ \bibinfo  {school} {université Paris-Saclay, IJCLab, Orsay} (\bibinfo {year} {2020}),\ \bibinfo {note} {{NNT:2020UPASS147}}\BibitemShut {NoStop}%
\bibitem [{\citenamefont {{S. Boffi and B. Pasquini}}(2007)}]{GPDproperties}%
  \BibitemOpen
  \bibfield  {author} {\bibinfo {author} {\bibnamefont {{S. Boffi and B. Pasquini}}},\ }\href@noop {} {\bibfield  {journal} {\bibinfo  {journal} {Riv. Nuovo Cimento.}\ }\textbf {\bibinfo {volume} {30}},\ \bibinfo {pages} {387} (\bibinfo {year} {2007})}\BibitemShut {NoStop}%
\bibitem [{\citenamefont {{D. Müller {\it et al}.}}(1994)}]{DD1}%
  \BibitemOpen
  \bibfield  {author} {\bibinfo {author} {\bibnamefont {{D. Müller {\it et al}.}}},\ }\href@noop {} {\bibfield  {journal} {\bibinfo  {journal} {Fortschritte der Physik/Progress of Physics}\ }\textbf {\bibinfo {volume} {42}},\ \bibinfo {pages} {101} (\bibinfo {year} {1994})}\BibitemShut {NoStop}%
\bibitem [{\citenamefont {{A. V. Radyushkin}}(1997)}]{DD2}%
  \BibitemOpen
  \bibfield  {author} {\bibinfo {author} {\bibnamefont {{A. V. Radyushkin}}},\ }\href@noop {} {\bibfield  {journal} {\bibinfo  {journal} {Phys. Rev. D}\ }\textbf {\bibinfo {volume} {56}},\ \bibinfo {pages} {5524} (\bibinfo {year} {1997})}\BibitemShut {NoStop}%
\bibitem [{\citenamefont {{L. L. Frankfurt{\it et al}.}}(1999)}]{PiPol1}%
  \BibitemOpen
  \bibfield  {author} {\bibinfo {author} {\bibnamefont {{L. L. Frankfurt{\it et al}.}}},\ }\href@noop {} {\bibfield  {journal} {\bibinfo  {journal} {Phys. Rev. D}\ }\textbf {\bibinfo {volume} {60}},\ \bibinfo {pages} {014010} (\bibinfo {year} {1999})}\BibitemShut {NoStop}%
\bibitem [{\citenamefont {{M. Penttinen {\it et al}.}}(2000)}]{PiPol2}%
  \BibitemOpen
  \bibfield  {author} {\bibinfo {author} {\bibnamefont {{M. Penttinen {\it et al}.}}},\ }\href@noop {} {\bibfield  {journal} {\bibinfo  {journal} {Phys. Rev. D}\ }\textbf {\bibinfo {volume} {62}},\ \bibinfo {pages} {014024} (\bibinfo {year} {2000})}\BibitemShut {NoStop}%
\bibitem [{\citenamefont {{B. Berthou {\it et al}.}}(2018)}]{PARTONS}%
  \BibitemOpen
  \bibfield  {author} {\bibinfo {author} {\bibnamefont {{B. Berthou {\it et al}.}}},\ }\href {https://doi.org/10.1140/Eur. Phys. J.c/s10052-018-5948-0} {\bibfield  {journal} {\bibinfo  {journal} {Eur. Phys. J. C}\ }\textbf {\bibinfo {volume} {78}},\ \bibinfo {pages} {478} (\bibinfo {year} {2018})},\ \Eprint {https://arxiv.org/abs/1512.06174} {arXiv:1512.06174 [hep-ph]} \BibitemShut {NoStop}%
\bibitem [{\citenamefont {{K. Kumerički}}(2006)}]{gepard}%
  \BibitemOpen
  \bibfield  {author} {\bibinfo {author} {\bibnamefont {{K. Kumerički}}},\ }\href@noop {} {\bibinfo {title} {Gepard: Tool for studying the 3d quark and gluon distributions in the nucleon}},\ \bibinfo {howpublished} {\url{https://gepard.phy.hr/credits.html}} (\bibinfo {year} {2006})\BibitemShut {NoStop}%
\bibitem [{\citenamefont {{M. Vanderhaeghen {\it et al}.}}(1998)}]{VGG1}%
  \BibitemOpen
  \bibfield  {author} {\bibinfo {author} {\bibnamefont {{M. Vanderhaeghen {\it et al}.}}},\ }\href@noop {} {\bibfield  {journal} {\bibinfo  {journal} {Phys. Rev. Lett.}\ }\textbf {\bibinfo {volume} {80}},\ \bibinfo {pages} {5064} (\bibinfo {year} {1998})}\BibitemShut {NoStop}%
\bibitem [{\citenamefont {{M. Vanderhaeghen {\it et al}.}}(1999)}]{VGG2}%
  \BibitemOpen
  \bibfield  {author} {\bibinfo {author} {\bibnamefont {{M. Vanderhaeghen {\it et al}.}}},\ }\href@noop {} {\bibfield  {journal} {\bibinfo  {journal} {Phys. Rev. D}\ }\textbf {\bibinfo {volume} {60}},\ \bibinfo {pages} {094017} (\bibinfo {year} {1999})}\BibitemShut {NoStop}%
\bibitem [{\citenamefont {{K. Goeke {\it et al}.}}(2001)}]{VGG3}%
  \BibitemOpen
  \bibfield  {author} {\bibinfo {author} {\bibnamefont {{K. Goeke {\it et al}.}}},\ }\href@noop {} {\bibfield  {journal} {\bibinfo  {journal} {Prog. Part. Nucl. Phys.}\ }\textbf {\bibinfo {volume} {47}},\ \bibinfo {pages} {401} (\bibinfo {year} {2001})}\BibitemShut {NoStop}%
\bibitem [{\citenamefont {{M. Guidal {\it et al}.}}(2005)}]{VGG4}%
  \BibitemOpen
  \bibfield  {author} {\bibinfo {author} {\bibnamefont {{M. Guidal {\it et al}.}}},\ }\href@noop {} {\bibfield  {journal} {\bibinfo  {journal} {Phys. Rev. D}\ }\textbf {\bibinfo {volume} {72}},\ \bibinfo {pages} {054013} (\bibinfo {year} {2005})}\BibitemShut {NoStop}%
\bibitem [{\citenamefont {{R. S. Thorne {\it et al}.}}(2009)}]{MRST}%
  \BibitemOpen
  \bibfield  {author} {\bibinfo {author} {\bibnamefont {{R. S. Thorne {\it et al}.}}},\ }\href@noop {} {\bibfield  {journal} {\bibinfo  {journal} {arXiv:0907.2387}\ } (\bibinfo {year} {2009})}\BibitemShut {NoStop}%
\bibitem [{\citenamefont {{ T. Hou {\it et al}.}}(2021)}]{CTEQ18}%
  \BibitemOpen
  \bibfield  {author} {\bibinfo {author} {\bibnamefont {{ T. Hou {\it et al}.}}},\ }\href@noop {} {\bibfield  {journal} {\bibinfo  {journal} {Phys. Rev. D}\ }\textbf {\bibinfo {volume} {103}},\ \bibinfo {pages} {014013} (\bibinfo {year} {2021})}\BibitemShut {NoStop}%
\bibitem [{\citenamefont {{M. Guidal {\it et al}.}}(2013)}]{GPDrev}%
  \BibitemOpen
  \bibfield  {author} {\bibinfo {author} {\bibnamefont {{M. Guidal {\it et al}.}}},\ }\href@noop {} {\bibfield  {journal} {\bibinfo  {journal} {Rep. Prog. Phys.}\ }\textbf {\bibinfo {volume} {76}},\ \bibinfo {pages} {066202} (\bibinfo {year} {2013})}\BibitemShut {NoStop}%
\bibitem [{\citenamefont {{S. Goloskokov {\it et al}.}}(2007)}]{GK1}%
  \BibitemOpen
  \bibfield  {author} {\bibinfo {author} {\bibnamefont {{S. Goloskokov {\it et al}.}}},\ }\href@noop {} {\bibfield  {journal} {\bibinfo  {journal} {Eur. Phys. J. C}\ }\textbf {\bibinfo {volume} {50}},\ \bibinfo {pages} {829} (\bibinfo {year} {2007})}\BibitemShut {NoStop}%
\bibitem [{\citenamefont {{S. Goloskokov and P. Kroll}}(2008)}]{GK2}%
  \BibitemOpen
  \bibfield  {author} {\bibinfo {author} {\bibnamefont {{S. Goloskokov and P. Kroll}}},\ }\href@noop {} {\bibfield  {journal} {\bibinfo  {journal} {Eur. Phys. J. C}\ }\textbf {\bibinfo {volume} {53}},\ \bibinfo {pages} {367} (\bibinfo {year} {2008})}\BibitemShut {NoStop}%
\bibitem [{\citenamefont {{J. Bl{\"u}mlein and H. B{\"o}ttcher}}(2002)}]{BB}%
  \BibitemOpen
  \bibfield  {author} {\bibinfo {author} {\bibnamefont {{J. Bl{\"u}mlein and H. B{\"o}ttcher}}},\ }\href@noop {} {\bibfield  {journal} {\bibinfo  {journal} {Nucl. Phys. B}\ }\textbf {\bibinfo {volume} {636}},\ \bibinfo {pages} {225} (\bibinfo {year} {2002})}\BibitemShut {NoStop}%
\bibitem [{\citenamefont {{A. Airapetian {\it et al}.}}(2008)}]{HERMES1}%
  \BibitemOpen
  \bibfield  {author} {\bibinfo {author} {\bibnamefont {{A. Airapetian {\it et al}.}}},\ }\href@noop {} {\bibfield  {journal} {\bibinfo  {journal} {Physics Lett. B}\ }\textbf {\bibinfo {volume} {659}},\ \bibinfo {pages} {486} (\bibinfo {year} {2008})}\BibitemShut {NoStop}%
\bibitem [{\citenamefont {{A. Airapetian {\it et al}.}}(2010)}]{HERMES2}%
  \BibitemOpen
  \bibfield  {author} {\bibinfo {author} {\bibnamefont {{A. Airapetian {\it et al}.}}},\ }\href@noop {} {\bibfield  {journal} {\bibinfo  {journal} {Physics Lett. B}\ }\textbf {\bibinfo {volume} {682}},\ \bibinfo {pages} {345} (\bibinfo {year} {2010})}\BibitemShut {NoStop}%
\bibitem [{\citenamefont {{P. Kroll {\it et al}.}}(2013)}]{GKcoeff}%
  \BibitemOpen
  \bibfield  {author} {\bibinfo {author} {\bibnamefont {{P. Kroll {\it et al}.}}},\ }\href@noop {} {\bibfield  {journal} {\bibinfo  {journal} {Eur. Phys. J. C}\ }\textbf {\bibinfo {volume} {73}},\ \bibinfo {pages} {1} (\bibinfo {year} {2013})}\BibitemShut {NoStop}%
\bibitem [{\citenamefont {{K. Kumeri{\v{c}}ki and D. Mueller}}(2010{\natexlab{b}})}]{KM1}%
  \BibitemOpen
  \bibfield  {author} {\bibinfo {author} {\bibnamefont {{K. Kumeri{\v{c}}ki and D. Mueller}}},\ }\href@noop {} {\bibfield  {journal} {\bibinfo  {journal} {Nucl. Phys. B}\ }\textbf {\bibinfo {volume} {841}},\ \bibinfo {pages} {1} (\bibinfo {year} {2010}{\natexlab{b}})}\BibitemShut {NoStop}%
\bibitem [{\citenamefont {{K. Kumeri{\v{c}}ki {\it et al}.}}(2008)}]{MellinBarnes}%
  \BibitemOpen
  \bibfield  {author} {\bibinfo {author} {\bibnamefont {{K. Kumeri{\v{c}}ki {\it et al}.}}},\ }\href@noop {} {\bibfield  {journal} {\bibinfo  {journal} {Nucl. Phys. B}\ }\textbf {\bibinfo {volume} {794}},\ \bibinfo {pages} {244} (\bibinfo {year} {2008})}\BibitemShut {NoStop}%
\bibitem [{\citenamefont {{K. Kumerički {\it et al}.}}(2008)}]{DR_DDVCS}%
  \BibitemOpen
  \bibfield  {author} {\bibinfo {author} {\bibnamefont {{K. Kumerički {\it et al}.}}},\ }\href@noop {} {\bibfield  {journal} {\bibinfo  {journal} {Eur. Phys. J. C}\ }\textbf {\bibinfo {volume} {58}},\ \bibinfo {pages} {193} (\bibinfo {year} {2008})}\BibitemShut {NoStop}%
\bibitem [{\citenamefont {{D. S. Hwang and D. Mueller}}(2008)}]{spectator}%
  \BibitemOpen
  \bibfield  {author} {\bibinfo {author} {\bibnamefont {{D. S. Hwang and D. Mueller}}},\ }\href@noop {} {\bibfield  {journal} {\bibinfo  {journal} {Physics Lett. B}\ }\textbf {\bibinfo {volume} {660}},\ \bibinfo {pages} {350} (\bibinfo {year} {2008})}\BibitemShut {NoStop}%
\bibitem [{\citenamefont {{E. Aschenauer {\it et al}.}}(2013)}]{AFKM12}%
  \BibitemOpen
  \bibfield  {author} {\bibinfo {author} {\bibnamefont {{E. Aschenauer {\it et al}.}}},\ }\href@noop {} {\bibfield  {journal} {\bibinfo  {journal} {JHEP}\ }\textbf {\bibinfo {volume} {2013}}\bibinfo  {number} { (9)},\ \bibinfo {pages} {1}}\BibitemShut {NoStop}%
\bibitem [{\citenamefont {{E. Aschenauer {\it et al}.}}(2022)}]{EpIC}%
  \BibitemOpen
\bibfield  {number} {  }\bibfield  {author} {\bibinfo {author} {\bibnamefont {{E. Aschenauer {\it et al}.}}},\ }\href@noop {} {\bibfield  {journal} {\bibinfo  {journal} {Eur. Phys. J. C}\ }\textbf {\bibinfo {volume} {82}},\ \bibinfo {pages} {1} (\bibinfo {year} {2022})}\BibitemShut {NoStop}%
\bibitem [{\citenamefont {{M. Ungaro}}(2016)}]{gemc}%
  \BibitemOpen
  \bibfield  {author} {\bibinfo {author} {\bibnamefont {{M. Ungaro}}},\ }\href@noop {} {\bibinfo {title} {Clas12 geant4 simulation package gemc}},\ \bibinfo {howpublished} {\url{http://gemc.jlab.org}} (\bibinfo {year} {2016})\BibitemShut {NoStop}%
\bibitem [{\citenamefont {{HERMES Collaboration}}(2011)}]{transverse}%
  \BibitemOpen
  \bibfield  {author} {\bibinfo {author} {\bibnamefont {{HERMES Collaboration}}},\ }in\ \href@noop {} {\emph {\bibinfo {booktitle} {J. Phys. Conf. Ser.}}},\ Vol.\ \bibinfo {volume} {295}\ (\bibinfo {organization} {IOP Publishing},\ \bibinfo {year} {2011})\ p.\ \bibinfo {pages} {012024}\BibitemShut {NoStop}%
\bibitem [{\citenamefont {{Y. Furletova and S. Mantry}}(2018)}]{EICpos1}%
  \BibitemOpen
  \bibfield  {author} {\bibinfo {author} {\bibnamefont {{Y. Furletova and S. Mantry}}},\ }in\ \href@noop {} {\emph {\bibinfo {booktitle} {AIP Conf. Proc.}}},\ Vol.\ \bibinfo {volume} {1970}\ (\bibinfo {organization} {AIP Publishing},\ \bibinfo {year} {2018})\BibitemShut {NoStop}%
\bibitem [{\citenamefont {{A. Accardi {\it et al}.}}(2016{\natexlab{b}})}]{EICpos2}%
  \BibitemOpen
  \bibfield  {author} {\bibinfo {author} {\bibnamefont {{A. Accardi {\it et al}.}}},\ }\href@noop {} {\bibfield  {journal} {\bibinfo  {journal} {Eur. Phys. J. A}\ }\textbf {\bibinfo {volume} {52}},\ \bibinfo {pages} {1} (\bibinfo {year} {2016}{\natexlab{b}})}\BibitemShut {NoStop}%
\bibitem [{\citenamefont {{W. Melnitchouk}}(2020)}]{EICpos3}%
  \BibitemOpen
  \bibfield  {author} {\bibinfo {author} {\bibnamefont {{W. Melnitchouk}}},\ }\href@noop {} {\bibinfo {title} {{EW physics with positrons at the EIC}}} (\bibinfo {year} {2020}),\ \bibinfo {note} {talk, presented at the Electroweak \& BSM Physics at the EIC (CFNS) workshop}\BibitemShut {NoStop}%
\bibitem [{\citenamefont {{A. Hobart {\it et al}.}}(2024)}]{Mostafa}%
  \BibitemOpen
  \bibfield  {author} {\bibinfo {author} {\bibnamefont {{A. Hobart {\it et al}.}}},\ }\href@noop {} {\bibfield  {journal} {\bibinfo  {journal} {Phys. Rev. Lett.}\ }\textbf {\bibinfo {volume} {133}},\ \bibinfo {pages} {211903} (\bibinfo {year} {2024})}\BibitemShut {NoStop}%
\bibitem [{\citenamefont {{R. A. Khalek {\it et al}}}(2022)}]{ePIC_det}%
  \BibitemOpen
  \bibfield  {author} {\bibinfo {author} {\bibnamefont {{R. A. Khalek {\it et al}}}},\ }\href@noop {} {\bibfield  {journal} {\bibinfo  {journal} {Nuc. Phys. A}\ }\textbf {\bibinfo {volume} {1026}},\ \bibinfo {pages} {122447} (\bibinfo {year} {2022})}\BibitemShut {NoStop}%
\bibitem [{\citenamefont {{C. Muñoz}}(2022)}]{carlos}%
  \BibitemOpen
  \bibfield  {author} {\bibinfo {author} {\bibnamefont {{C. Muñoz}}},\ }\href {https://indico.cern.ch/event/1152971/contributions/4841393/attachments/2430661/4162119/ECCE_Muons.pdf} {\bibinfo {title} {{ECCE}}} (\bibinfo {year} {2022}),\ \bibinfo {note} {talk, presented at the 'Muon detection and quarkonium reconstruction at the EIC' meeting}\BibitemShut {NoStop}%
\bibitem [{\citenamefont {{R. Boussarie and Y. Mehtar-Tani}}(2024)}]{gluonic}%
  \BibitemOpen
  \bibfield  {author} {\bibinfo {author} {\bibnamefont {{R. Boussarie and Y. Mehtar-Tani}}},\ }\href@noop {} {\bibfield  {journal} {\bibinfo  {journal} {JHEP}\ }\textbf {\bibinfo {volume} {2024}}\bibinfo  {number} { (10)},\ \bibinfo {pages} {1}}\BibitemShut {NoStop}%
\bibitem [{\citenamefont {{M. J. Schlosser}}(2013)}]{F1prop}%
  \BibitemOpen
\bibfield  {number} {  }\bibfield  {author} {\bibinfo {author} {\bibnamefont {{M. J. Schlosser}}},\ }\href@noop {} {\bibfield  {journal} {\bibinfo  {journal} {Computer algebra in quantum field theory: integration, summation and special functions}\ ,\ \bibinfo {pages} {305}} (\bibinfo {year} {2013})}\BibitemShut {NoStop}%
\end{thebibliography}%

\appendix

\section{Extension of the EKM model in the \texorpdfstring{$\xi\neq \xi'$}{xi/=xip} region.} \label{app2}

Following the results on \cite{KM1}, GPD $H$ and $\widetilde{H}$ are computed using the DD representation of Eq. (\ref{DD}), with support on the $|\xi'|<\xi$ region, given by the following integral:
\begin{align}
    F(x,\xi,t)=\int_{0}^{\frac{x+\xi}{1+\xi}}\frac{dy}{\xi}f\left(y,\frac{x-y}{\xi},t\right). \label{DDapp}
\end{align}

At $t=0$ the DD is factorized into a PDF and a profile function
\begin{align}
    f&(y,z,0)=\frac{\Gamma(3/2 + b)}{\Gamma(1/2)\Gamma(1+b)}\frac{q(y)}{1-y}\left(1-\frac{z^{2}}{(1-y)^{2}}\right)^{b}. \nonumber
\end{align}

To introduce the $t$-dependence, we use the following Ansatz inspired by the functional form of the spectator model \cite{spectator}

\begin{align}
    f(y,z,t)=&\frac{\Gamma(3/2 + b)}{\Gamma(1/2)\Gamma(1+b)}\frac{q(y)}{1-y}\left(1-\frac{z^{2}}{(1-y)^{2}}\right)^{b} \nonumber \\
    &\left(\frac{-(1-y)^{2}}{-(1-y)^{2} + \frac{t}{4M^{2}}((1-y)^{2}-z^{2})}\right)^{p}.
\end{align}

To make the DD analytically integrable, we limit ourselves to the small $y$ case by preserving only linear terms. Then, we apply the variable change $y=u(x+\xi)/(1+\xi)$ so the integration limits on Eq. \ref{DDapp} become $x$-independent,

\begin{align}
    F&(x,\xi,t)= \label{DDu}\\
    &\frac{x+\xi}{1+\xi}\int_{0}^{1}\frac{du}{\xi}f\left(\frac{x+\xi}{1+\xi}u,\frac{x(1+\xi)-u(x+\xi)}{\xi(1+\xi)},t\right).\nonumber
\end{align}

\noindent
with the double distribution given by:
\begin{align}
    f(x,\vartheta,t)=&\left(\frac{x+ \frac{x}{\vartheta}}{1+\frac{x}{\vartheta}}u\right)^{-\alpha(t)}\nonumber \\
    &\left(1-\vartheta^{2} - \frac{2\vartheta^{2}(1+\vartheta)(-1+x)}{x+\vartheta}u\right)^{b} \label{DDfin}\\
    &\left(\frac{1}{1-\frac{t}{4M^{2}}(1-\vartheta^{2}) - \frac{t}{2M^{2}}\frac{\vartheta^{2}(1+\vartheta)(1-x)}{x + \vartheta}u}\right)^{p}, \nonumber
\end{align}

\noindent
where $\vartheta=x/\xi$. The values of $\alpha(t)$, $b$ and $p$ for $H$ and $\widetilde{H}$ can be consulted on \cite{KM1}. Then, integration in Eq. (\ref{DDu}) can be easily written in terms of the Appell Hypergeometric Function $F_{1}$:

\begin{align}
    F&(x,\vartheta,t)\nonumber \\
    =&\frac{(1-\vartheta^{2})^{b}}{\left(1-\frac{t}{4M^{2}}(1-\vartheta^{2})\right)^{p}}\left(\frac{x+ \frac{x}{\vartheta}}{1+\frac{x}{\vartheta}}\right)^{-\alpha(t)} \nonumber \\
    &\frac{\Gamma(a)\Gamma(c-a)}{\Gamma(c)}F_{1}(a,b_{1},b_{2},c,x',z').    \\
    =&\frac{\Gamma(a)\Gamma(c-a)}{\Gamma(c)}F_{1}(c-a,b_{1},b_{2},c,\frac{x'}{x'-1},\frac{z'}{z'-1}) \nonumber \\
    &\left(\frac{x+ \frac{x}{\vartheta}}{1+\frac{x}{\vartheta}}\right)^{-\alpha(t)} \left(1-\vartheta^{2} - \frac{2\vartheta^{2}(1+\vartheta)(-1+x)}{x+\vartheta} \right)^{b} \nonumber \\
    &\left(1-\frac{t}{4M^{2}}(1-\vartheta^{2}) - \frac{1}{\frac{t}{2M^{2}}\frac{\vartheta^{2}(1+\vartheta)(1-x)}{x + \vartheta}}\right)^{p}\label{DDinteg}
    \end{align}

\noindent 
where $a=1-\alpha$, $b_{1}=-b$, $b_{2}=p$, $c=2-\alpha$ while $x'$ and $z'$ are given by:

\begin{align}
 x'&=\frac{2\vartheta^{2}(1+\vartheta)(-1+x)}{(x+\vartheta)(1-\vartheta^{2})}   \\
 z'&=\frac{t}{2M^{2}}\frac{\vartheta^{2}(1+\vartheta)(1-x)}{(x + \vartheta)\left(1-\frac{t}{4M^{2}}(1-\vartheta^{2})\right)}.
\end{align}

To obtain the result of Eq. \ref{DDinteg} we use the following property \cite{F1prop}.
\begin{align}
&F_{1}[a;b_{1},b_{2};c;x',z']= \\
&(1-x')^{-b_{1}}(1-z')^{-b_{2}}F_{1}\left[c-a;b_{1},b_{2};c;\frac{x'}{x'-1},\frac{z'}{z'-1}\right]. \nonumber
\end{align}

Based on Eq. (\ref{DDinteg}), we define our final GPDs by taking the resulting profile function and introducing the respective normalization factors from \cite{KM1} to ensure the correct DVCS ($\vartheta=1$) limit. The final GPDs read:

\begin{align}
     H^{+}&(x,\vartheta,t)=nr\frac{1+\vartheta}{2}\frac{\vartheta}{\vartheta+x} \\
     &\left(\frac{x+ \frac{x}{\vartheta}}{1+\frac{x}{\vartheta}}\right)^{-\alpha(t)} \left(\frac{1-\vartheta^{2}}{4} + \frac{\vartheta^{2}(1+\vartheta)}{2}\frac{(1-x)}{\vartheta + x} \right)^{b} \nonumber \\
    &\left(\frac{1}{1-\frac{t}{4M^{2}}(1-\vartheta^{2}) - \frac{t}{M^{2}}\frac{\vartheta^{2}(1+\vartheta)}{2}\frac{(1-x)}{x + \vartheta}}\right)^{p}. \nonumber
\end{align}

\begin{figure}[H]
    \centering
    \includegraphics[width=0.45\textwidth]{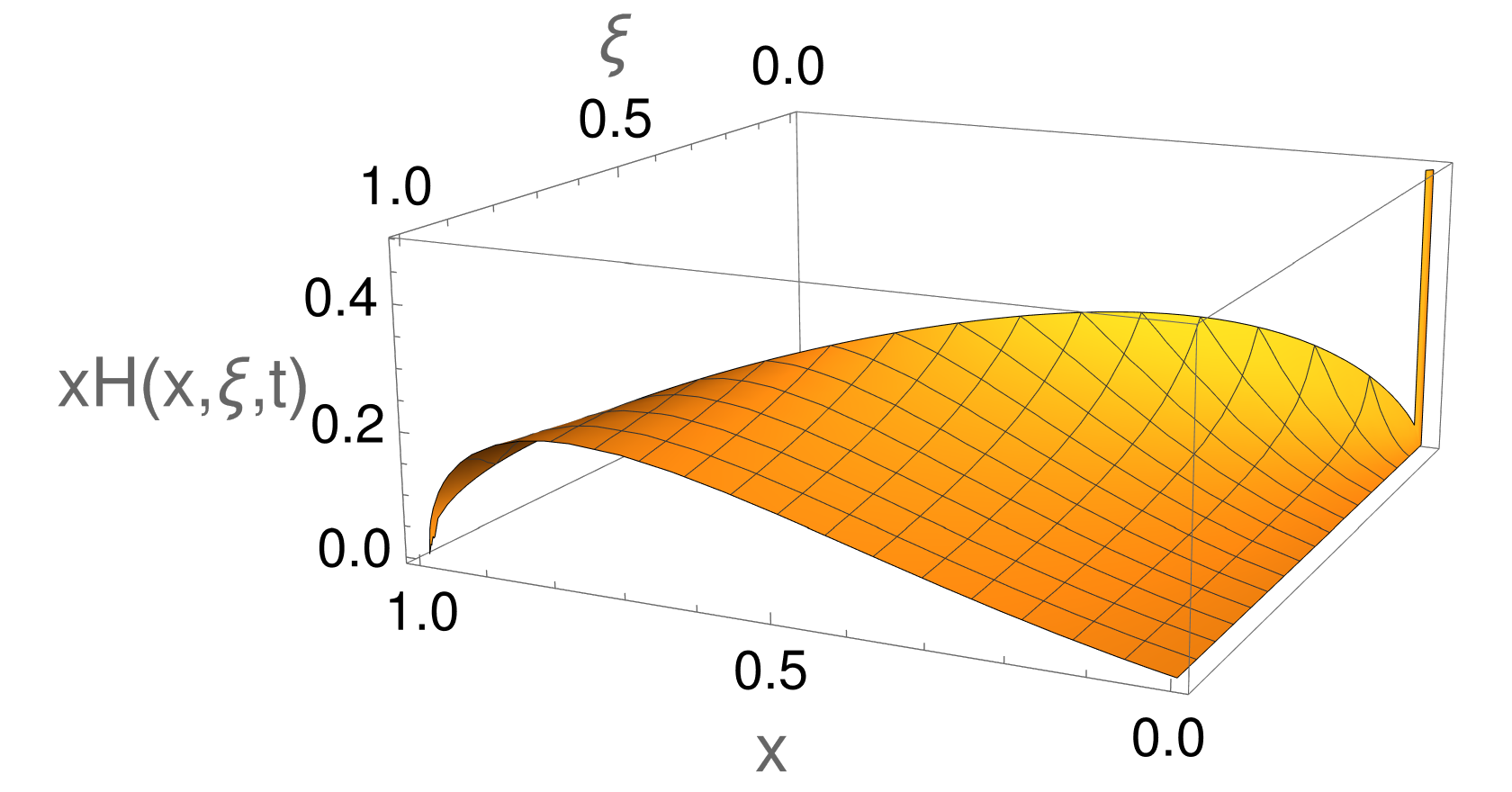}
    \caption{3D contour of $xH(x,\xi,t)$ at $t=-0.15$ GeV$^{2}$.}
    \label{GPD_2D}
\end{figure}

\begin{figure}[H]
    \centering
    \includegraphics[width=0.4\textwidth]{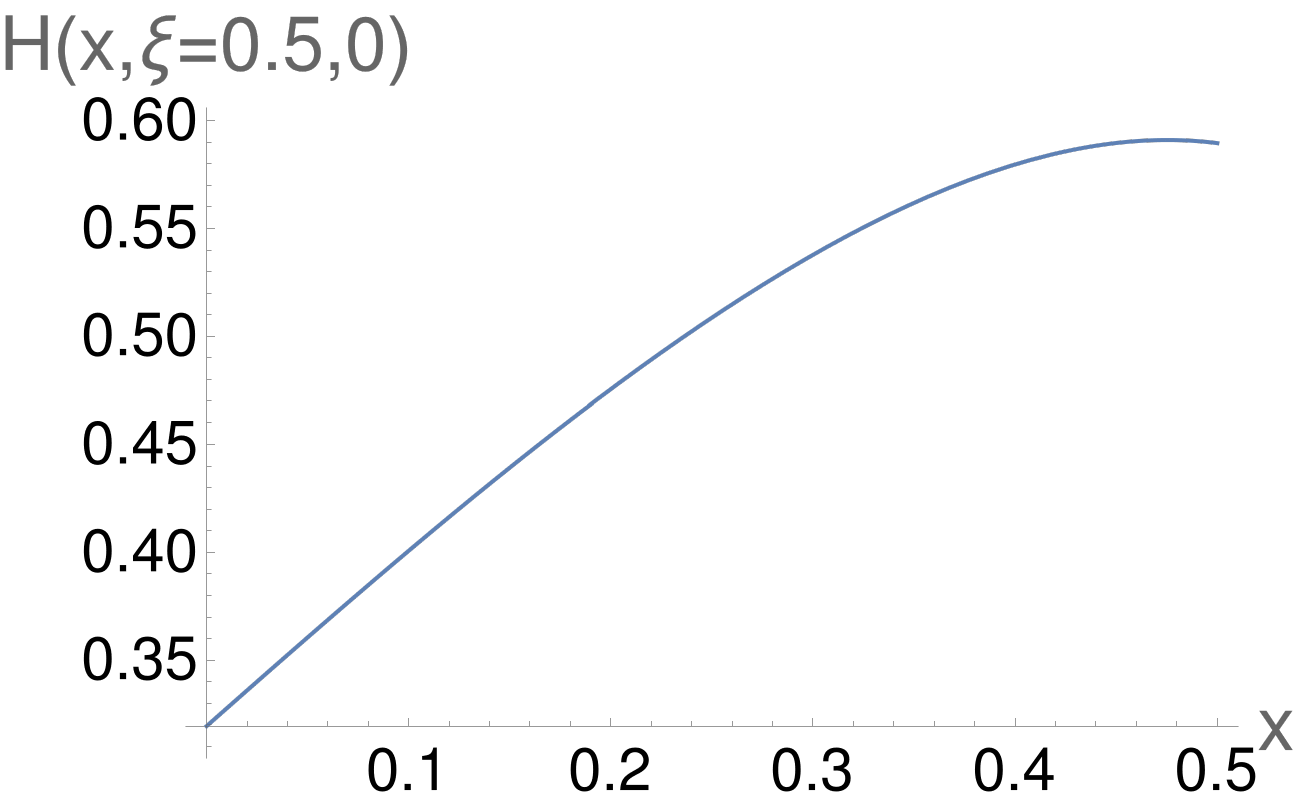}
    \caption{GPD $H$ as a function of $x$ at $t=0$ and $\xi=0.5$.}
    \label{GPD_t0}
\end{figure}

GPD $\widetilde{H}$ is obtained by replacing $nr$ by $\pi \left(2\frac{4}{9} + \frac{1}{9}\right) \widetilde{n}$ together with the respective values of $b$ and $p$ available on \cite{KM1}. We show in Fig. \ref{GPD_2D} a three-dimensional representation of $xH(x,\xi,t)$ at $t=-0.15$ GeV$^{2}$. Besides,  we present Fig. \ref{GPD_t0} the behavior of $H$ at $t=0$ and $\xi=0.5$, not indicating fast oscillations near $x=0$.

\end{document}